\documentclass[10pt,preprint]{aastex}
\usepackage{color}

\newcommand{\av}{$A_V$}

\newcommand{\etal}{et~al.}

\newcommand{\ks}{$K_{\rm s}$}

\newcommand{\mum}{$\mu$m}

\begin{document}

\title{New Young Star Candidates in BRC 27 and BRC 34}

\slugcomment{Accepted to AJ}

\author{L.\ M.\ Rebull\altaffilmark{1},
C.\ H.\ Johnson\altaffilmark{2},
J.\ C.\ Gibbs\altaffilmark{3},
M.\ Linahan\altaffilmark{4},
D.\ Sartore\altaffilmark{5},
R.\ Laher\altaffilmark{1},
M.\ Legassie\altaffilmark{1,6},
J.\ D.\ Armstrong\altaffilmark{7},
L.\ E.\ Allen\altaffilmark{8},
P.\ McGehee\altaffilmark{9}
D.~L.~Padgett\altaffilmark{10}
S.\ Aryal\altaffilmark{3},
K.\ S.\ Badura\altaffilmark{5}, 
T.\ S.\ Canakapalli\altaffilmark{3},
S.~Carlson\altaffilmark{2}, 
M.~Clark\altaffilmark{2}, 
N.~Ezyk\altaffilmark{4},
J.~Fagan\altaffilmark{4},
N.~Killingstad\altaffilmark{2}, 
S.~Koop\altaffilmark{2}, 
T.~McCanna\altaffilmark{2},
M.~M.~Nishida\altaffilmark{3},
T.~R.~Nuthmann\altaffilmark{3},
A.~O'Bryan\altaffilmark{2}, 
A.~Pullinger\altaffilmark{4},
A.~Rameswaram\altaffilmark{4},
T.~Ravelomanantsoa\altaffilmark{2},
H.~Sprow\altaffilmark{4},
C.\ M.\ Tilley\altaffilmark{5}
}

\altaffiltext{1}{Spitzer Science Center/Caltech, M/S 220-6, 1200
E.\ California Blvd., Pasadena, CA  91125
(rebull@ipac.caltech.edu)}
\altaffiltext{2}{Breck School, 123 Ottawa Ave. N.,
Golden Valley, MN 55422 USA}
\altaffiltext{3}{Glencoe High School, 2700 NW Glencoe Rd.,
Hillsboro, OR 97124 USA}
\altaffiltext{4}{Carmel Catholic High School, One Carmel Parkway,
Mundelein, IL 60060}
\altaffiltext{5}{Pine Ridge High School, 926 Howland Blvd., Deltona,
FL 32738 USA}
\altaffiltext{6}{Raytheon Mission Operations and Services, 299 N.
Euclid Ave, Pasadena, CA, 91101 USA}
\altaffiltext{7}{Las Cumbres Observatory Global Telescope (LCOGT) \&
University of Hawaii, HI}
\altaffiltext{8}{National Optical Astronomy Observatory (NOAO), Tucson, AZ}
\altaffiltext{9}{Infrared Processing and Analysis Center (IPAC), M/S
220-6, 1200 E.\ California Blvd., Pasadena, CA  91125}
\altaffiltext{10}{NASA's Goddard Space Flight Center (GSFC), 8800
Greenbelt Rd., Greenbelt, MD, 20771}

\begin{abstract}

We used archival Spitzer Space Telescope mid-infrared data to search
for young stellar objects (YSOs) in the immediate vicinity of two
bright-rimmed clouds, BRC 27 (part of CMa R1) and BRC 34 (part of the
IC 1396 complex). These regions both appear to be actively forming
young stars, perhaps triggered by the proximate OB stars. In BRC 27,
we find clear infrared excesses around 22 of the 26  YSOs or YSO
candidates identified in the literature, and identify 16 new YSO
candidates that appear to have IR excesses.  In BRC 34, the one
literature-identified YSO has an IR excess, and we suggest 13 new YSO
candidates in this region, including a new Class I object. Considering
the entire ensemble, both BRCs are likely of comparable ages, within
the uncertainties of small number statistics and without spectroscopy
to confirm or refute the YSO candidates.  Similarly, no clear
conclusions can yet be drawn about any possible age gradients that may
be present across the BRCs. 

\end{abstract}


\section{Introduction}
\label{sec:intro}

In 1991, Sugitani, Fukui, \& Ogura presented a catalog of
bright-rimmed clouds (BRCs) identified from a comparison of the
Palomar Observatory Sky Survey (POSS) plates and the Infrared
Astronomy Satellite (IRAS) point source catalog (PSC; Beichman \etal\
1998). Sugitani \etal\ (1991) were limited to the northern hemisphere;
Sugitani \& Ogura (1994) continued the analysis (and their BRC numbering
scheme) into the southern hemisphere. These BRCs were selected via a
systematic search of the POSS regions for dark clouds edged by curved,
bright rims, which also coincided with an IRAS source clearly within
the rim, as opposed to on the rim's edge. In order to illuminate the
rim, these BRCs had to be in or around \ion{H}{2} regions. The
``heads'' of the clouds are probably dense portions of the parental
molecular cloud, which have been disturbed by the OB stars powering the
\ion{H}{2} region. Sugitani \etal\ interpreted these BRCs as likely
results of radiation-driven implosions and consequently hosts to
triggered star formation, as in, likely hosts to actively forming
stars. Subsequently, Ogura \etal\ (2002) searched for young stellar
objects (YSOs) around 28 of these BRCs by looking for stars bright in
H$\alpha$ emission. They found many YSOs (or candidate YSOs),
suggesting that, in fact, there was triggered star formation near
these sites, and even pointed to a gradient in star formation, e.g.,
older stars further from the BRC (closer to the OB stars) and younger
stars closer to the BRC.  Similar structures with similar apparent
waves of star formation have been found elsewhere as well (e.g.,
Smith \etal\ 2010).

Surveys in H$\alpha$ in \ion{H}{2} regions, as noted by Ogura \etal\
(2002), can be difficult because of the brightness of the region,  the
distance of the complex, contamination from foreground and background
stars (since most \ion{H}{2} regions are located in or near the
Galactic plane), and, more locally to the BRCs, the high density of
the dust behind the bright rim.  Since it is now commonly believed
that every low-mass star goes through a period of having a
circumstellar disk, young stars can be identified via an infrared (IR)
excess, assumed to be due to circumstellar matter (an envelope
and/or disk). A survey in the IR can be used to identify objects
having an IR excess and thus distinguish candidate young stars from
most foreground or background objects which do not themselves have an
IR excess.   The IR radiation also more easily penetrates the dusty
environs of star-forming regions, particularly dark globules such as
these BRCs.  The Spitzer Space Telescope (Werner \etal\ 2004) has
proven to be a very efficient machine for finding YSO candidates,
particularly around some of these BRCs and their larger environs. For
example, BRC 31 is a small part of the North America Nebula complex
(see, e.g., Rebull \etal\ 2011a and references therein), and BRC 48 is
part of the CG4+Sa101 region in the Gum Nebula complex (see, e.g.,
Rebull \etal\ 2011b and references therein).  

Some of the less-well-known complexes where some BRCs are located were
not observed with Spitzer as extensively as others. L.~E.~Allen,
R.~Gutermuth, G.~Fazio and collaborators initiated a small-field
($\sim5\arcmin\times5\arcmin$), guaranteed time (GTO) survey of most of
the BRCs with the Infrared Array Camera (IRAC; Fazio \etal\ 2004) at
3.6, 4.5, 5.8, and 8 \mum, and with the Multiband Imaging Photometer
for Spitzer (MIPS; Rieke \etal\ 2004) at 24 and 70 \mum. Their goals
were to attempt to locate a statistically unbiased sample of YSOs via
a Spitzer color selection and attempt to better quantify whether or
not the apparent age gradient reported by Ogura \etal\ (2002) could
persist with Spitzer-identified sources. That work is still in
preparation, but as a result of that project, there are at least
small-field observations of several BRCs in the Spitzer Heritage
Archive, even if the entire corresponding complex surrounding each of
these BRCs has not necessarily been observed with Spitzer. For
example, BRC 31 is included in the GTO survey, and is within the North
America Nebula complex; it was re-observed in the context of mapping
$\sim$7 square degrees of the entire complex, and the YSO candidates
in the vicinity of BRC 31 were identified as part of the analysis of
this larger map (Rebull \etal\ 2011a and references therein).  In
contrast, BRC 38 is included in the GTO survey, so therefore small
IRAC and MIPS maps exist (Choudhury \etal\ 2010), but BRC~38 is part
of the rim of the IC~1396 \ion{H}{2} complex, which, most likely due
to its large angular extent, has not been mapped in its entirety by
Spitzer. 

For the present study, we selected two of these relatively unstudied
small fields, BRC 27 and BRC 34, to investigate in detail, looking at
the IR properties of YSOs and YSO candidates identified in the
literature and identifying new YSO candidates from their apparent IR
excess. Our goal was to obtain as complete and reliable a list of YSOs
in our two regions as possible. We obtained ground-based optical
photometric data, combined them with these Spitzer data, and with data
from the near-infrared (NIR) Two-Micron All-Sky Survey (2MASS;
Skrutskie \etal\ 2006).  We used the resulting multi-wavelength
catalog to assemble our list of YSOs from the literature, YSO
candidates from the literature, and new YSO candidates primarily
identified via our mid-IR (MIR) Spitzer data.

BRC 27 is part of the CMa R1 molecular cloud.  The source of the shock
front that triggered the star formation in this region is still uncertain
(Gregorio-Hetem \etal\ 2009). Soares \& Bica (2002, 2003) estimated a
distance of $\sim$1.2 kpc and age $\sim$1.5 Myr. This distance
measurement is consistent with the findings of Shevchenko \etal\
(1999), who placed the distance at 1.05$\pm$0.15 kpc.  

BRC 34 is one of several BRCs located along the rim of the IC 1396
\ion{H}{2} complex, and is relatively unstudied at Spitzer bands or
any other band. It is thought to be at about 800 pc, based on the
distance to the OB stars powering the complex (e.g., Nakano \etal\
2012).

We summarize the details of the literature studies of young stars in
these regions in Section~\ref{sec:litsrcs}, and define some samples
with which we will work through the rest of the paper. Our new
observations and data reduction are described in  Section
\ref{sec:obs}.  We select YSO candidates using Spitzer colors in
Section \ref{sec:findthem}, and discuss their overall properties in
Section \ref{sec:properties}.  Finally, we summarize our main points
in Section \ref{sec:concl}.

\section{Literature Sources}
\label{sec:litsrcs}

We now review in detail the literature for each of our BRCs. To set
the stage for this, first we review briefly the evolution of a YSO,
define some pertinent terms, and establish a star-forming region to
which we will be comparing later in the paper
(Section~\ref{sec:evol}). Then, we describe what we did to resolve
source identifications for each literature catalog for BRC 27
(Section~\ref{sec:litbrc27}) and BRC 34 (Section~\ref{sec:litbrc34}),
with the former being far more complicated than the latter.  We focus
on the region of four-band IRAC coverage in each BRC (see
Section~\ref{sec:obsirac} below). A summary of this section appears in
Section~\ref{sec:litsumm}, where we define the samples of
``(literature-identified) YSOs,'' ``literature candidate YSOs'', and
``new candidate YSOs.''  We note here for completeness that some of
the literature-identified YSOs and candidate YSOs are identified using
wavelengths other than the MIR and as such may not have MIR excesses
suggestive of circumstellar dust. They may, however, still be
legitimate YSOs.  

Cross-identifications, J2000 coordinates, and literature photometry
and spectral types for the literature YSOs and literature YSO
candidates are in Table~\ref{tab:knownysos}.  

\subsection{Context: Definitions and Nomenclature}
\label{sec:evol}

With the explosion of recently available tools, particularly in the
infrared, with which we can study young stars, there has been an
explosion of terminology. Most of the various terms have been
collected in a ``Diskionary'' (Evans \etal\ 2009b), where it is noted
that the same terms used by different teams can have different
meanings. Here we briefly summarize the process of star formation and
the relevant terms as it applies to our discussion here. For this
paper, we will use the term  ``young stellar object'' (YSO) to
encompass all stages of star formation prior to hydrogen burning.

Early studies (e.g., Wilking \etal\ 2001; see also Lada \& Wilking
1984,  Lada 1987,  Greene \etal\ 1994, and Bachiller 1996) of the
low-mass star formation process developed terminology based on the
shape of the observed spectral energy distribution (SED).  The
nomenclature we use here is also tied to the shape of the observed
SED, and is consistent with (if not actually identical to) the
definitions presented in Evans \etal\ (2009b). The earliest stage of
star formation, Class 0, is defined as an object where most of the
energy is being emitted at wavelengths longer than the infrared. The
peak of the SED corresponds roughly to a temperature of $\sim$30 K. At
this stage, there is thought to be a central mass concentration,
entirely embedded within an envelope of gas and dust.  This is also
likely to be the shortest-lived phase; one of the most recent
timescale estimates sets the timescale at $\sim$0.1-0.16 Myr (Evans
\etal\ 2009a).  The next stage of star formation, Class I, is likely
to last $\sim$0.5 Myr (Evans \etal\ 2009a). This phase is again
defined with respect to the shape of the SED -- the slope
of the SED between $\sim$2 and $\sim$20 \mum, $\alpha$, is $\geq$0.3.
In this phase, the energy emitted is still dominated by that from the
envelope, but it is possible to still detect some evidence of a
`photosphere' of the YSO at the shortest bands (see, e.g., Figure 11
in Bachiller 1996).  The next stage, the `Flat' class, arises from the
group of objects whose SED is in transition from an SED with a
positive slope (where the peak of the energy distribution is due to
the circumstellar material) to a negative slope (where the peak of the
energy distribution is due to the YSO photosphere). For these objects,
$-0.3 \leq \alpha < 0.3$. The next phase, Class II objects, may last
$\sim$2 Myr (Evans \etal\ 2009a), and physically corresponds to a
phase in which there is no more circumstellar envelope, but an
optically thick circumstellar disk remains. The SED indicates that
most of the energy comes from the YSO photosphere, though there is
still a substantial contribution from the circumstellar accretion
disk: $-1.6 \leq \alpha < -0.3$. Finally, Class III objects have
little or no excess emission in the infrared due to a circumstellar
disk; for these objects, $\alpha < -1.6$. These objects may have
tenuous dust disks but substantial gas disks from which they are still
accreting; they may have no disk at all, but their youth is suggested
by, e.g., fast rotation, or bright X-ray emission.  Class III objects
cannot be completely identified using only the IR; since so many of
them have little or no disk, other wavelengths must be employed. 

Complications to this scheme include the following. (a) Strictly
speaking, the SED `class' is an entirely empirical definition tied to
the shape of the SED between $\sim$2 and $\sim$20 \mum. The connection
between the SED slope and the physical interpretation of `degree of
embeddedness' is a separate logical step, one replete with
uncertainties such as the inclination of the system. An edge-on Class
II object can resemble a flat or even a Class I object (see, e.g.,
Robitaille \etal\ 2007). In part because of this uncertainty, some
authors (e.g., Smith \etal\ 2010, Evans \etal\ 2009b) have grouped
objects into `stages' rather than `classes,' where `Class I' objects
are often but not always also `Stage I' objects. (b) Older
circumstellar disks may disperse ``inside out'' (e.g., Su \etal\
2006), meaning an inner disk hole begins close to the YSO and widens
outwards, or ``homologously'' (e.g., Currie \& Kenyon 2009), meaning
that the whole disk essentially evenly dissipates at all radii more or
less simultaneously. (c) As protoplanets form in the circumstellar
disk, they will sweep up matter, creating gaps in the disk. They will
also collide, producing a second generation of dust. This second
generation dust disk is a so-called `debris disk'. A late stage disk
broadband SED is not necessarily readily distinguishable from a
primordial disk with a large inner disk hole.  (d) Timescales for all
of these stages are statistical determinations from ensembles of
stars; individual stars may retain or disperse disks at different
rates such that, e.g., Class IIs and IIIs can be found at the same
age, often within close physical proximity ($\sim$0.1 pc; e.g., Rebull
\etal\ 2007). (e) Finally, this evolutionary scheme as described has
been developed for low-mass stars in isolated environments. Brown
dwarfs are likely to follow a similar evolutionary path, just more
slowly (e.g., Apai \etal\ 2005). More massive stars may also follow a
similar path (e.g., Wright \etal\ 2012, Zapata \etal\ 2008), though
faster. Stars embedded within an \ion{H}{2} region, close to O and B
stars, may have their disk ablated away on shorter timescales than if
they were further away from the O and B stars (e.g., Balog \etal\
2007).

In the context of this paper, we will assume that the IR excess we
observe for our YSOs and candidates is in fact due to circumstellar
dust (in a disk or envelope) around the YSO, and we will identify YSO
candidates from that IR excess (Section~\ref{sec:findthem}).  We will
use SED slope fitting between 2 and 24 \mum\ to place our objects in
bins of Class 0, I, flat, II, and III (Section~\ref{sec:sedslopes}).
We will use relative fractions of objects in these bins as a very
rough proxy for age. Our targets all likely possess primordial, rather
than debris, disks, though follow-up observations are needed to
determine this. We are likely to have detected YSOs as massive as B
stars (see known B star listed in Table~\ref{tab:knownysos}) down to
possibly proto-brown dwarfs; we need follow-up spectroscopy to obtain
spectral types for all but 2 objects in our sample. All of these stars
are near or within an \ion{H}{2} region; they have not formed in
isolation, but they are likely at least $\sim$15 pc from the OB star
cluster powering the \ion{H}{2} region. The YSOs in each BRC all
formed within $\sim$1.2-1.5 pc, given our region of interest
(5$\arcmin$ on a side), and distance estimates to our BRCs (800-1000
pc).

Throughout this paper, we make comparisons of BRC 27 and 34 to another
BRC -- BRC 48 is identified as part of the CG4+Sa101 region in the Gum
Nebula complex (Rebull \etal\ 2011b). It is taken to be between 300
and 500 pc away.  The CG4 portion (cometary globule 4) is the region
formally identified as the main portion of BRC 48 (Sugitani \& Ogura
1994); the Sa101 portion is slightly further back from the rim of the
globule, and appears to have been shadowed, at least partially, by CG4
from the ionization front. The two regions are often analyzed together
as one region: CG4+Sa101.  There are several reasons we have selected
this region for comparison rather than any other star-forming region.
First, the fact that CG4+Sa101 is also a BRC suggests that its
formation mechanism is similar to that of BRC 27 and 34; all three of
these regions are part of \ion{H}{2} complexes with nearby O and B
stars disturbing the gas and dust in the parent molecular cloud.
Triggered star formation could thus be occurring in any of these BRCs.
Moreover, the age distribution may be roughly comparable in each of
them, just because they are morphologically similar. Star formation
in, for example, the Taurus Molecular Cloud, would not be a good
physical comparison, since Taurus does not host an \ion{H}{2} region. 
Second, the Spitzer-selected YSO candidates in CG4+Sa101 were selected
and analyzed in a very similar fashion to the YSO candidates selected
here; in all three BRCs, we primarily use Spitzer to select YSO
candidates, with additional information used from NIR $JHK_s$ and
optical photometry. Also, in all three regions, there have been some
efforts in the literature to identify YSOs using a variety of
wavelengths. While it is true that other star forming regions (like
Taurus) have more of the follow-up spectroscopy needed to confirm
youth and that the YSO candidates identified in CG4+Sa101 are still
candidates, the fact that the selection mechanism is very similar
between the regions suggests that any systematics between regions due
to the selection mechanism are minimized, and that contamination rates
may be comparable. Third, the IRAC and MIPS maps in the CG4+Sa101 region
are not very large. They cover $\sim$0.5 square degrees, which is
large compared to the $\sim$25 square arcminutes in each of the BRCs
analyzed here, but they are small compared to Spitzer maps of, say,
Taurus ($\sim$44 square degrees; Rebull \etal\ 2010) or the North
America Nebula ($\sim$7 square degrees; Rebull \etal\ 2011a). Even
Serpens, one of the smaller maps of star forming regions obtained by
one of the Spitzer Legacy teams, has a $\sim$0.9 square degree IRAC
map (Harvey \etal\ 2006). Thus, while the CG4+Sa101 region maps are
larger than the BRC maps we consider here, they are still closer in
angular size to our BRCs than many other Spitzer maps of star-forming
regions, and thus the overall star count should be somewhat
comparable. While there are about half a million point sources in both
the Taurus and North America Nebula maps, there are ``only'' several
thousand in CG4+Sa101, to be compared with several hundred in the BRCs
under consideration here. There are also $\sim$25 YSO candidates found
in CG4+Sa101, so roughly comparable to the ``yield'' of YSOs found here.
As for BRC 27 and 34, the sample of literature YSOs for CG4+Sa101
consists both of high-confidence YSOs and candidate YSOs (see Rebull
\etal\ 2011b and references therein).  In terms of distance, however,
CG4+Sa101 is likely less than $\sim$half the distance to BRC 27 or BRC
34; we are, as a result, more likely to find lower-mass YSO candidates
in the closer CG4+Sa101 than the further BRC 27 or BRC 34.

\begin{deluxetable}{rlllcccccccc}
\tablecaption{YSOs and YSO candidates from the literature in 
BRC 27 and BRC 34\tablenotemark{a}\label{tab:knownysos}}
\rotate
\tabletypesize{\tiny}
\tablewidth{0pt}
\tablehead{
\colhead{prior name}  &  
\colhead{why identified as YSO\tablenotemark{b}}& 
\colhead{current status\tablenotemark{c}}&  
\colhead{catalog name\tablenotemark{d}}&  
\colhead{row\tablenotemark{e}} &
\colhead{Position (J2000)} & \colhead{$U$ (mag)}& \colhead{$B$ (mag)} 
& \colhead{$V$ (mag)} & \colhead{$R$ (mag)} &  \colhead{$I$ (mag)} &
SpTy }
\startdata
\cutinhead{BRC 27}
               Chauhan109&                                    NIR excess&     lit.~YSO~cand.& 070352.2-112100& 1&  07 03 52.3 -11 21 01&        \nodata&15.62$\pm$ 0.06&14.55$\pm$ 0.07&        \nodata&13.79$\pm$ 0.05& \nodata\\
         Ogura2,Chauhan81&     H$\alpha$ emission ($<$10\AA), NIR excess&     lit.~YSO~cand.& 070352.7-112313& 2&  07 03 52.7 -11 23 13&        \nodata&18.95$\pm$ 0.02&17.47$\pm$ 0.01&        \nodata&15.31$\pm$ 0.03& \nodata\\
                   Ogura3&                 H$\alpha$ emission ($<$10\AA)&     lit.~YSO~cand.& 070353.2-112403& 3&  07 03 53.2 -11 24 04&        \nodata&        \nodata&        \nodata&        \nodata&        \nodata& \nodata\\
             Shevchenko90&                                    early type&                YSO& 070353.5-112350& 4&  07 03 53.5 -11 23 51&          10.97&          10.97&          10.89&          10.78&        \nodata&   A0   \\
         Ogura4,Chauhan82&     H$\alpha$ emission ($<$10\AA), NIR excess&     lit.~YSO~cand.& 070353.7-112428& 5&  07 03 53.7 -11 24 29&        \nodata&        \nodata&20.02$\pm$ 0.01&        \nodata&16.76$\pm$ 0.00& \nodata\\
               Chauhan108&                                    NIR excess&     lit.~YSO~cand.& 070354.6-112011& 7&  07 03 54.7 -11 20 11&        \nodata&15.87$\pm$ 0.07&14.95$\pm$ 0.08&        \nodata&14.39$\pm$ 0.07& \nodata\\
         Ogura5,Chauhan94&     H$\alpha$ emission ($<$10\AA), NIR excess&     lit.~YSO~cand.& 070354.9-112514& 8&  07 03 55.0 -11 25 15&        \nodata&20.35$\pm$ 0.05&18.77$\pm$ 0.00&        \nodata&16.15$\pm$ 0.01& \nodata\\
         Ogura7,Chauhan83&     H$\alpha$ emission ($<$10\AA), NIR excess&     lit.~YSO~cand.& 070357.1-112432& 9&  07 03 57.1 -11 24 33&        \nodata&20.76$\pm$ 0.07&19.14$\pm$ 0.00&        \nodata&16.48$\pm$ 0.00& \nodata\\
             Chauhan-anon&                                    MIR excess&     lit.~YSO~cand.& 070401.2-112233&14&  07 04 01.3 -11 22 33&        \nodata&        \nodata&        \nodata&        \nodata&        \nodata& \nodata\\
  Gregorio74,Chauhan-anon&         ROSAT X-ray detection, NIR+MIR excess&                YSO& 070401.3-112334&15&  07 04 01.4 -11 23 35&        \nodata&        \nodata&        \nodata&          12.60&        \nodata& \nodata\\
  Shevchenko99,Gregorio75&                             \tablenotemark{f}&                YSO& 070402.3-112539&20&  07 04 02.3 -11 25 39&          10.23&          10.60&          10.45&          10.80&        \nodata&   B3-5 \\
       Ogura8+9,Chauhan84&     H$\alpha$ emission ($<$10\AA), NIR excess&     lit.~YSO~cand.& 070402.9-112337&22&  07 04 02.9 -11 23 38&        \nodata&20.68$\pm$ 0.09&19.01$\pm$ 0.01&        \nodata&16.33$\pm$ 0.01& \nodata\\
        Ogura10,Chauhan85&     H$\alpha$ emission ($>$10\AA), NIR excess&                YSO& 070403.0-112350&23&  07 04 03.1 -11 23 50&        \nodata&        \nodata&20.18$\pm$ 0.01&        \nodata&17.40$\pm$ 0.00& \nodata\\
               Chauhan107&                                    NIR excess&     lit.~YSO~cand.& 070403.1-112327&24&  07 04 03.1 -11 23 28&        \nodata&12.92$\pm$ 0.04&11.53$\pm$ 0.04&        \nodata&10.71$\pm$ 0.03& \nodata\\
            Shevchenko102&    $E(B-V)>$0.16, coincident with IRAS source&     lit.~YSO~cand.& 070403.9-112609&25&  07 04 03.9 -11 26 10&           9.45&          10.05&           9.93&           9.78&        \nodata&        \\
        Ogura12,Chauhan86&     H$\alpha$ emission ($>$10\AA), NIR excess&                YSO& 070404.2-112355&27&  07 04 04.3 -11 23 56&        \nodata&20.90$\pm$ 0.07&19.62$\pm$ 0.01&        \nodata&16.72$\pm$ 0.00& \nodata\\
                  Ogura13&                 H$\alpha$ emission ($<$10\AA)&     lit.~YSO~cand.& 070404.5-112555&28&  07 04 04.6 -11 25 55&        \nodata&        \nodata&        \nodata&        \nodata&        \nodata& \nodata\\
        Ogura14,Chauhan87&     H$\alpha$ emission ($<$10\AA), NIR excess&     lit.~YSO~cand.& 070404.7-112339&29&  07 04 04.7 -11 23 40&        \nodata&20.03$\pm$ 0.04&18.32$\pm$ 0.00&        \nodata&15.97$\pm$ 0.00& \nodata\\
        Ogura15,Chauhan88&     H$\alpha$ emission ($>$10\AA), NIR excess&                YSO& 070405.1-112313&30&  07 04 05.2 -11 23 13&        \nodata&20.51$\pm$ 0.05&19.09$\pm$ 0.00&        \nodata&16.55$\pm$ 0.00& \nodata\\
        Ogura16,Chauhan89&     H$\alpha$ emission ($<$10\AA), NIR excess&     lit.~YSO~cand.& 070405.9-112358&32&  07 04 05.9 -11 23 59&        \nodata&19.82$\pm$ 0.03&18.22$\pm$ 0.01&        \nodata&15.93$\pm$ 0.00& \nodata\\
        Ogura17,Chauhan90&     H$\alpha$ emission ($<$10\AA), NIR excess&     lit.~YSO~cand.& 070406.0-112315&34&  07 04 06.0 -11 23 16&        \nodata&        \nodata&20.05$\pm$ 0.01&        \nodata&17.31$\pm$ 0.00& \nodata\\
        Ogura18,Chauhan91&     H$\alpha$ emission ($>$10\AA), NIR excess&                YSO& 070406.4-112336&35&  07 04 06.4 -11 23 36&        \nodata&        \nodata&20.58$\pm$ 0.01&        \nodata&16.84$\pm$ 0.00& \nodata\\
        Ogura19,Chauhan92&     H$\alpha$ emission ($<$10\AA), NIR excess&     lit.~YSO~cand.& 070406.5-112316&38&  07 04 06.6 -11 23 16&        \nodata&19.70$\pm$ 0.03&18.08$\pm$ 0.00&        \nodata&15.74$\pm$ 0.00& \nodata\\
                  Ogura21&                 H$\alpha$ emission ($<$10\AA)&     lit.~YSO~cand.& 070407.9-112311&39&  07 04 08.0 -11 23 11&        \nodata&        \nodata&        \nodata&        \nodata&        \nodata& \nodata\\
        Ogura22,Chauhan97&     H$\alpha$ emission ($>$10\AA), NIR excess&                YSO& 070408.0-112354&40&  07 04 08.0 -11 23 55&        \nodata&17.17$\pm$ 0.01&15.95$\pm$ 0.00&        \nodata&14.35$\pm$ 0.00& \nodata\\
        Ogura23,Chauhan98&    H$\alpha$ emission ($>>$10\AA), NIR excess&                YSO& 070408.1-112309&42&  07 04 08.2 -11 23 10&        \nodata&21.78$\pm$ 0.14&20.34$\pm$ 0.01&        \nodata&17.41$\pm$ 0.00& \nodata\\
\cutinhead{BRC 34}
          Ogura1,Nakano17&   H$\alpha$ emission ($>$10\AA) at two epochs&                YSO& 213329.2+580250&48&  21 33 29.2 +58 02 51&        \nodata&        \nodata&        \nodata&        \nodata&        \nodata& \nodata\\
\enddata
\tablenotetext{a}{Information tabulated here comes largely from the
literature, as described in the text, with positions updated to be
J2000 and tied to the Spitzer and 2MASS coordinate system.  If not
specified, we assumed the errors on the photometry to be $\sim$20\%
when plotting them in the SEDs
(Figures~\ref{fig:seds1}--\ref{fig:seds7}.}
\tablenotetext{b}{This column notes why this object was identified in
the literature as a possible YSO.}
\tablenotetext{c}{This column notes whether we regard this object as a
fairly high-confidence literature YSO, or still a
(literature-identified) candidate YSO, awaiting follow-up
spectroscopy.}
\tablenotetext{d}{This column lists the IAU-compliant position-based
catalog name, used throughout the rest of the paper.}
\tablenotetext{e}{This column lists the row number from
Table~\ref{tab:ourysos} and Table~\ref{tab:ourysonotes}, used
throughout the rest of the paper.}
\tablenotetext{f}{early type, $E(B-V)>$0.16, coincident with IRAS
source, ROSAT and XMM X-ray detection}
\end{deluxetable}

\subsection{BRC 27}
\label{sec:litbrc27}

There are five prior studies of note of BRC 27 YSOs in the literature.
There are a total of 26 unique objects identified as YSOs or YSO
candidates in the literature in our region of interest in BRC 27,
which we now discuss.

Wiramihardja \etal\ (1986) used the Kiso Observatory Schmidt telescope
to survey $\sim$58 square degrees for stars bright in H$\alpha$ in the
vicinity of the CMa R1 association; see their Figure~1 for an
indication of the region they observed. They also obtained
photographic $UBV$ for some objects. They obtained objective prism
observations, and that, combined with a Q-value analysis (Johnson
1958) of their broadband photometry, yielded coarse spectral types for
the brightest, earliest-type stars. They report coordinates of their
targets in 1950 coordinates, as determined off their photographic
plates. For each of the targets in the vicinity of our region of
interest, we examined 2MASS images near the same location on the sky,
taking the nearest bright object as the best possible updated
coordinates for the object in question. There are only two objects
from Wiramihardja \etal\ (1986) close to our region of interest. One
was their number 23, whose coordinates we have updated to 
07:04:09.95, $-$11:23:16.4, and identified it as also Ogura 25 and
Chauhan 100, and it is on the edge of our region of interest such that
photometry is not likely to be reliable. Number 22 from Wiramihardja
\etal\ (1986) (also number 162 from Shevchenko \etal\ 1999 and number
20 from Ogura \etal\ 2002) is also just off the edge of the IRAC
observations (see Section \ref{sec:obs} below), such that the
point-spread-function (PSF) wings from a big, bright source can be
seen in the dithers closest to that object. 

Shevchenko \etal\ (1999) obtained photoelectric $UBVR$ photometry and
objective prism spectroscopy of several stars over $\sim$4 square
degrees of CMa R1, additionally comparing their results with the IRAS
catalog to check for bright infrared emission in the region. They also
report 1950 coordinates for their targets based on their photographic
plates, but a finding chart is provided. Again, for each of the
objects in our region of interest, we examined 2MASS images of the
immediate vicinity, taking the nearest bright 2MASS sources as the
correct match, comparing to the provided finding chart in any
confusing cases. Three objects from this Shevchenko \etal\ paper are
in our region of interest observed with IRAC: 90, 99, and 102. Number
90 has two possible 2MASS counterparts, where the slightly more
distant one is brighter. However, the assembled SED, when the optical
data from Shevchenko \etal\ is merged with 2MASS+IRAC (see
\S\ref{sec:bandmerging} below), makes it more likely that the closer
one is, in fact, the true match. Those coordinates are reported in 
Table~\ref{tab:knownysos}. Similarly, number 99 has two possible 2MASS
matches, but a match to the brighter, closer one provides a better SED
and is most likely the true match. Shevchenko \etal\ (1999) report
spectral type estimates for two of these three objects; Shevchenko 90
is an A0, and Shevchenko 99 is reported to be B3-5.  

Sugitani \etal\ (1995), using $JHK$, identify a cluster of young stars
approximately on the bright rim of this BRC, but do not list individual
sources in that paper. It is the same apparent cluster that we
rediscover in Section \ref{sec:locationonthesky} below; by comparison
of star patterns with Figure 3 from Sugitani \etal\ (1995), we have
not identified all of the same objects, but many of them are in
common. No spectroscopic follow-up was reported in Sugitani \etal\
(1995). 

Ogura \etal\ (2002) report on sources bright in H$\alpha$ detected via
a wide field grism spectrograph. They report J2000 coordinates, and
they provide finding charts. As above, for each of the objects in our
region of interest, we examined 2MASS images of the immediate
vicinity, taking the nearest bright 2MASS sources as the correct
match, comparing to the provided finding chart in any confusing cases.
There were, in fact, several confusing cases. Ogura \etal\ (2002)
report two sources very close together, their number 8 and 9. 2MASS
and IRAC do not resolve this source, though the 2MASS source is
slightly extended in the direction expected from the Ogura \etal\
finding charts. We report the net flux from both these objects as tied
to ``Ogura 8+9'' in Table~\ref{tab:knownysos}.  Given the finding
chart from Ogura \etal, numbers 21 and 23 are close to each other, and
both just north of a third, brighter source. In the tabulated list of
coordinates, Ogura \etal\ cite the coordinates of 21 and 23 as
uncertain. 2MASS and IRAC are both able to successfully identify all
three objects as unique sources. Nineteen sources from Ogura \etal\
appear in our region of interest, and also in
Table~\ref{tab:knownysos}. Ogura \etal\ (2002) report H$\alpha$
equivalent widths based on their grism observations. However,
measurements were not possible for all of the objects, and moreover, M
stars that are not young stars but possess typical levels of activity
for M stars can also have H$\alpha$ in emission. Many investigators
have reported estimates of dividing lines between just an active star
and a star actively accreting (e.g., Slesnick \etal\ 2008, Barrado y
Navascu\'es \& Mart\'in 2003). Such a limit was not imposed in Ogura
\etal\ (2002), who may also have been effectively (due to the relative
depths of their survey) considering only types earlier than M. No
spectral classifications are reported by Ogura \etal\ (2002). Six of
the BRC 27 objects in our region of interest have unambiguous
H$\alpha$ equivalent widths $>$10\AA.  Despite the lack of
classification spectroscopy, we suspect that most of these with
equivalent width of H$\alpha>$10\AA\ are likely legitimate young
stars.  (All of these also turn out to have a MIR excess -- see
Section~\ref{sec:findthem} and Table~\ref{tab:ourysonotes}).

Using an early release of the 2MASS catalog ($JHK_s$), Soares \& Bica
(2002, 2003) identified YSO candidates in the region we consider here,
but did not report them in a table.  They used these objects to
determine a distance of $\sim$1.2 kpc and age $\sim$1.5 Myr.

Gregorio-Hetem \etal\ (2009) used Roentgen Satellite (ROSAT) Position
Sensitive Proportional Counters (PSPC) images, followed by X-ray
Multi-Mirror Mission (XMM-Newton) and Chandra X-ray Observatory (CXO)
data where possible, United States Naval Observatory (USNO) $R$, 2MASS
$JHK_s$, and new $VRI$ data to search for YSOs in a $\sim$5 square
degree region of the CMa R1 region. They report fairly high accuracy
J2000 coordinates; we had no issues in finding counterparts in our
images for their objects. Their numbers 74 and 75 both appear in our
region of interest.  These objects are relatively bright; the 2MASS
$JHK_s$ photometry for their number 74 is flagged as bad using the
2MASS photometric quality flags. However, it seems quite consistent
with the rest of the SED as obtained below (see \S\ref{sec:seds}), so
we retained it, albeit with larger errors. No spectroscopic follow-up
was reported in this paper. However, source number 74 is identified as
having an $H-K$ excess, as well as an X-ray detection; number 75 is
identified as just having an X-ray detection, but with ROSAT as well
as XMM. Despite the lack of classification spectroscopy, we strongly
suspect that these are most likely legitimate young stars.  (Both of
these sources also turn out to have a MIR excess -- see
Section~\ref{sec:findthem} and Table~\ref{tab:ourysonotes}). 

Chauhan \etal\ (2009) studied BRC 27 with new $BRI_c$ photometry
combined with 2MASS $JHK_s$ and archival IRAC observations (the same
IRAC data set as we are using for BRC 27).  Chauhan \etal\  identify
YSO candidates, first using NIR color-color diagrams, then using MIR
color-color diagrams to classify YSOs. Their YSO candidate sources
appear numbered in their Table~4 with $BVI_c$ magnitudes and
unnumbered (but with RA/Dec) in their Table~6 with IRAC magnitudes.
There are three sources that appear in the BRC 27 tables for IRAC that
do not appear in the source list for $JHK_s$. They may have been
identified by the Allen \etal\ (2004) method for selecting YSO
candidates (see \S\ref{sec:findthem} below). We have identified them
as ``Chauhan-anon'' in our catalog, and there are two such sources in
our region of interest.  The 2MASS counterparts (with 2MASS
coordinates) are listed in their Table~3. We took the 2MASS
coordinates as ``truth''; the IRAC coordinates are tied to the 2MASS
coordinate system, so they should match within an arcsecond of the
2MASS coordinate system. Including the two orphan objects, there are
21 sources from Chauhan \etal\ in our region of interest. For Chauhan
107, extended emission can be seen in the 2MASS image; Chauhan 108 and
109 both are faint and possibly marginally extended in the 2MASS
image. We examined these sources in detail because, for all three of
these sources, the 2MASS portion of the SED is inconsistent with the
Chauhan \etal\ $BVI_c$, but when using our optical data (see Sections
\ref{sec:opticaldata} and \ref{sec:seds}), the SED seems consistent
with the 2MASS photometry, so the 2MASS photometry is most likely
correct. No spectroscopic follow-up was reported in Chauhan \etal\
(2009).

\subsection{BRC 34}
\label{sec:litbrc34}

This region has much less discussion in the literature than BRC 27. 
The only survey for YSOs in BRC 34 that we identified before beginning
our work was Ogura \etal\ (2002), which identified two
H$\alpha$-bright sources in this vicinity. Concurrently with our work,
two studies were published searching for H$\alpha$ emission-line stars
in the entire complex, Barentsen \etal\ (2011), which used the
$r^{\prime} i^{\prime} H\alpha$ bands from the Isaac Newton Telescope
(INT) Photometric H-Alpha Survey (IPHAS), and Nakano \etal\ (2012),
which used slitless grism spectroscopy, also using primarily H$\alpha$
with $i^{\prime}$.

The two sources listed in Ogura \etal\ (2002) as having bright
H$\alpha$ were numbered 1 (reported at position 21 33 29.4,  +58 02 50
in J2000 coordinates, and having an H$\alpha$ width of $\sim$12\AA)
and 2 (reported at position 21 33 55.8,  +58 01 18, also in J2000
coordinates). Source 1 is very close to 2MASS source 21332921+5802508,
and was also independently recovered by Nakano \etal\ (2012), also in
H$\alpha$ (with an equivalent width $>$10 \AA), though no $i^{\prime}$
magnitude was reported for it. We have identified that object as the
2MASS source, and it falls within the perimeters of our IRAC data.  We
suspect that this is a legitimate young star.

The position of Ogura source 2, however, has no counterpart in 2MASS
within $\sim$10$\arcsec$, and does not appear to have been recovered
by Nakano \etal\ (2012). While young stars are well known to be
variable in H$\alpha$, as well as $JHK_s$, it seems unlikely that a
young star would be bright enough to be detected by Ogura \etal\
(2002) in H$\alpha$ in the late 1990s, but too faint for 2MASS
(limiting magnitude $J\sim16$, $K_s\sim14$), also obtained in the late
1990s. We suspect that source 2 may not be recoverable. 

Barentsen \etal\ (2011) did not identify any objects in our region of
interest.

\subsection{Summary: On the Reliability of These Literature YSOs}
\label{sec:litsumm}

Many (26) objects in our region of BRC 27 are identified in the
literature as YSOs (or candidate YSOs), and only one object in our
region of BRC 34 is identified in the literature as a YSO. Some of
these objects are identified as young stars from optical wavelengths
(H$\alpha$, $UBVR$), or from X-rays, and some use the near-IR ($JHK_s$
from 2MASS).  Some of these objects are identified as young stars from
the same mid-IR data we are also using here to identify young stars --
independently identified, using different methods, but still using the
same data.  Each of the objects listed in Table~\ref{tab:knownysos}
also has notes about why that object was identified in the literature
as a YSO. 

Very few of these literature YSOs have spectroscopy, at least at
classification resolution, to obtain a reliable spectral type and
distinguish them from active M stars (which would also have bright
H$\alpha$ and bright X-rays) or other foreground/background stars.
Just two of the literature-identified YSOs have spectral types; these
are from objective prism observations (Shevchenko \etal\ 1999).   The
objects we identify below (\S\ref{sec:findthem}) as possible YSOs, of
course, also need follow-up spectroscopy. Ideally, such a spectrum
(for either the newly identified or literature-identified YSO
candidates) would also obtain a measure of H$\alpha$ as an indicator
of active accretion. However, in most cases, we are still lacking such
a spectrum.

Thus, we have identified three categories of YSOs: (1) very likely
YSOs identified in the literature, which we refer to as ``known
YSOs,'' or ``likely YSOs,'' or simply ``YSOs''; (2) YSO candidates
identified in the literature, awaiting follow-up spectroscopy, which
we refer to as ``literature YSO candidates'' or ``literature
candidates''; (3) YSO candidates newly identified here, awaiting
follow-up spectroscopy, which we refer to as ``our YSO candidates'' or
``our candidates.''   

The objects we have tagged as likely YSOs are listed as such in
Table~\ref{tab:knownysos}. They are identified as likely YSOs because
they have been identified as an early spectral type (Shevchenko 90 and
99, or rows 4 and 20 in Table~\ref{tab:knownysos}), detected in X-rays
(Gregorio 74 and 75, or rows 15 and 20 in Table~\ref{tab:knownysos}), or
have H$\alpha$ equivalent widths measured to be $>$10\AA\ (Ogura 10,
12, 15, 18, 22, and 23 from BRC 27, and Ogura 1 from BRC 34, or rows
23, 27, 30, 35, 40, 42, and 48 in Table~\ref{tab:knownysos}).  All the
other objects identified either in the literature or in the present
paper need follow-up spectroscopy to confirm their youth.

\section{New Observations, Data Reduction, and Ancillary Data}
\label{sec:obs}

In this Section, we discuss the IRAC, MIPS, and optical data
acquisition and reduction.   We also discuss merging the photometric
data between IRAC and MIPS, with the 2MASS near-IR catalog ($JHK_s$),
with the optical data, and with the literature. 

We note for completeness that the four channels of IRAC are 3.6, 4.5,
5.8, and 8 \mum, and that the three channels of MIPS are 24, 70, and
160 \mum. These bands can be referred to equivalently by their channel
number or wavelength; the bracket notation, e.g., [24], denotes the
measurement in magnitudes rather than flux density units (e.g., Jy).
Further discussion of the bandpasses can be found in, among other
places, the Instrument Handbooks, available from the Spitzer Science
Center (SSC) or the Infrared Science Archive (IRSA) Spitzer Heritage
Archive (SHA)  websites\footnote{http://ssc.spitzer.caltech.edu/ ,
http://irsa.ipac.caltech.edu/data/SPITZER/docs/ ,
http://sha.ipac.caltech.edu/applications/Spitzer/SHA/ }.

\subsection{IRAC Data}
\label{sec:obsirac}

\begin{figure*}[tbp]
\epsscale{0.8}
\plotone{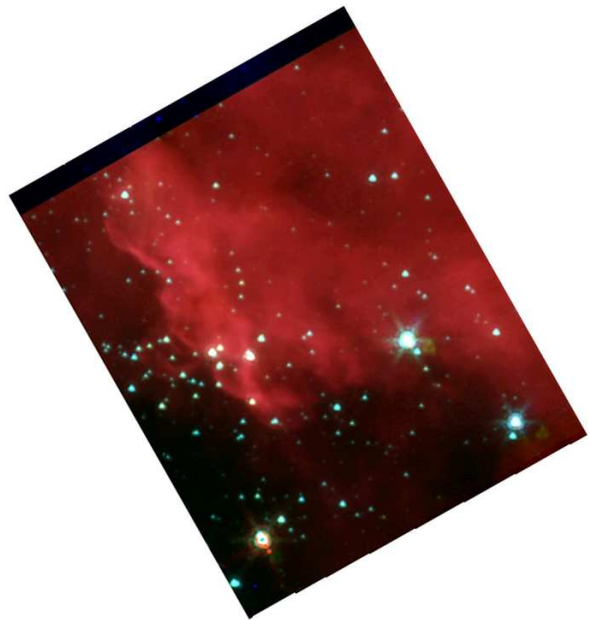}
\caption{BRC 27, in the region of four-band IRAC coverage: IRAC-1 (3.6
\mum; blue), IRAC-2 (4.5 \mum; green), and IRAC-4 (8 \mum; red).   The
pointing is a single dithered IRAC field of view, so about 5$\arcmin$ on a
side. North is up. Both nebulosity and point sources can be seen.}
\label{fig:brc27-irac}
\end{figure*}

\begin{figure*}[tbp]
\epsscale{0.8}
\plotone{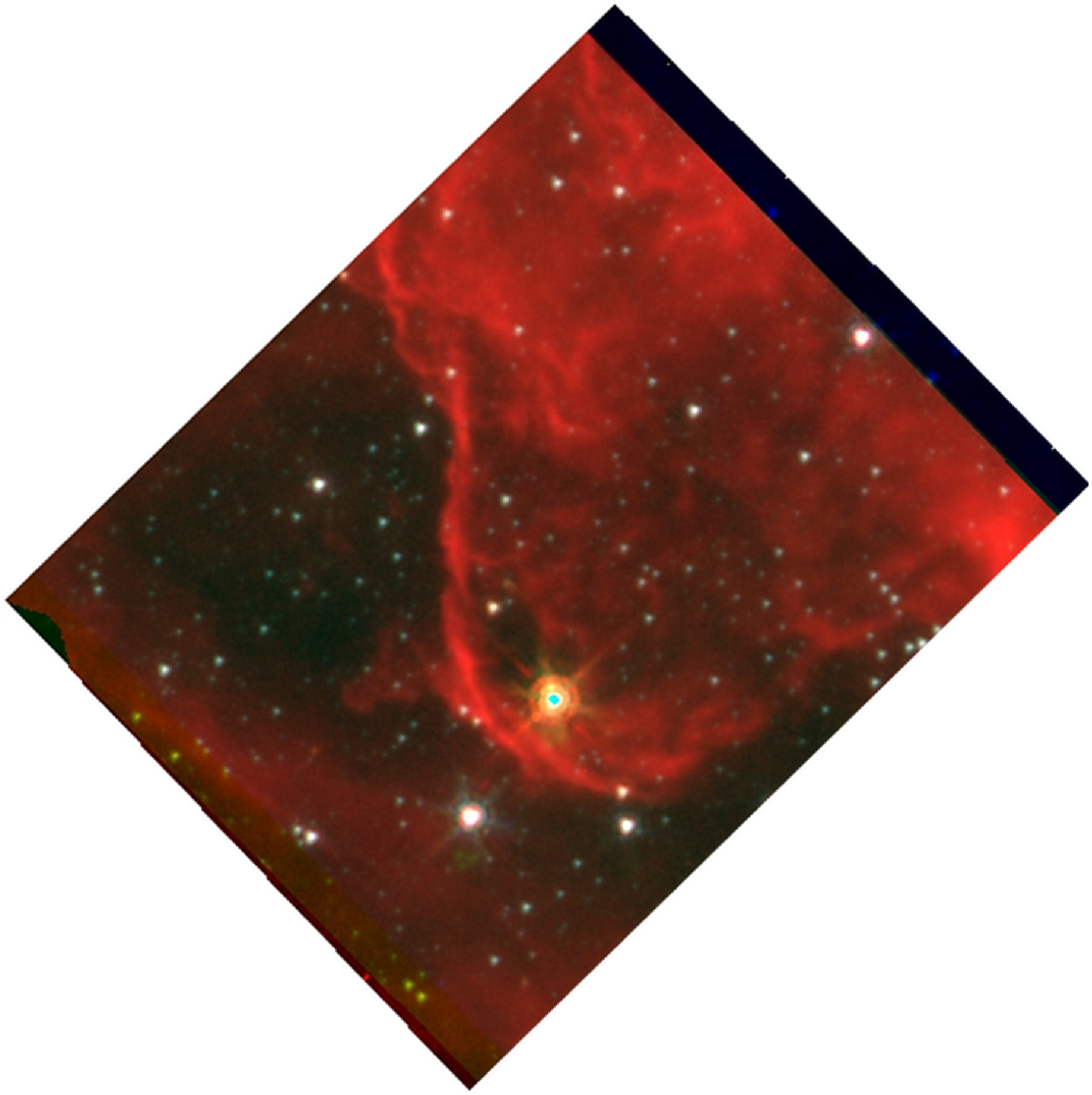}
\caption{BRC 34, in the region of four-band IRAC coverage: IRAC-1 (3.6
\mum; blue), IRAC-2 (4.5 \mum; green), and IRAC-4 (8 \mum; red). The
pointing is a single dithered IRAC field of view, so about 5$\arcmin$ on a
side. North is up. Both nebulosity and point sources can be seen.}
\label{fig:brc34-irac}
\end{figure*}

We used the IRAC data for BRC 27 from Spitzer program 30050,
AORKEY\footnote{An AOR is an Astronomical Observation Request, the
fundamental unit of Spitzer observing. An AORKEY is the unique 8-digit
integer identifier for the AOR, which can be used to retrieve these
data from the Spitzer Archive.} 17512192; for BRC 34, we used data
from Spitzer program 202, AORKEY 6031616. The BRC 27 data were
obtained on 2006-11-22, and were centered on $\alpha$=07:03:59,
$\delta$=$-$11:23:09 (J2000); the BRC 34 data were obtained on
2004-07-04 and were centered on $\alpha$=21:33:32,
$\delta$=+58:16:12.8 (J2000).  Both AORs were 12 sec
high-dynamic-range (HDR) frames, so there are two exposures at each
pointing, 0.6 and 12 s, with 5 small-scale dithers per position, for a
total integration time of 60 s (on average).

We started with the corrected basic calibrated data (CBCDs) processed
using SSC pipeline version 18.18.  We reprocessed the IRAC data, using
MOPEX (Makovoz \& Marleau 2005) to calculate frame-to-frame background
matching (overlap) corrections and create mosaics with reduced
instrumental artifacts compared to the pipeline mosaics. A 3-color
mosaic for BRC 27 is shown in Fig.~\ref{fig:brc27-irac}, and for BRC
34 in Fig.~\ref{fig:brc34-irac}.  The pixel size for our mosaics  was
0.6$\arcsec$, identical to the SSC pipeline mosaics. This is half of
the native pixel scale. We created separate mosaics for the long and
the short exposures at each channel for photometric analysis.  

In both BRCs, we focused our analysis on the region covered by all
four IRAC bands, i.e., a region $\sim5\arcmin\times5\arcmin$ centered
on the coordinates above. As a result of the way the
instrument+telescope is designed, serendipitous data are obtained at
two bands in each of two non-overlapping $\sim5\arcmin\times5\arcmin$
fields whose centers are offset $\sim6.5\arcmin$ from the target
field, in opposite directions; see the IRAC Instrument Handbook for
more details. These regions with serendipitous data will be discussed
in a forthcoming paper. 

To obtain photometry of sources in each BRC region, we used the
APEX-1frame module from MOPEX to perform source detection on the
resultant long and short mosaics for each observation separately.  We
took those source lists and used the aper.pro routine in IDL to
perform aperture photometry on each of these source detections in the
corresponding mosaics with an aperture of 3 native pixels (6 resampled
pixels), and an annulus of 3-7 native pixels (6-14 resampled pixels).
The corresponding (multiplicative) aperture corrections are, for the
four IRAC channels,  1.124, 1.127, 1.143, \& 1.234, respectively, as
listed in the IRAC Instrument Handbook.  As a check on this automatic
photometry, the educators and students associated with this project
used the Aperture Photometry Tool (APT; Laher \etal\ 2012a,b) to
confirm by hand the measurements for all the targets of interest
(i.e., they inspected and clicked individually on each of the objects
in each of the bands).  To convert the flux densities to magnitudes,
we used the zero points as provided in the IRAC Instrument Handbook:
280.9, 179.7, 115.0, and 64.13 Jy, respectively, for the four
channels. (No array-dependent color corrections nor regular color
corrections were applied.) We took the errors as produced by IDL to be
the best possible internal error estimates; to compare to flux
densities from other sources, we took a flat error estimate of 5\%
added in quadrature. 

At this point in the process, for each BRC, we have one source list
for each exposure time, for each channel, so a total of 8 source lists
per BRC target.  To obtain one source list per channel per BRC
observation, we then merged the short and the long exposure source
lists for each channel separately. We performed this merging via a
strict by-position search, looking for the closest match within
1$\arcsec$.  This maximum radius for matching was determined via
experience with this analysis step in other star-forming regions
(e.g., Rebull \etal\ 2010).  If a match between the source lists was
found, if the source is brighter than a threshold, the photometry was
used from the short frame, and if it was fainter, then the photometry
was taken from the long frame. The brightness thresholds beyond which
photometry was obtained from the IRAC short exposures follows from the
IRAC study of the Taurus star-forming region (Rebull \etal\ 2010,
Padgett \etal, in prep). They are 9.5, 9.0, 8.0, \&  7.0 mag for the
four IRAC channels respectively.   The limiting magnitudes of these
final source lists are the same for both observations, and are
[3.6]$\sim$14 mag, [4.5]$\sim$14 mag, [5.8]$\sim$12 mag, and
[8]$\sim$10.5 mag.

\subsection{MIPS Data}
\label{sec:obsmips}

\begin{figure*}[tbp]
\epsscale{0.5}
\plotone{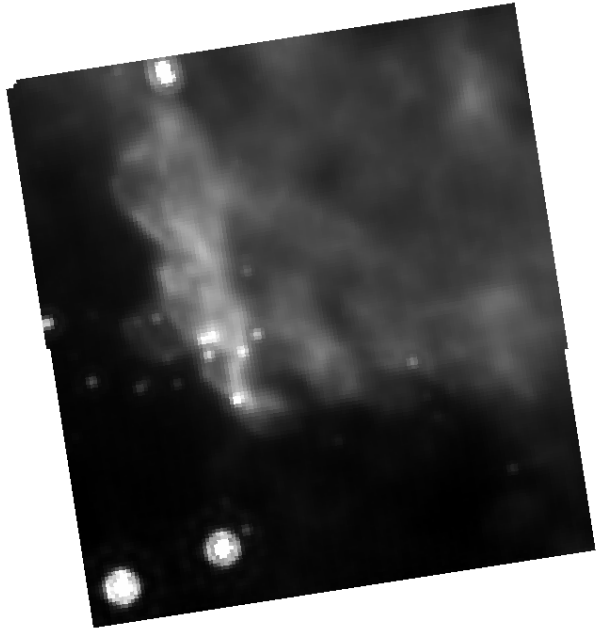}
\caption{BRC 27 in 24 \mum. The 70 \mum\ image (not shown) is more or less
featureless nebulosity. North is up; the image is $\sim$7.5$\arcmin$
on a side. Both nebulosity and point sources can be seen here.}
\label{fig:brc27mips1}
\end{figure*}

\begin{figure*}[tbp]
\epsscale{0.5}
\plotone{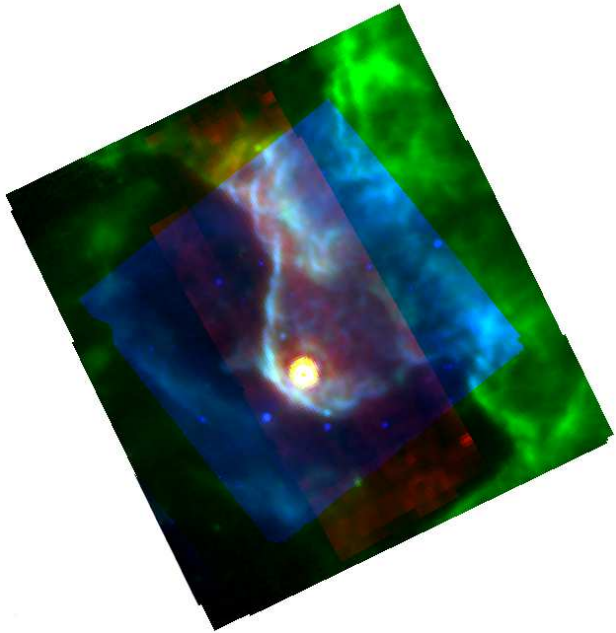}
\caption{BRC 34 in 8 \mum\ (blue), 24 \mum\ (green) and 70 \mum\
(red); note that all three bands have a slightly different footprint
on the sky. North is up. The bright point source near the center has
counterparts at all bands. The 24 \mum\ band mosaic (the largest here)
is $\sim$7.5$\arcmin$ on a side.    Both nebulosity and point sources
can be seen here.} 
\label{fig:brc34mips12}
\end{figure*}

We again used the MIPS data for BRC 27 from Spitzer program 30050,
AORKEY 17512448, obtained on 2006-11-04; for BRC 34, we used data from
Spitzer program 202, AORKEY  6031872, obtained on 2004-10-19. 

In BRC 27, the AOR was designed to obtain two cycles of large-field
photometry mode observations at 24 \mum\ (with a 10$\arcsec$ sky
offset), with 3 s per exposure.  In BRC 34, the AOR was designed to
obtain three cycles of small-field photometry, 3 s per exposure, at 24
\mum.  For both BRCs, the AORs also obtained one cycle of small-field
default-scale mode observations at 70 \mum. These observations are
centered on the same location as the 4-band IRAC data. The final 24
\mum\ coverage is $\sim$7.5$\arcmin$ on a side, so slightly larger
than the four-band IRAC coverage (i.e., the region covered by all four
IRAC bands), with 3 s per pointing and a maximum integration of 42 s
only in the center $\sim3\arcmin\times3\arcmin$ portion for BRC 27,
and 126 seconds in the center portion for BRC 34. The final 70 \mum\
coverage is $\sim3\arcmin\times7.5\arcmin$, with a total of 30 s
integration (3 s per pointing, one cycle). We note for completeness
that serendipitous data with a field center about 12$\arcmin$ offset
from of our target were obtained at 24 \mum\ during the 70 \mum\
photometry integration (see the MIPS Instrument Handbook for more
information); since these fall outside of the 4-band IRAC coverage,
they are beyond the scope of this paper and will be included in a
forthcoming paper.

The data for 24 \mum\ required additional processing beyond what the
MIPS pipelines provided. We started with S18.12 enhanced BCDs (eBCDs)
from the pipeline and created mosaics using MOPEX, as for IRAC. Our
mosaics were constructed to have the same pixel size as the pipeline
mosaics, 2.45$\arcsec$.   The MIPS data for BRC 27 appear in
Figure~\ref{fig:brc27mips1}.  The 70 \mum\ data for BRC 27 revealed no
point sources, and appears to be essentially featureless nebulosity,
so we did not process it beyond the pipeline. The 70 \mum\ data for
BRC 34, on the other hand, had point sources and texture in the
nebulosity.  The pipeline produces both filtered and unfiltered
mosaics; the filtering preserves the flux densities of the point
sources and improves their signal-to-noise, especially for faint
sources, but destroys the flux density information for the extended
emission. The unfiltered mosaic is shown in
Figure~\ref{fig:brc34mips12}, but we performed photometry on the
filtered mosaic. The pipeline mosaics have resampled 4$\arcsec$ pixels
(as opposed to 5.3$\arcsec$ native pixels). 

To obtain photometry at 24 \mum\ (and 70 \mum\ for BRC 34), we ran
APEX-1frame on each of the mosaics and performed
point-response-function (PRF) fitting photometry using the
SSC-provided PRF.   We used the signal-to-noise ratio (SNR) value
returned by APEX-1frame as the best estimate of the internal
(statistical) errors, adding a 4\% flux density error in quadrature as
a best estimate of the absolute uncertainty.  For some sources of
interest below, an upper limit at 24 \mum\ was obtained at the given
position by laying down an aperture as if a source were there, and
taking 3 times that value for the 3$\sigma$ limit.  For the single 70
\mum\ detection, we assumed a conservative, flat 20\% flux density
error. (A second, fainter 70 \mum\ source is visible in
Fig.~\ref{fig:brc34mips12}; this source is literally on the edge of
the mosaic, and as such we cannot obtain reliable photometry for it.)
To convert the flux densities to magnitudes, we used the zero points
as found in the MIPS Instrument Handbook, 7.14 Jy and 0.775 Jy for 24
and 70 \mum, respectively.

\subsection{Optical Data}
\label{sec:opticaldata}

Based on experience with other star forming regions (e.g., Rebull
\etal\ 2010, 2011c, Guieu \etal\ 2010), we know that optical data
(either just photometric points or higher spatial resolution images,
or both) can be tremendously helpful in discriminating between true
YSOs and background galaxies. We obtained observations of our target
region from the 2-m Las Cumbres Observatory Global Telescope (LCOGT)
Network member telescope, Faulkes Telescope North (FTN), on Haleakala.
The Faulkes telescope has a $\sim10\arcmin$ field of view, easily
encompassing our region of interest in both BRCs. The spatial
resolution of the telescope is $\sim$1.1$\arcsec$, most often
seeing-limited; this is well-matched to our $\sim$1.5$\arcsec$
resolution IRAC data. The pixel scale is  0.3$\arcsec$ pixel$^{-1}$. 

The filters that we used were Sloan $r$ and $i$ bands. While there is
no Sloan Digital Sky Survey (SDSS; Abazajian \etal\ 2009) coverage of
our target, there are some reasonably nearby (on the sky) SDSS
observations that we used to ``bootstrap'' our calibration.  We
obtained calibration images close in time (and in airmasses) to our
science images; for BRC 27, the calibration image was at
$\alpha\sim7.6^h, \delta\sim-11\arcdeg$, and for BRC 34, the
calibration image was at $\alpha\sim20.7^h, \delta\sim+58\arcdeg$. 
The BRC 34 data were obtained on 2011-10-21, and the BRC 27 data were
obtained on 2012-01-02. The images are initially processed through the
LCOGT pipeline, which performs the bias and flatfield corrections. For
BRC 34, it attached a world coordinate system (WCS) to the header, but
it failed to do so in the case of BRC 27. We used astrometry.net to
attach a WCS in that case. The science exposures were all 120 sec. The
calibration frames for BRC 27 were also all 120 sec; the initial
calibration frames for BRC 34 were 360 sec in $r$ and 240 sec in $i$,
and then the final calibration frames were 120 sec. We took this into
account in our data reduction. The observations were all obtained
through about 1.3 airmasses.  We used APT to obtain source detections
for the calibration fields, and matched the sources in the calibration
fields to the existing SDSS data sets using a conservative source
matching of 1$\arcsec$ radius. There were between 177 and 303 sources
of moderate brightness, depending on field and filter band, that were
used to establish calibration for the science target fields. We used
an 8 pixel aperture radius and a sky annulus from 9 to 15 pixels.   

As for the calibration fields, we used APT using the same settings to
detect and measure photometry for objects in the science fields, and
applied our calibration solutions to the science fields.  We again
used an 8 pixel aperture radius and a sky annulus from 9 to 15
pixels.  To match the much smaller subset of science target sources of
interest between the two optical bands, in each pointing, we used a
2$\arcsec$ matching radius; empirically, this provided the best
results. The completeness limits of these observations were $r\sim20$,
and $i\sim19$, using Sloan (AB) magnitudes (Oke \& Gunn 1983). We took
the errors as produced by APT to be the best possible internal error
estimates; to compare to flux densities from other sources, we took a
flat, conservative absolute error estimate of 6\% added in
quadrature.  If the objects are legitimately young, their intrinsic
variability (due to cool star spots or accretion hot spots) at these
wavelengths is likely to be larger than these error estimates.  To
convert these magnitudes to fluxes (for inclusion in the SEDs in
Section \ref{sec:seds}), we used the standard 3631 Jy  as a zero point
(see, e.g., Finkbeiner \etal\ 2004). If an object of interest was not
automatically detected in the images, we examined the images at the
location of the object, and obtained by hand either an upper limit or
a measurement of the photometry of the detection. Limits as reported
in Table~\ref{tab:ourysos} are 3$\sigma$ limits.

\subsection{Bandmerging and the Final Catalog}
\label{sec:bandmerging}

In summary, to bandmerge our data, we first merged the photometry from
all four IRAC channels together with near-IR 2MASS  data within each
BRC observation, followed by MIPS data, and then the optical data.  
We now discuss each of these steps in more detail.  We then compare
our catalog to the literature catalog which we established in
\S\ref{sec:litsrcs}.

To merge the photometry from all four IRAC channels together, we
started with a source list from 2MASS.  This 2MASS source list
includes $JHK_s$ photometry and limits, with high-quality
astrometry.   We merged this 2MASS source list by position to the
IRAC-1 source list, using a search radius of 1$\arcsec$, a value 
empirically determined via experience with other star-forming regions
(e.g., Rebull \etal\ 2010).  Objects appearing in the IRAC-1 list but
not the $JHK_s$ list were retained as new potential sources.  The
master catalog was then merged, in succession, to IRAC-2, 3, and 4,
again each using a matching radius of 1$\arcsec$.  Because the source
detection algorithm we used can erroneously detect instrumental
artifacts as point sources, we explicitly dropped objects seen only in
one IRAC band as likely artifacts. 

The MIPS 24 \mum\ source list was then combined into the merged
2MASS+IRAC catalog, using a positional source match radius of
2$\arcsec$, again determined via experience with other star-forming
regions (e.g., Rebull \etal\ 2010).  The MOPEX source detection
algorithm can erroneously report structure in the nebulosity as a
chain of point sources found in the image, and by inspection, this was
the case for these data. To weed out these false `sources', we dropped
objects from the catalog that were detections only at 24 \mum\ and no
other bands.   There is only one 70 \mum\ source (in BRC 34), so that
was added by hand into the master catalog, matched to the appropriate
source.

Finally, to merge the $J$ through 70 \mum\ catalog to the optical
($ri$) catalog, we looked for nearest neighbors within 2$\arcsec$. 
That matching radius was determined empirically to be the best via
comparison of these images and catalogs.


To put these observations in context with other similar surveys (e.g.,
Rebull \etal\ 2011b), BRC 27 has $\sim$220 sources with IRAC-1,
$\sim$120 sources with IRAC-4, and  $\sim$24 sources with MIPS 24
\mum\ (and $\sim$180 sources with 2MASS data). BRC 34 has $\sim$580
sources with IRAC-1,  $\sim$120 sources with IRAC-4, and only 5
sources with MIPS 24 \mum\ (and $\sim$200 sources with 2MASS data). 

The strong falloff of source numbers with increasing wavelength is
typical for these bandpasses for the following reasons. The SED for
stars without dust can be approximated by a blackbody curve, 
\begin{math}\lambda B_{\lambda} =
\left(\frac{2hc^2/\lambda^4}{\exp(hc/\lambda kT)-1)}\right) \end{math}
where $h$ is Planck's constant, $c$ is the speed of light, $\lambda$
is wavelength, $k$ is Boltzmann's constant, and $T$ is the temperature
of the blackbody (or $T_{\rm eff}$ for a stellar approximation); 
$\lambda B_{\lambda}$  (rather than $B_{\lambda}$) is in units of
energy density (e.g., erg s$^{-1}$ cm$^{-2}$) for the SED. At the
wavelengths in the mid-infrared (between roughly 3 and 70 microns as
considered here), the SED falls off as $\lambda^{-3}$. For six
theoretical, equally sensitive channels at 3.6, 4.5, 5.8, 8, 24, and
70 microns, the expected brightness from a dust-free star would fall
through these bands, and one would expect many fewer sources, say, at
70 \mum\ compared to 3.6 \mum. In reality, IRAC-1 and 2 (3.6 and 4.5
\mum) are comparably sensitive. However, for these observations as
conducted, IRAC-3 (5.8 \mum) is very roughly 4 times less sensitive
than IRAC-1 and 2, and IRAC-4 (8 \mum) is very roughly 3 times less
sensitive.  The two different MIPS 24 \mum\ observations have
different integration times, so a sensitivity calculation for these
observations indicates that these observations are very roughly 2-4
times less sensitive than IRAC-1 and 2. The MIPS 70 \mum\ observations
integrate to 30 s; the resulting sensitivity of these observations is
nearly 200 times less sensitive than IRAC-1 and 2. Most of the sources
seen in any given Spitzer image are photospheres (stars without dust),
and as such their expected flux density falls steadily through the
2-70 \mum\ range.  Moreover, the sensitivity worsens essentially
steadily through this same wavelength region, very roughly as
$\lambda^{2.5}$ in an SED plot.  So, the expected net source counts
fall rapidly with increasing wavelength due both to the intrinsic
falloff of the stellar SEDs with wavelength and the decreasing
sensitivity of the observations with wavelength.


The difference in source count rates in IRAC-1 between the two BRCs
can be traced to the number density of Galactic foreground and
background stars. The Galactic coordinates of BRC 27 are $(l,b)$ =
(224$\arcdeg$, $-2\arcdeg$) and for BRC 34, they are
(99$\arcdeg$,$+5\arcdeg$).  Given these positions, there are more
foreground and/or background objects for BRC 34. Most
foreground/background objects do not have IR excesses, and, thus, do
not have counterparts detected in the longer Spitzer bandpasses. Most
of the objects seen in these images are not young stars, but instead
contaminants (background or foreground objects).

\section{Selection of YSO Candidates with Infrared Excess}
\label{sec:findthem}

With our new multi-wavelength view of the two BRC regions, we can
begin to look for young stars. We focus on finding sources having an
infrared excess characteristic of YSOs surrounded by a dusty envelope
and/or disk.  In this Section, first we provide an overview of the
color selection we used primarily to identify young stars (Section
\ref{sec:coloroverview}). Then we discuss the IRAC color-color diagram
(Section \ref{sec:iracysos}), the IRAC and MIPS color-magnitude
diagram (Section \ref{sec:3324ysos}), the IRAC color-magnitude diagram
(Section \ref{sec:338ysos}), and remaining literature objects without
apparent IR excesses (Section \ref{sec:otherobj}).  We summarize the
entire process in Section~\ref{sec:irxsummary}. Two tables are
provided in this section: Table~\ref{tab:ourysos} provides multi-band
measurements of the YSOs and YSO candidates discussed here, and
Table~\ref{tab:ourysonotes} summarizes notes about specific objects
called out in the text.

\subsection{Overview of Color Selection}
\label{sec:coloroverview}

There is no single Spitzer color selection criterion (or set of
criteria) that is 100\% reliable in separating members from non-member
contaminants.  Many have been considered in the literature (e.g.,
Allen \etal\ 2004, Rebull \etal\ 2007, Harvey \etal\ 2007, Gutermuth
\etal\ 2008, 2009, Rebull \etal\ 2010, 2011a). Some make use of just
MIPS bands, some make use of just IRAC bands, most use a series of
many color criteria, and where possible, they make use of (sometimes
substantial) ancillary data. One of the earliest methods was presented
in Allen \etal\ (2004), which marked out regions of IRAC color-color
space as most likely to harbor objects of various classes, but likely
also include contaminants.  The best general choice for selecting YSO
candidates from Spitzer+2MASS data is the approach developed by
Gutermuth \etal\ (2008, 2009). This selection method starts from the
set of objects detected at all four IRAC bands and uses 2MASS and MIPS
data where possible. It implements a series of (many) color cuts to
attempt to remove contaminants such as background galaxies (usually
red and faint) and knots of nebulosity. The most common contaminants
left by any of these color selections are active galactic nuclei (AGN)
and asymptotic giant branch (AGB) stars, both of which can have
similar colors to legitimate YSOs (see, e.g., Stern \etal\ 2005 for
AGN and Blum \etal\ 2006 for AGBs). YSOs generally are bright and red,
though, depending on distance, mass, and degree of reddening and/or
embeddedness, they can also be faint and red (see, e.g., Rebull \etal\
2010, 2011a and references therein).

The regions of interest for our study are small areas on the sky, and
we do not have reliable high-resolution extinction maps for these
regions. We have used the Gutermuth method as adapted by Guieu \etal\
(2009, 2010) for the case in which no extinction map is available.  
In these BRC cases, the lack of an extinction map and subsequent lack
of reddening-corrected steps in the selection process may erroneously
include, in particular, bright background AGB stars. To attempt to
compensate, once we have identified potential YSO candidates, we
inspect each of these candidates in all available images, check their
position in color-color and color-magnitude diagrams, and construct
and inspect their SEDs. On the basis of this inspection, we drop
objects that are most likely bright foreground or distant background
objects, have insignificant IR excesses once errors are incorporated,
are evidently contaminated by an image artifact or bright nearby
source, or are clearly not point sources. In the process of doing
this, when we construct color-color and color-magnitude diagrams, we
identify objects that are worth investigating as additional YSO
candidates due to their location in the diagram but were not picked up
by the color cuts. Such objects were either originally missing a
detection in a band that would have enabled automatic identification
as a YSO candidate by the method we have implemented, or have such
subtle excesses that the method didn't identify them {\it a priori}.
As such, our sample is not a statistically unbiased sample, but our
goal was to obtain a complete sample of YSOs rather than an unbiased
sample.

Table~\ref{tab:ourysos} includes all of the measurements for all of
the literature YSOs, literature YSO candidates, and new YSO candidates
that survived this selection and weeding process. 
Table~\ref{tab:ourysonotes} collects notes on the objects (as in, if an
object is called out elsewhere in the paper, it is noted in
Table~\ref{tab:ourysonotes}), including identifying those literature
YSOs or literature YSO candidates that do not seem to have an IR
excess. In some cases where our individual inspection and evaluation
suggests the IR excess may be marginal, we have identified the IR
excess as uncertain in Table~\ref{tab:ourysonotes}. We next discuss
the distribution of these objects in several color-color and
color-magnitude diagrams, highlighting some objects as necessary. In
each case, we discuss where YSOs and contaminants are most likely to
fall.

\subsection{IRAC Color-Color Diagram}
\label{sec:iracysos}

\begin{figure*}[tbp]
\epsscale{0.9}
\plotone{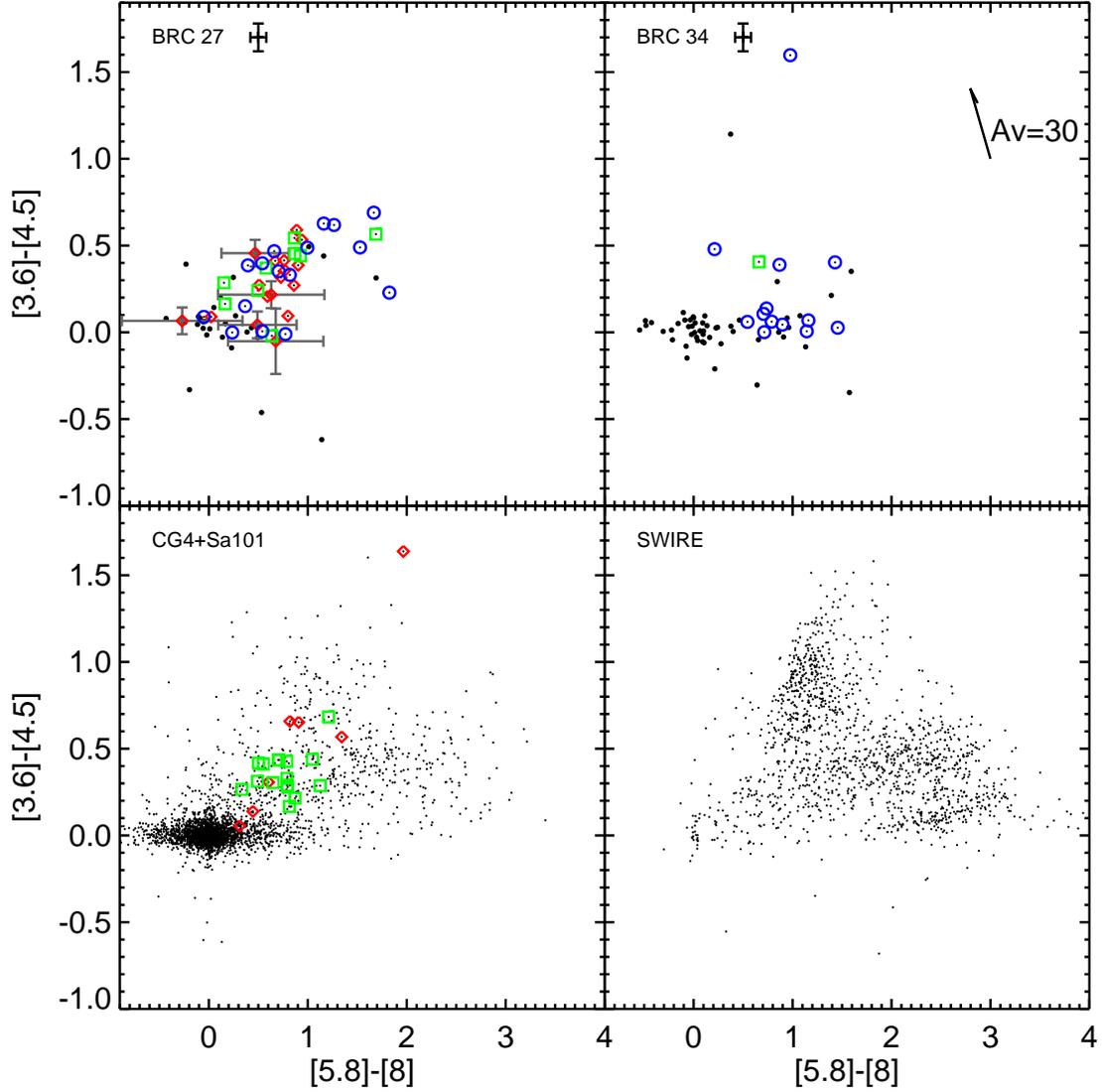}
\caption{[3.6]$-$[4.5] vs.\ [5.8]$-$[8] color-color diagram for BRC 27
(upper left) and BRC 34 (upper right), with CG4+Sa101 (lower left) and
SWIRE (lower right) for comparison. In each panel, small dots are
objects in the catalog (i.e., objects seen in the image); green
squares are literature high-confidence YSOs, red diamonds  are
literature candidate YSO, and blue circles are our new candidate YSOs
presented here.  Typical errors on the BRC data are $\sim$0.08 mags,
and are indicated by the sample error bars in the top left of each of
the BRC plots.  Objects with exceptionally large error bars have error
bars overplotted in grey. A sample \av=30 vector is included in the
BRC 34 panel for reference. The CG4+Sa101 and SWIRE data are provided
for comparison to other fields, where there are foreground and
background contaminants (stars and galaxies) in addition to YSOs; see
text for more discussion. All of the YSOs (known, literature
candidates, and new candidates) have colors in this diagram consistent
with known YSOs, but many contaminants do too. Some of the YSOs and
YSO candidates in BRC 27 have colors consistent with no IR excess.
This is discussed further in the text. }
\label{fig:iracysos}
\end{figure*}

Figure~\ref{fig:iracysos} shows the IRAC color-color diagram
([3.6]$-$[4.5] vs.\ [5.8]$-$[8]) for both BRC 27 and BRC 34, as well
as two other fields for comparison. The lower left is the CG4+Sa101
data (see Section~\ref{sec:evol}), and consists of background
galaxies, foreground and background stars without IR excesses, and
young stars (both high-confidence and candidate YSOs) with IR
excesses. The lower right, for comparison, contains data from the 6.1
deg$^2$ Spitzer Wide-area Infrared Extragalactic Survey (SWIRE;
Lonsdale \etal\ 2003) ELAIS (European Large Area ISO Survey) N1
extragalactic field\footnote{VizieR Online Data Catalog, II/255 (J.
Surace et al., 2004)} (the Cores-to-Disks [c2d; Evans \etal\ 2003,
2009a] reduction is used here, as in Rebull \etal\ 2011a). This sample
by its nature is expected to contain primarily galaxies, though likely
includes some foreground stars not expected to have IR excesses. The
SWIRE survey was relatively shallow compared to many extragalactic
surveys, and as such provides a good comparison to these relatively
shallow maps of galactic star forming regions.

By comparison of these panels in Figure~\ref{fig:iracysos}, we can
demonstrate where common objects are found. Ordinary stellar
photospheres (likely foreground or background stars) are found near 0
in both IRAC colors; objects like this can be found most prominently
in the CG4+Sa101 field, where there are many foreground or background
stars.  Galaxies are found throughout this diagram, but are often red
in one or both colors, as can be seen in the SWIRE panel. YSOs with
substantial IR excesses will be red in both [3.6]$-$[4.5] and
[5.8]$-$[8]; YSOs with inner disk holes will have small [3.6]$-$[4.5]
and red [5.8]$-$[8].  Objects with colors similar to YSOs can be seen
in both BRC 27 and BRC 34.

Of the known YSOs, literature candidate YSOs, and new candidate YSOs 
considered here in the BRCs, most are red in both [3.6]$-$[4.5] and
[5.8]$-$[8]. This is consistent with where we expect them to appear,
based on YSOs studied elsewhere (e.g., in CG4+Sa101 in the lower left
panel of the figure; Rebull \etal\ 2011b). However, many more known
YSOs and literature YSO candidates are found in BRC 27 than in BRC 34.
These previously identified YSOs and candidates were obtained via a
variety of means not necessarily involving the IR, including X-rays
(see \S\ref{sec:litsrcs}).  It is known that YSOs can be young without
having circumstellar disks or envelopes (see \S\ref{sec:evol} or,
e.g., Rebull \etal\ 2010). Thus, YSOs may be legitimately young even
though they do not have IR excesses. They may also have IR excesses at
wavelengths longer than the longest wavelength used in this specific
Figure, 8 \mum. The YSO and YSO candidate objects with near 0 color in
Fig.~\ref{fig:iracysos} are exactly these kinds of objects -- possibly
legitimately young, though not having an IR excess, or not having a
detectable IR excess using these data and data reduction. We now
discuss these objects because they are different than the rest of the
ensemble of YSOs and YSO candidates in this diagram; notes on these
objects appear in Table~\ref{tab:ourysonotes}.

In BRC 27, five objects have [3.6]$-$[4.5]$<$0.1, and three of those
also have the lowest [5.8]$-$[8] values, $<$ 0.03. Object
070352.2-112100 (=Chauhan 109=row 1 in the Tables) has a very small
[3.6]$-$[4.5] but a larger [5.8]$-$[8]=0.5, so it appears to have a
small excess at the longer bands, consistent with an inner disk hole.
It is a literature YSO candidate, having been identified from an
apparent NIR excess, which would be inconsistent with an inner disk
hole, though we too identify it as having a small NIR excess (see
\S\ref{sec:nearirproperties} below). Follow-up spectroscopy might
clarify this issue. Object 070352.7-112313 (=Ogura 2, Chauhan 81=row 2
in the Tables) has a very small [3.6]$-$[4.5] and a large, negative
[5.8]$-$[8], suggesting that it does not have a disk; this object is
quite faint at 8 \mum, and as such, has a large error. The error is
large enough that it could move it to [5.8]$-$[8]$\sim$0, consistent
with other disk-free YSOs. It is also a literature YSO candidate, but
it does not appear to have a measurable IR excess. Object
070353.8-112341 (=row 6 in the Tables) is a new YSO candidate. It has
[3.6]$-$[4.5]=0.09 and [5.8]$-$[8]=$-$0.05, so it does not have much
of an IRAC excess; it does, however, have a small excess at 24 \mum\
(see Section~\ref{sec:3324ysos}). The fourth object, 070403.9-112609
(=Shevchenko 102=row 25 in the Tables), again has small [3.6]$-$[4.5]
and [5.8]$-$[8], but with [3.6]$-$[4.5]$>$[5.8]$-$[8]. This one does
not appear to have a measurable IR excess; it is a literature YSO
candidate, so additional spectra would be particularly useful to
determine if it is a foreground star or truly a member of BRC 27.
Finally, object 070406.0-112128 (=row 33 in the Tables) is a new
candidate YSO. It has [3.6]$-$[4.5]=0 and [5.8]$-$[8]=0.23, so it
appears to have a small excess at the longer bands, consistent with an
inner disk hole. It will be discussed again in
Section~\ref{sec:3324ysos} below, where it appears to have a quite
significant [3.6]$-$[24] excess. 

In BRC 34, one new YSO candidate object (213334.8+580409=row 51 in the
Tables) has [5.8]$-$[8]$\sim$0.2 and [3.6]$-$[4.5]$\sim$0.5; it is
somewhat unusual to have [3.6]$-$[4.5]$>$[5.8]$-$[8]; for inner disk
holes, one generally expects [3.6]$-$[4.5]$<$[5.8]$-$[8]. This object
also has the most extreme values of $J-H$ and $H-K_s$ (see
\S\ref{sec:nearirproperties} below). It is probably subject to
considerable reddening, which accounts for the observation that
[3.6]$-$[4.5]$>$[5.8]$-$[8] (see the reddening vector in
Figure~\ref{fig:iracysos}). 

All the remaining YSOs and YSO candidates in these two BRCs have
significant IRAC excesses. The largest [3.6]$-$[4.5] is found in BRC
34, 213332.2+580329 (=row 49 in the Tables); it also has a large 24
\mum\ excess (see Section~\ref{sec:3324ysos} below). It most likely is
subject to significant reddening, which accounts for the fact that
[3.6]$-$[4.5]$>$[5.8]$-$[8] (see the reddening vector in
Figure~\ref{fig:iracysos}). It is also the only source in either BRC
detected at 70 \mum.

\subsection{IRAC \& MIPS Color-Magnitude Diagram}
\label{sec:3324ysos}

\begin{figure*}[tbp]
\epsscale{0.9}
\plotone{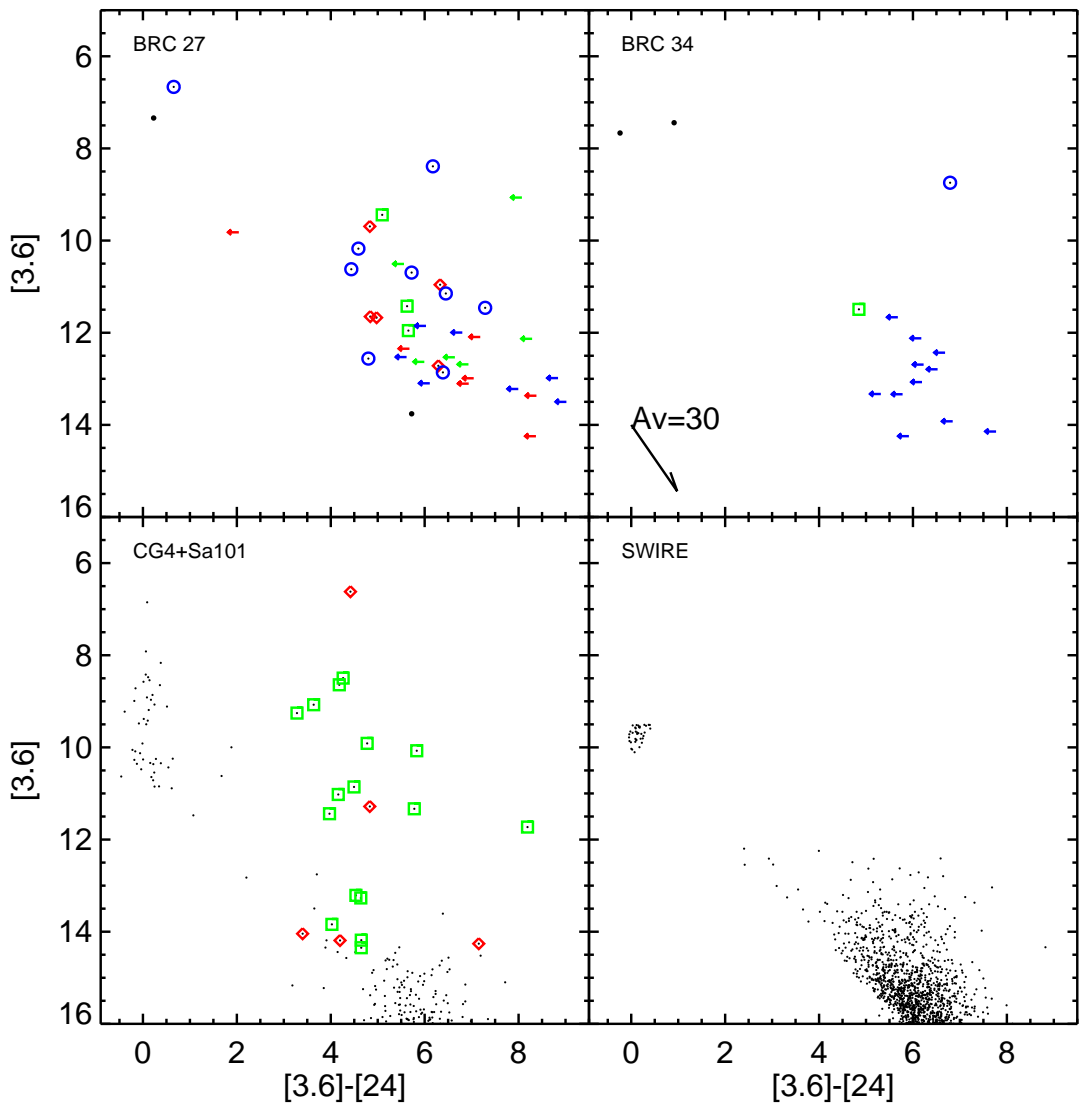}
\caption{[3.6] vs.\ [3.6]$-$[24] color-magnitude diagram for BRC 27
(upper left) and BRC 34 (upper right), with CG4+Sa101 (lower left) and
SWIRE (lower right) for comparison. In each panel, small dots are
objects in the catalog (i.e., objects seen in the image); green
squares are literature high-confidence YSOs (green arrows are limits
for objects undetected in the 24 \mum\ map), red diamonds are
literature candidate YSO (red arrows are limits), and blue circles are
our new candidate YSOs presented here (blue arrows are limits).  Error
bars are typically smaller than the symbol -- 0.06 mags in [3.6] and
0.07 mags in [3.6]$-$[24]. No objects here have exceptionally large
error bars. A sample \av=30 vector is included in the BRC 34 panel for
reference. There are very few sources seen at 24 \mum\ in either of
these BRC regions; the lower two panels give a better sense of the
distribution of objects seen in general.  All of the BRC 27 and BTRC
34 YSOs (known, literature candidates, and new candidates) have colors
in this diagram consistent with known YSOs; see text for more
discussion.}
\label{fig:3324ysos}
\end{figure*}

Young stars having inner disk holes and thus excesses at only the
longest bands can be revealed in particular via comparison of the 24
\mum\ or 8 \mum\ measurement to a shorter band, such as 3.6 \mum.  
Figure~\ref{fig:3324ysos} shows [3.6] vs.\ [3.6]$-$[24] for both BRC
27 and BRC 34, as well as the CG4+Sa101 and SWIRE samples for
comparison.  

Since there are very few 24 \mum\ sources detected that are not YSOs
or YSO candidates in either BRC, it is easier to understand the
expected distribution of sources by inspection of the CG4+Sa101 and
SWIRE samples. Foreground or background stars have $[3.6]-[24]\sim$0,
and galaxies likely make up the source concentration near
[3.6]$-$[24]$\sim$6, [3.6]$\sim$16. Objects that are red and bright
can be YSOs; since we are comparing [3.6] and [24] here, even disks
with large inner disk holes will appear here as red. However, AGB
stars can also occupy this part of the parameter space. As in
Fig.~\ref{fig:iracysos}, most of the YSOs and candidate  YSOs (from
the literature or new) that are detected are in fact in the location
in this diagram expected for YSOs. Many objects are not detected at 24
\mum, and those are indicated as limits for comparison.  Many more
objects are detected at 24 \mum\ in CG4+Sa101 because the CG4+Sa101
[24] observation  had a longer exposure time, and the sources that are
members of CG4+Sa101 are brighter because CG4+Sa101 is considerably
closer to us (3-500 pc vs.\ 800-1000 pc for the BRCs).

In the BRC 27 plot, one object (070406.0-112128 =row 33 in the Tables)
appeared as having [3.6]$-$[24]$\sim$4.6 (and [3.6]$\sim$10.2),
comfortably within the distribution of YSOs and candidates (it has a
reasonably large [3.6]$-$[24]), but not having been picked {\it a
priori} as a YSO using the Gutermuth method and the IRAC colors.   
This object was mentioned in Section~\ref{sec:iracysos} as having a
very small IRAC excess at the longer wavelength bands. The object is
in a region of relatively bright nebulosity at 8 and 24 \mum; though
it is clearly detected as a point source at 8 \mum, its detection is
less certain at 24 \mum. Had this object only had an excess at 24
\mum, we might attribute the apparent excess to nebular contamination.
However, Fig.~\ref{fig:iracysos} and the SED (see \S\ref{sec:sednotes}
below) suggests that there might be a small excess at the two longest
IRAC bands. In order to formally calculate the significance of any
excess, we need a spectral type and model fitting to the SED, but we
can extend a line with a Rayleigh-Jeans (RJ) slope (to approximate a
blackbody) from the 2.2 \mum\ or 3.6 \mum\ point to get an approximate
guess as to what the expected photospheric flux density might be, and
then compare that to the measured flux density (including its error
estimate). Performing this calculation suggests that the measured 8
\mum\ detection of this object has a marginal significance of
$\sim4\sigma$. This measurement, while not significant on its own, is
independent of the excess at 24 \mum; the fact that it might be a
excess suggests that nebula might not be the only contributor to any
excess, and that there might be circumstellar dust around this object.
This object could have a large inner disk hole, resulting in excess
only at the longest wavelengths sampled here. We have included this
object in our list as having a possible IR excess (see
Table~\ref{tab:ourysonotes}), but follow-up spectroscopy would be
particularly important in this object's case, because of the potential
for contamination by the nebulosity (or an unresolved background
object).

Also in the BRC 27 plot, the brightest YSO candidate object
([3.6]$\sim$6.5) is 070353.8-112341 (=row 6 in the Tables). This
object was mentioned in Section~\ref{sec:iracysos} as having small
IRAC colors. It is included in the set of new YSO candidates because
it has an apparently marginally significant 24 \mum\ excess, with
[3.6]$-$[24]=0.65, which is not large in comparison to many of the
other [3.6]$-$[24] values seen in Figure~\ref{fig:3324ysos}, but given
the uncertainties on the photometry, and the approach above extending
an RJ slope from 2.2 \mum, the 24 \mum\ excess is $\sim$11$\sigma$. It
does not have a significant 8 \mum\ excess. It is bright at IRAC
bands, and clearly detected in the 24 \mum\ image (see
Figure~\ref{fig:brc27-3color} below), and not obviously contaminated
by nebulosity. There is a reasonable chance that this is a foreground
star and that the photometry is compromised, or it is a very
interesting object with a very large inner disk hole (potentially
containing protoplanets); follow up spectroscopy is necessary. We have
tagged it as an uncertain IR excess in Table~\ref{tab:ourysonotes}.

The bluest limit in BRC 27 (at [3.6]$\sim$10 and [3.6]$-$[24]$\sim$2)
is 070403.9-112609 (=Shevchenko 102=row 25). This object was
mentioned in Section~\ref{sec:iracysos} above as not having a
significant IRAC excess. It is not detected at 24 \mum, which is
consistent with either a small excess at 24 \mum\ or no excess at all.
We do not detect a MIR excess in this object.

The brightest YSO candidate in BRC 34 also has the largest detectable
[3.6]$-$[24] in BRC 34, and it is 213332.2+580329 (=row 49 in the
Tables). This object is also the only object in either BRC detected at
70 \mum.

\subsection{IRAC Color-Magnitude Diagram}
\label{sec:338ysos}

\begin{figure*}[tbp]
\epsscale{0.9}
\plotone{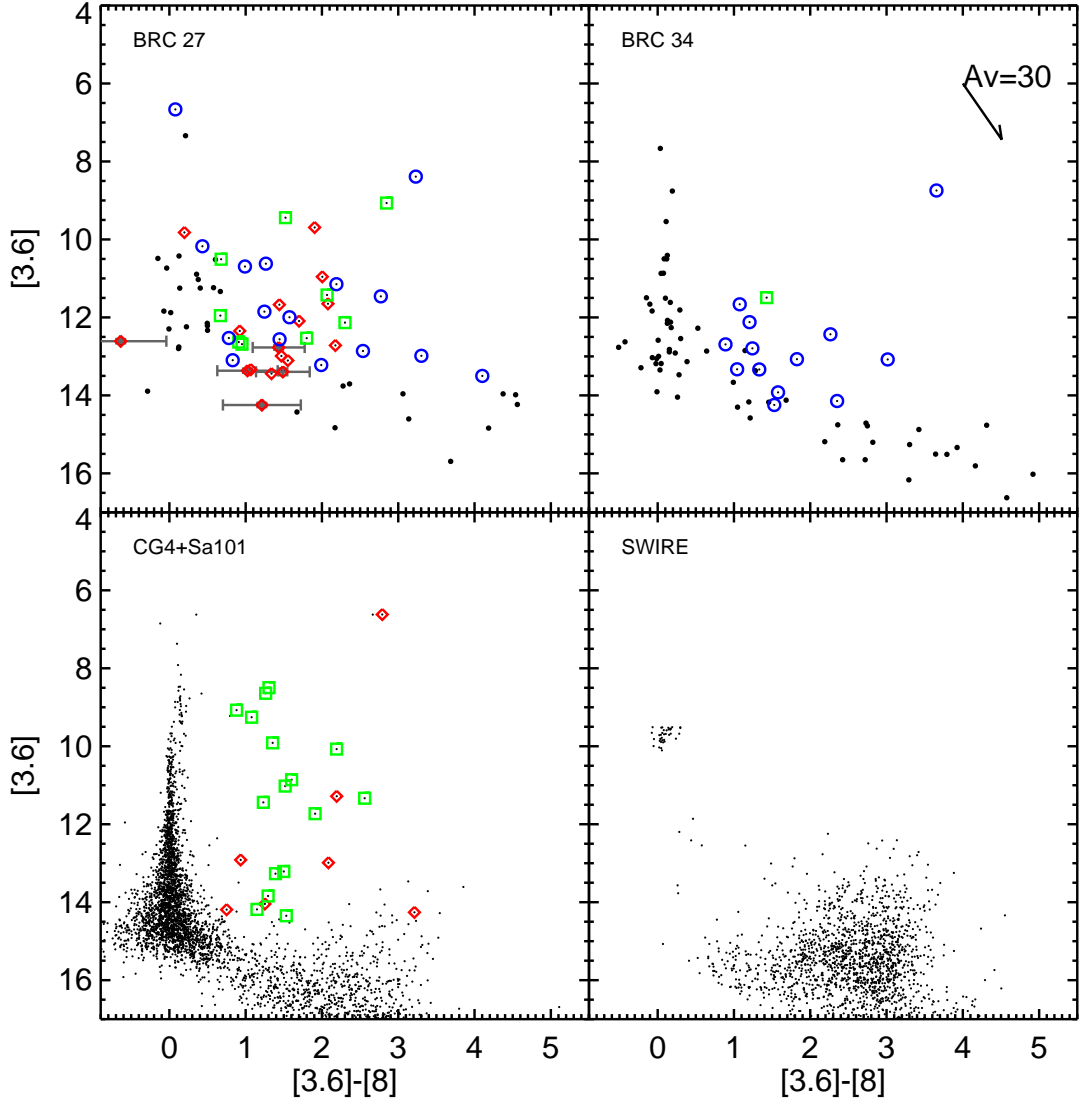}
\caption{[3.6] vs.\ [3.6]$-$[8] color-magnitude diagram for BRC 27
(upper left) and BRC 34 (upper right), with CG4+Sa101 (lower left) and
SWIRE (lower right) for comparison. In each panel, small dots are
objects in the catalog (i.e., objects seen in the image); green
squares are literature high-confidence YSOs, red diamonds  are
literature candidate YSO, and blue circles are our new candidate YSOs
presented here.  Error bars are typically smaller than the symbol --
0.06 mags in [3.6] and 0.07 mags in [3.6]$-$[8]. Objects with
exceptionally large error bars have error bars overplotted in grey. A
sample \av=30 vector is included in the BRC 34 panel for reference.
SWIRE data saturate at about [3.6]$\sim$9.5 because their IRAC data
were not obtained in HDR mode (see \S\ref{sec:obsirac}). Most of the
BRC 27 and BRC 34 YSOs (known, literature candidates, and new
candidates) have colors in this diagram consistent with known YSOs;
one of the previously-known YSOs in BRC 27 has an anomalously blue
color but also large errors at [8]. See text for more discussion.}
\label{fig:338ysos}
\end{figure*}

We also examined the [3.6] vs.\ [3.6]$-$[8] color-magnitude diagram;
see Figure~\ref{fig:338ysos}. Data from CG4+Sa101 and SWIRE are again
included for context.  As for Figure~\ref{fig:3324ysos}, foreground
and background photospheres have [3.6]$-$[8]$\sim$0, and galaxies
populate the clump at [3.6]$-$[8]$\sim$2 and [3.6]$\sim$16. Young
stars are bright and red; since we are comparing [3.6] and [8] here,
even disks with relatively large inner disk holes will appear here as
red. However, AGB stars can also occupy this part of the parameter
space.  We investigated all of the objects that have YSO-like colors
but were not identified in the color-selection method above. Several
of these objects appear to have small excesses at 8 \mum, which we now
discuss.

From BRC 27, there are three objects worth considering that we
identify from this Figure (as opposed to being selected via the
Gutermuth-style color cuts above in Section~\ref{sec:coloroverview}).
Object 070354.9-112514 (=Ogura 5, Chauhan 94=row 8 in the Tables) has
no 24 \mum\ data (it is off the edge of the MIPS-24 map). It appears
in the grouping of sources near [3.6]$-$[8]$\sim$1 and [3.6]$\sim$13.
It has only a $\sim$7$\sigma$ excess at 8 \mum; it has a very weak
possible 5.8 \mum\ excess as well (see SED in
Section~\ref{sec:sednotes}).  It is a literature YSO candidate.  We
have identified this object as having a very uncertain IR excess.  

Object 070406.5-112227 (=row 36 in the Tables) has a nominal 8 \mum\
excess at the $\sim$6$\sigma$ level, and no 24 \mum\ detection.  Like
the prior object, it too has a very weak possible 5.8 \mum\ excess
(see SED in Section~\ref{sec:sednotes}), and is grouping of sources
near [3.6]$-$[8]$\sim$0.8 and [3.6]$\sim$13. We have also identified
this object as having a very uncertain IR excess. 

Object 070407.9-112311 (=Ogura 21=row 39 in the Tables) also appears
somewhat red in this diagram, near [3.6]$-$[8]$\sim$0.9 and
[3.6]$\sim$12. It has an 8 \mum\ excess at the 11$\sigma$ level. The
[5.8] point also does not appear to be photospheric (see
Section~\ref{sec:sednotes}); it is not detected at [24], and it may be
subject to high \av. We have identified this object as having an IR
excess in Table~\ref{tab:ourysonotes}.

A fourth object from BRC 27, object 070408.1-112313 (=row 41 in the
Tables) is identified via the Gutermuth-style color cuts. It appears
as somewhat red in this diagram, near [3.6]$-$[8]$\sim$0.8 and
[3.6]$\sim$12.5. Like the other similar objects above, it  has an 8
\mum\ excess at the $\sim$7$\sigma$ level, and no 24 \mum\ detection.
However, it seems to have an excess beginning at [4.5], so it is
identified as having a more confident IR excess in
Table~\ref{tab:ourysonotes}.

In BRC 34, six objects appear as red in Figure~\ref{fig:338ysos}, none
of which were selected via the Gutermuth-style color cuts above in
Section~\ref{sec:coloroverview}. All of them have upper limits at 24
\mum, all of them have 8 \mum\ excesses $>$10$\sigma$, and all of them
have possible low-significance 5.8 \mum\ excesses. Most of them are
near [3.6]$-$[8]$\sim$1-2 and [3.6]$\sim$12-14, as for similar objects
from BRC 27 above. We have identified all of them as having IR
excesses, though a case could be made for them being uncertain because
of the contamination possible at [8]. They are: 213314.5+580351 (=row
43 in the Tables), 213315.6+580407 (=row 44), 213319.4+580406 (=row
45), 213323.8+580632 (=row 46; this is the reddest of this set at
[3.6]$-$[8]$\sim$2 and [3.6]$\sim$14), 213332.2+580558 (=row 50), and
213334.8+580409 (=row 51); this last may also be subject to high \av,
as noted in Section~\ref{sec:iracysos}, and it is also the brightest
of this set, at [3.6]$-$[8]$\sim$1 and [3.6]$\sim$12.

\subsection{Remaining Literature Objects Without Excesses}
\label{sec:otherobj}

Object 070353.5-112350 (=Shevchenko 90=row 4 in the Tables) is a
likely YSO from the literature, with an A0 spectral type. It is not
detected at [24], the [8] point is only about 6$\sigma$ above the
photosphere, and all the rest of the 2-5.8 \mum\ points appear to be
detecting the photosphere. It does not appear to have a significant IR
excess.

We inspected object 070404.5-112555 (=Ogura 13=row 28 in the Tables)
because it is identified in the literature as a candidate YSO.
However, this object is not detected at [24], and it is only detected
with very large errors at [5.8] and [8]. Taking into account
the errors, the 8 \mum\ excess is only significant at the 2.5$\sigma$
level. We do not identify this in Table~\ref{tab:ourysonotes} as
having an excess.

For completeness, we mention here that 070403.9-112609 (=Shevchenko
102=row 25 in the Tables) and 070352.7-112313 (=Ogura 2, Chauhan
81=row 2 in the Tables) were discussed in Section~\ref{sec:iracysos}
(and \ref{sec:3324ysos} for 070403.9-112609) as not having a
measurable IR excess.

\subsection{Summary of IR Excess Selection}
\label{sec:irxsummary}

In summary, we have 42 YSOs or YSO candidates (new or from the
literature) in BRC 27, and 14 YSOs or new YSO candidates in BRC 34. 

Of the 26 literature YSOs or literature YSO candidates in BRC 27,  we
find some indication of IR excess around 22 of them (one of those has
an uncertain IR excess).  Of the 9 high-confidence literature YSOs, 8
have high-confidence IR excesses, and one has no apparent IR excess.
Of the 17 literature YSO candidates, 14 have IR excesses, one has an
uncertain IR excess, and 3 have no detectable IR excess. There are 16
new YSO candidates presented here, 13 of which have IR excesses (and 3
of which have uncertain IR excesses). 

BRC 34 is again simpler; the one literature YSO also has an IR excess.
There are 13 new objects with IR excesses presented here.

We move ahead from here with this set of YSOs and YSO candidates, and
now investigate their multi-band properties.

\begin{deluxetable}{cllccccccccccccccc}
\tablecaption{Multiband measurements of known YSOs, literature YSO
candidates, and new Spitzer-identified YSO candidates in BRC 27 and
BRC 34\label{tab:ourysos}}
\rotate
\tabletypesize{\tiny}
\tablewidth{0pt}
\tablehead{
\colhead{row}  & \colhead{name}  &\colhead{Alt.~name} & 
\colhead{$r$\tablenotemark{a}}  &
\colhead{$i$\tablenotemark{a}} &  
\colhead{$J$} &  \colhead{$H$} &\colhead{$K_s$} & 
\colhead{[3.6]} & \colhead{[4.5]} &
\colhead{[5.8]} & \colhead{[8.0]} & \colhead{[24]}  & \colhead{[70]}}
\startdata
\cutinhead{BRC 27}
 1&   070352.2-112100&              Chauhan109&   20.92$\pm$   0.22&   19.50$\pm$   0.09&   15.71$\pm$   0.06&   14.59$\pm$   0.07&   13.78$\pm$   0.05&   13.37$\pm$   0.05&   13.32$\pm$   0.05&   12.83$\pm$   0.06&   12.34$\pm$   0.39&          $>$   4.99&             \nodata\\
 2&   070352.7-112313&        Ogura2,Chauhan81&   16.88$\pm$   0.04&   15.84$\pm$   0.04&   13.83$\pm$   0.04&   13.03$\pm$   0.05&   12.83$\pm$   0.04&   12.61$\pm$   0.05&   12.54$\pm$   0.05&   12.98$\pm$   0.13&   13.25$\pm$   0.60&             \nodata&             \nodata\\
 3&   070353.2-112403&                  Ogura3&   20.50$\pm$   0.27&   19.02$\pm$   0.09&   15.92$\pm$   0.07&   14.60$\pm$   0.05&   13.84$\pm$   0.05&   12.77$\pm$   0.05&   12.31$\pm$   0.05&   11.81$\pm$   0.05&   11.34$\pm$   0.34&             \nodata&             \nodata\\
 4&   070353.5-112350&            Shevchenko90&   11.26$\pm$   0.04&   11.44$\pm$   0.04&   10.75$\pm$   0.20&   10.71$\pm$   0.20&   10.67$\pm$   0.20&   10.51$\pm$   0.05&   10.53$\pm$   0.05&   10.47$\pm$   0.06&    9.83$\pm$   0.09&          $>$   4.95&             \nodata\\
 5&   070353.7-112428&        Ogura4,Chauhan82&   19.23$\pm$   0.07&   17.54$\pm$   0.04&   15.04$\pm$   0.04&   14.22$\pm$   0.05&   13.94$\pm$   0.06&   13.44$\pm$   0.05&   13.17$\pm$   0.05&   12.61$\pm$   0.06&   12.10$\pm$   0.06&             \nodata&             \nodata\\
 6&   070353.8-112341&                 \nodata&   11.83$\pm$   0.30&   11.20$\pm$   0.30&    8.17$\pm$   0.01&    7.22$\pm$   0.04&    6.84$\pm$   0.02&    6.66$\pm$   0.05&    6.58$\pm$   0.05&    6.54$\pm$   0.05&    6.59$\pm$   0.05&    6.01$\pm$   0.04&             \nodata\\
 7&   070354.6-112011&              Chauhan108&   20.37$\pm$   0.14&   19.10$\pm$   0.07&   15.93$\pm$   0.07&   14.98$\pm$   0.08&   14.37$\pm$   0.07&   14.25$\pm$   0.05&   14.03$\pm$   0.06&   13.66$\pm$   0.18&   13.03$\pm$   0.51&          $>$   5.88&             \nodata\\
 8&   070354.9-112514&        Ogura5,Chauhan94&   18.00$\pm$   0.05&   16.74$\pm$   0.04&   14.62$\pm$   0.03&   13.83$\pm$   0.04&   13.59$\pm$   0.05&   13.34$\pm$   0.05&   13.24$\pm$   0.05&   13.07$\pm$   0.07&   12.27$\pm$   0.09&             \nodata&             \nodata\\
 9&   070357.1-112432&        Ogura7,Chauhan83&   18.40$\pm$   0.05&   17.12$\pm$   0.04&   14.82$\pm$   0.03&   13.98$\pm$   0.02&   13.74$\pm$   0.05&   13.11$\pm$   0.05&   12.75$\pm$   0.05&   12.32$\pm$   0.06&   11.55$\pm$   0.06&          $>$   6.17&             \nodata\\
10&   070358.4-112325&                 \nodata&   13.39$\pm$   0.04&   13.19$\pm$   0.04&   12.32$\pm$   0.02&   11.97$\pm$   0.03&   11.92$\pm$   0.03&   11.85$\pm$   0.05&   11.86$\pm$   0.05&   11.38$\pm$   0.07&   10.61$\pm$   0.05&          $>$   5.82&             \nodata\\
11&   070400.7-112323&                 \nodata&          $>$  20.60&          $>$  20.68&   15.71$\pm$   0.06&   13.16$\pm$   0.03&   11.80$\pm$   0.02&   10.69$\pm$   0.05&   10.31$\pm$   0.05&   10.10$\pm$   0.05&    9.70$\pm$   0.05&    4.97$\pm$   0.04&             \nodata\\
12&   070401.2-112531&                 \nodata&   18.74$\pm$   0.05&   17.53$\pm$   0.05&   14.26$\pm$   0.03&   12.96$\pm$   0.02&   11.98$\pm$   0.02&   10.62$\pm$   0.05&   10.22$\pm$   0.05&    9.90$\pm$   0.05&    9.36$\pm$   0.05&    6.18$\pm$   0.04&             \nodata\\
13&   070401.2-112242&                 \nodata&          $>$  22.55&          $>$  20.47&   16.27$\pm$   0.09&   14.94$\pm$   0.08&   13.82$\pm$   0.05&   12.86$\pm$   0.05&   12.23$\pm$   0.05&   11.49$\pm$   0.06&   10.32$\pm$   0.06&    6.47$\pm$   0.06&             \nodata\\
14&   070401.2-112233&            Chauhan-anon&          $>$  23.01&          $>$  19.16&   15.91$\pm$   0.07&   14.50$\pm$   0.07&   13.72$\pm$   0.05&   12.99$\pm$   0.05&   12.67$\pm$   0.05&   12.25$\pm$   0.06&   11.52$\pm$   0.06&          $>$   5.94&             \nodata\\
15&   070401.3-112334& Gregorio74,Chauhan-anon&   13.99$\pm$   0.04&   13.48$\pm$   0.04&   11.45$\pm$   0.20&   10.84$\pm$   0.20&   10.33$\pm$   0.20&    9.44$\pm$   0.05&    9.07$\pm$   0.05&    8.50$\pm$   0.06&    7.92$\pm$   0.05&    4.35$\pm$   0.04&             \nodata\\
16&   070401.6-112406&                 \nodata&   18.52$\pm$   0.15&   18.38$\pm$   0.15&   16.59$\pm$   0.12&   14.89$\pm$   0.12&   13.65$\pm$   0.06&   11.46$\pm$   0.05&   10.84$\pm$   0.05&    9.95$\pm$   0.07&    8.69$\pm$   0.05&    4.17$\pm$   0.04&             \nodata\\
17&   070401.6-112132&                 \nodata&          $>$  20.25&   19.47$\pm$   0.09&   15.36$\pm$   0.04&   13.85$\pm$   0.03&   12.98$\pm$   0.03&   12.00$\pm$   0.05&   11.64$\pm$   0.05&   11.13$\pm$   0.06&   10.42$\pm$   0.07&          $>$   5.19&             \nodata\\
18&   070402.1-112512&                 \nodata&          $>$  20.23&   19.79$\pm$   0.14&   15.95$\pm$   0.08&   14.72$\pm$   0.11&   14.05$\pm$   0.07&   12.56$\pm$   0.05&   12.09$\pm$   0.05&   11.78$\pm$   0.06&   11.12$\pm$   0.05&    7.76$\pm$   0.09&             \nodata\\
19&   070402.2-112542&                 \nodata&   12.06$\pm$   0.04&   12.09$\pm$   0.04&   11.31$\pm$   0.03&   10.75$\pm$   0.03&    9.94$\pm$   0.05&    8.39$\pm$   0.05&    7.70$\pm$   0.05&    6.83$\pm$   0.05&    5.16$\pm$   0.05&    2.21$\pm$   0.04&             \nodata\\
20&   070402.3-112539& Shevchenko99,Gregorio75&   11.09$\pm$   0.04&   11.23$\pm$   0.04&   10.40$\pm$   0.04&   10.32$\pm$   0.07&   10.26$\pm$   0.02&    9.07$\pm$   0.05&    8.50$\pm$   0.05&    7.91$\pm$   0.05&    6.22$\pm$   0.05&          $>$   1.00&             \nodata\\
21&   070402.7-112325&                 \nodata&          $>$  22.44&          $>$  19.55&             \nodata&   13.94$\pm$   0.10&   12.58$\pm$   0.04&   11.15$\pm$   0.05&   10.66$\pm$   0.05&    9.95$\pm$   0.06&    8.96$\pm$   0.08&    4.70$\pm$   0.07&             \nodata\\
22&   070402.9-112337&      Ogura8+9,Chauhan84&   17.51$\pm$   0.04&   16.26$\pm$   0.04&   13.56$\pm$   0.04&   12.43$\pm$   0.05&   11.86$\pm$   0.03&   10.96$\pm$   0.05&   10.54$\pm$   0.05&    9.71$\pm$   0.05&    8.95$\pm$   0.05&    4.63$\pm$   0.04&             \nodata\\
23&   070403.0-112350&       Ogura10,Chauhan85&   19.04$\pm$   0.06&   17.82$\pm$   0.05&   15.68$\pm$   0.06&   14.34$\pm$   0.04&   13.55$\pm$   0.04&   12.13$\pm$   0.05&   11.69$\pm$   0.05&   10.76$\pm$   0.08&    9.83$\pm$   0.05&          $>$   3.84&             \nodata\\
24&   070403.1-112327&              Chauhan107&   18.38$\pm$   0.05&   16.94$\pm$   0.05&   13.03$\pm$   0.04&   11.57$\pm$   0.04&   10.69$\pm$   0.02&    9.69$\pm$   0.05&    9.31$\pm$   0.05&    8.69$\pm$   0.06&    7.79$\pm$   0.07&    4.86$\pm$   0.15&             \nodata\\
25&   070403.9-112609&           Shevchenko102&   10.62$\pm$   0.04&   10.74$\pm$   0.04&    9.76$\pm$   0.02&    9.72$\pm$   0.03&    9.63$\pm$   0.02&    9.82$\pm$   0.05&    9.73$\pm$   0.05&    9.65$\pm$   0.05&    9.62$\pm$   0.05&          $>$   7.78&             \nodata\\
26&   070403.9-112326&                 \nodata&          $>$  20.37&          $>$  19.80&   16.75$\pm$   0.14&   15.31$\pm$   0.10&   14.22$\pm$   0.06&   12.98$\pm$   0.05&   12.50$\pm$   0.06&   11.21$\pm$   0.05&    9.68$\pm$   0.07&          $>$   4.14&             \nodata\\
27&   070404.2-112355&       Ogura12,Chauhan86&   18.61$\pm$   0.05&   17.52$\pm$   0.04&   15.01$\pm$   0.04&   14.02$\pm$   0.04&   13.54$\pm$   0.05&   12.53$\pm$   0.05&   12.08$\pm$   0.05&   11.60$\pm$   0.05&   10.73$\pm$   0.05&          $>$   5.89&             \nodata\\
28&   070404.5-112555&                 Ogura13&   18.11$\pm$   0.05&   16.89$\pm$   0.04&   14.77$\pm$   0.03&   13.88$\pm$   0.02&   13.73$\pm$   0.04&   13.40$\pm$   0.09&   13.45$\pm$   0.17&   12.58$\pm$   0.34&   11.91$\pm$   0.34&             \nodata&             \nodata\\
29&   070404.7-112339&       Ogura14,Chauhan87&   17.61$\pm$   0.04&   16.61$\pm$   0.04&   14.16$\pm$   0.04&   13.09$\pm$   0.04&   12.51$\pm$   0.04&   11.67$\pm$   0.05&   11.26$\pm$   0.05&   10.90$\pm$   0.06&   10.23$\pm$   0.10&    6.70$\pm$   0.19&             \nodata\\
30&   070405.1-112313&       Ogura15,Chauhan88&   19.91$\pm$   0.11&   18.84$\pm$   0.07&   14.49$\pm$   0.07&   13.27$\pm$   0.07&   12.46$\pm$   0.04&   11.42$\pm$   0.05&   10.88$\pm$   0.05&   10.22$\pm$   0.05&    9.36$\pm$   0.05&    5.80$\pm$   0.04&             \nodata\\
31&   070405.7-112123&                 \nodata&          $>$  18.08&          $>$  17.82&   15.86$\pm$   0.20&   14.80$\pm$   0.20&   14.25$\pm$   0.11&   13.22$\pm$   0.05&   12.89$\pm$   0.06&   12.05$\pm$   0.10&   11.23$\pm$   0.06&          $>$   5.23&             \nodata\\
32&   070405.9-112358&       Ogura16,Chauhan89&   17.90$\pm$   0.04&   16.88$\pm$   0.04&   14.42$\pm$   0.03&   13.47$\pm$   0.03&   12.93$\pm$   0.03&   11.65$\pm$   0.05&   11.12$\pm$   0.05&   10.51$\pm$   0.05&    9.57$\pm$   0.05&    6.81$\pm$   0.04&             \nodata\\
33&   070406.0-112128&                 \nodata&   11.86$\pm$   0.04&   11.68$\pm$   0.06&   10.64$\pm$   0.03&   10.41$\pm$   0.02&   10.26$\pm$   0.02&   10.18$\pm$   0.05&   10.18$\pm$   0.05&    9.98$\pm$   0.06&    9.74$\pm$   0.07&    5.59$\pm$   0.04&             \nodata\\
34&   070406.0-112315&       Ogura17,Chauhan90&   19.14$\pm$   0.07&   17.92$\pm$   0.05&   15.11$\pm$   0.06&   13.97$\pm$   0.04&   13.25$\pm$   0.03&   12.72$\pm$   0.05&   12.13$\pm$   0.05&   11.43$\pm$   0.06&   10.54$\pm$   0.06&    6.43$\pm$   0.04&             \nodata\\
35&   070406.4-112336&       Ogura18,Chauhan91&   20.01$\pm$   0.13&   17.79$\pm$   0.05&   14.70$\pm$   0.05&   13.81$\pm$   0.05&   13.36$\pm$   0.07&   12.69$\pm$   0.05&   12.40$\pm$   0.05&   11.89$\pm$   0.07&   11.73$\pm$   0.06&          $>$   5.75&             \nodata\\
36&   070406.5-112227&                 \nodata&   18.04$\pm$   0.05&   17.02$\pm$   0.04&   14.51$\pm$   0.03&   13.60$\pm$   0.03&   13.30$\pm$   0.03&   13.10$\pm$   0.05&   13.09$\pm$   0.05&   12.81$\pm$   0.07&   12.27$\pm$   0.09&          $>$   6.99&             \nodata\\
37&   070406.5-112128&                 \nodata&   19.11$\pm$   0.15&   18.55$\pm$   0.15&   15.87$\pm$   0.14&   15.27$\pm$   0.14&   14.75$\pm$   0.11&   13.50$\pm$   0.05&   13.27$\pm$   0.05&   11.22$\pm$   0.08&    9.40$\pm$   0.08&          $>$   4.48&             \nodata\\
38&   070406.5-112316&       Ogura19,Chauhan92&   17.40$\pm$   0.04&   16.41$\pm$   0.04&   13.90$\pm$   0.06&   12.95$\pm$   0.04&   12.53$\pm$   0.03&   12.09$\pm$   0.05&   11.82$\pm$   0.05&   11.25$\pm$   0.07&   10.39$\pm$   0.09&          $>$   4.91&             \nodata\\
39&   070407.9-112311&                 Ogura21&   19.54$\pm$   0.13&   18.02$\pm$   0.06&          $>$  14.35&   14.23$\pm$   0.09&   13.68$\pm$   0.07&   12.35$\pm$   0.05&   12.14$\pm$   0.05&   12.02$\pm$   0.06&   11.42$\pm$   0.06&          $>$   6.67&             \nodata\\
40&   070408.0-112354&       Ogura22,Chauhan97&   15.50$\pm$   0.04&   14.91$\pm$   0.04&   13.12$\pm$   0.03&   12.44$\pm$   0.04&   12.20$\pm$   0.03&   11.95$\pm$   0.05&   11.79$\pm$   0.05&   11.45$\pm$   0.05&   11.29$\pm$   0.05&    6.30$\pm$   0.04&             \nodata\\
41&   070408.1-112313&                 \nodata&   18.06$\pm$   0.07&   16.97$\pm$   0.04&          $>$  14.29&   13.77$\pm$   0.09&   13.50$\pm$   0.09&   12.53$\pm$   0.06&   12.38$\pm$   0.06&   12.12$\pm$   0.06&   11.75$\pm$   0.07&          $>$   6.91&             \nodata\\
42&   070408.1-112309&       Ogura23,Chauhan98&   19.88$\pm$   0.13&   18.28$\pm$   0.07&             \nodata&   14.59$\pm$   0.06&   14.19$\pm$   0.07&   12.63$\pm$   0.06&   12.39$\pm$   0.05&   12.22$\pm$   0.06&   11.72$\pm$   0.06&          $>$   6.64&             \nodata\\
\tablebreak
\cutinhead{BRC 34}
43&   213314.5+580351&                 \nodata&          $>$  17.55&          $>$  16.62&   15.61$\pm$   0.07&   14.65$\pm$   0.07&   14.57$\pm$   0.09&   14.25$\pm$   0.05&   14.18$\pm$   0.08&   13.51$\pm$   0.06&   12.72$\pm$   0.06&          $>$   8.33&             \nodata\\
44&   213315.6+580407&                 \nodata&   19.09$\pm$   0.07&   17.93$\pm$   0.06&   14.98$\pm$   0.04&   13.96$\pm$   0.05&   13.60$\pm$   0.04&   13.33$\pm$   0.05&   13.33$\pm$   0.05&   13.00$\pm$   0.06&   12.29$\pm$   0.06&          $>$   8.02&             \nodata\\
45&   213319.4+580406&                 \nodata&          $>$  19.47&          $>$  18.14&   16.03$\pm$   0.09&   14.72$\pm$   0.09&   14.27$\pm$   0.07&   13.92$\pm$   0.05&   13.92$\pm$   0.05&   13.49$\pm$   0.06&   12.35$\pm$   0.06&          $>$   7.08&             \nodata\\
46&   213323.8+580632&                 \nodata&   19.35$\pm$   0.08&   18.31$\pm$   0.07&   15.63$\pm$   0.07&   14.75$\pm$   0.08&   14.28$\pm$   0.08&   14.14$\pm$   0.05&   14.08$\pm$   0.05&   12.95$\pm$   0.06&   11.79$\pm$   0.06&          $>$   6.37&             \nodata\\
47&   213327.2+580413&                 \nodata&   18.66$\pm$   0.07&   17.61$\pm$   0.06&   14.87$\pm$   0.04&   13.86$\pm$   0.04&   13.67$\pm$   0.04&   13.34$\pm$   0.05&   13.29$\pm$   0.05&   12.90$\pm$   0.06&   12.00$\pm$   0.06&          $>$   7.56&             \nodata\\
48&   213329.2+580250&         Ogura1,Nakano17&   18.61$\pm$   0.07&   16.89$\pm$   0.06&   14.11$\pm$   0.03&   12.98$\pm$   0.03&   12.42$\pm$   0.03&   11.49$\pm$   0.05&   11.09$\pm$   0.05&   10.72$\pm$   0.05&   10.06$\pm$   0.05&    6.64$\pm$   0.04&             \nodata\\
49&   213332.2+580329&                 \nodata&          $>$  20.67&          $>$  20.26&          $>$  18.22&          $>$  16.22&   13.93$\pm$   0.08&    8.74$\pm$   0.05&    7.15$\pm$   0.05&    6.07$\pm$   0.05&    5.09$\pm$   0.04&    1.95$\pm$   0.04&   -2.01$\pm$   0.01\\
50&   213332.2+580558&                 \nodata&          $>$  20.46&          $>$  18.31&   15.43$\pm$   0.06&   13.89$\pm$   0.04&   13.17$\pm$   0.03&   12.69$\pm$   0.05&   12.63$\pm$   0.05&   12.35$\pm$   0.06&   11.80$\pm$   0.06&          $>$   6.47&             \nodata\\
51&   213334.8+580409&                 \nodata&          $>$  21.68&          $>$  19.76&   16.22$\pm$   0.12&   13.99$\pm$   0.04&   12.91$\pm$   0.03&   11.66$\pm$   0.05&   11.18$\pm$   0.05&   10.80$\pm$   0.05&   10.59$\pm$   0.06&          $>$   5.99&             \nodata\\
52&   213335.3+580647&                 \nodata&   19.26$\pm$   0.08&   17.78$\pm$   0.06&   14.24$\pm$   0.02&   13.08$\pm$   0.04&   12.57$\pm$   0.02&   12.12$\pm$   0.05&   12.02$\pm$   0.05&   11.62$\pm$   0.05&   10.92$\pm$   0.05&          $>$   5.94&             \nodata\\
53&   213336.2+580324&                 \nodata&   19.98$\pm$   0.12&   18.24$\pm$   0.07&   14.91$\pm$   0.05&   14.21$\pm$   0.05&   13.69$\pm$   0.05&   13.08$\pm$   0.05&   12.68$\pm$   0.05&   11.49$\pm$   0.05&   10.06$\pm$   0.05&             \nodata&             \nodata\\
54&   213336.8+580329&                 \nodata&   17.48$\pm$   0.06&   16.50$\pm$   0.06&   13.96$\pm$   0.04&   13.00$\pm$   0.04&   12.66$\pm$   0.03&   12.43$\pm$   0.05&   12.41$\pm$   0.05&   11.63$\pm$   0.05&   10.17$\pm$   0.05&          $>$   5.75&             \nodata\\
55&   213340.8+580626&                 \nodata&          $>$  20.06&   20.47$\pm$   0.32&   16.54$\pm$   0.15&   14.79$\pm$   0.07&   14.20$\pm$   0.06&   13.07$\pm$   0.05&   12.68$\pm$   0.06&   12.11$\pm$   0.07&   11.25$\pm$   0.07&          $>$   6.88&             \nodata\\
56&   213340.8+580631&                 \nodata&          $>$  24.05&          $>$  21.60&   16.22$\pm$   0.13&   14.39$\pm$   0.08&   13.61$\pm$   0.06&   12.79$\pm$   0.05&   12.66$\pm$   0.05&   12.29$\pm$   0.06&   11.55$\pm$   0.09&          $>$   6.27&             \nodata\\
\enddata       
\tablenotetext{a}{Magnitudes for $r$ and $i$ bands are in AB
magnitudes; the rest of the magnitudes here are Vega magnitudes.}
\end{deluxetable}

\begin{deluxetable}{clllllp{10cm}}
\tablecaption{Notes on the known YSOs, literature YSO candidates, and
new Spitzer-identified YSO candidates in BRC 27 and BRC
34\label{tab:ourysonotes}}
\rotate
\tabletypesize{\tiny}
\tablewidth{0pt}
\tablehead{
\colhead{row}  & \colhead{name}  &\colhead{YSO status\tablenotemark{a}} & \colhead{IRx
status\tablenotemark{b}} & \colhead{$\alpha$\tablenotemark{c}} &
\colhead{class\tablenotemark{d}} & \colhead{notes\tablenotemark{e}}}
\startdata
\cutinhead{BRC 27}
  1&   070352.2-112100&     lit.~YSO~cand.&IRx&-1.81&  III&                                                               [3.6]$-$[4.5]=0.04, [5.8]$-$[8]=0.5, so small excess at longer bands (\S\ref{sec:iracysos}); among 14 reddest BRC 27 sources in $JHK_s$ diagram, high \av\ likely (\S\ref{sec:nearirproperties}); literature optical points discontinuous with respect to rest of SED, not clear exactly why (\S\ref{sec:sednotes})\\
  2&   070352.7-112313&  lit.~YSO~cand.&no IRx&-3.11&  III&                                                                                                                                                                                                                                                [3.6]$-$[4.5]=0.07, [5.8]$-$[8]$\sim-$0.3, but very faint at [8], so large error (\S\ref{sec:iracysos}); no significant IR excess.\\
  3&   070353.2-112403&     lit.~YSO~cand.&IRx&-1.01&   II&                                                                                                                                                                                                                                                                               among 14 reddest BRC 27 sources in $JHK_s$ diagram, high \av\ likely (\S\ref{sec:nearirproperties})\\
  4&   070353.5-112350&             YSO&no IRx&-2.30&  III&                                                                                                                                                                     {\em possible} very small [8] excess; no convincing indication of IR excess, despite likely YSO in literature; bright in $r$ (\S\ref{sec:opticalproperties}); blue in $JHK_s$ (\S\ref{sec:nearirproperties}).\\
  5&   070353.7-112428&     lit.~YSO~cand.&IRx&-1.48&   II&                                                                                                                                                                                                                                                                                                                                                            {\em no special notes}\\
  6&   070353.8-112341&    new YSO cand.& IRx:&-2.59&  III&                                                                                               small IRAC colors (\S\ref{sec:iracysos}) and small but apparently significant 24 \mum\ excess (\S\ref{sec:3324ysos}); contamination possible, so ``IRx:"; bright in $r$, though with large errors (\S\ref{sec:opticalproperties}); $r$ seems too high in SED (\S\ref{sec:sednotes})\\
  7&   070354.6-112011&     lit.~YSO~cand.&IRx&-1.87&  III&                                                                                                                                                                                                                                                                 literature optical points discontinuous with respect to rest of SED, not clear exactly why (\S\ref{sec:sednotes})\\
  8&   070354.9-112514&    lit.~YSO~cand.&IRx:&-1.96&  III&                                                                                                                                                                                                                                                                               {\em possible} very small [5.8],[8] excess (\S\ref{sec:338ysos}); very weak evidence for IR excess.\\
  9&   070357.1-112432&     lit.~YSO~cand.&IRx&-1.28&   II&                                                                                                                                                                                                                                                                                                                                                            {\em no special notes}\\
 10&   070358.4-112325&     new YSO cand.& IRx&-1.93&  III&                                                                                                                                                                                                                                                                                     bright in $r$ (\S\ref{sec:opticalproperties}); blue in $JHK_s$ (\S\ref{sec:nearirproperties})\\
 11&   070400.7-112323&     new YSO cand.& IRx&-0.33&   II&                                                                                                                                                                         among 4 reddest BRC 27 objects in $JHK_s$ diagram, high \av\ likely (\S\ref{sec:jhkysos}, \S\ref{sec:sednotes}); SED slope within 0.1 of borderline between SED class flat and II (\S\ref{sec:sedslopes})\\
 12&   070401.2-112531&     new YSO cand.& IRx&-0.74&   II&                                                                                                                                                                                                                                                                               among 14 reddest BRC 27 sources in $JHK_s$ diagram, high \av\ likely (\S\ref{sec:nearirproperties})\\
 13&   070401.2-112242&     new YSO cand.& IRx& 0.00& flat&                                                                                                                   among 4 reddest BRC 27 objects in $JHK_s$ diagram, high \av\ likely (\S\ref{sec:jhkysos}); reasonably deeply embedded SED (\S\ref{sec:sednotes}); when SED slope calculated between 2 and 8 \mum, class changes from a Flat to Class II (\S\ref{sec:sedslopes})\\
 14&   070401.2-112233&     lit.~YSO~cand.&IRx&-1.28&   II&                                                                                                                                                                                                                                                         among 14 reddest BRC 27 sources in $JHK_s$ diagram, high \av\ likely (\S\ref{sec:nearirproperties}, \S\ref{sec:sednotes})\\
 15&   070401.3-112334&                YSO&IRx&-0.59&   II&                                                                                                                                                                                                                                                                                                                                     bright in $r$ (\S\ref{sec:opticalproperties})\\
 16&   070401.6-112406&     new YSO cand.& IRx& 0.70&    I&                                          $r,i$ optical detections place below ZAMS (\S\ref{sec:opticalproperties}); among 4 reddest BRC 27 objects in $JHK_s$ diagram, high \av\ likely (\S\ref{sec:jhkysos}); somewhat unusually shaped SED (\S\ref{sec:sednotes}); on bright rim, bright source with nearby nebulosity and point sources in [3.6],[4.5] (\S\ref{sec:sednotes}).\\
 17&   070401.6-112132&     new YSO cand.& IRx&-1.01&   II&                                                                                                                                                                                                                                                                               among 14 reddest BRC 27 sources in $JHK_s$ diagram, high \av\ likely (\S\ref{sec:nearirproperties})\\
 18&   070402.1-112512&     new YSO cand.& IRx&-0.55&   II&                                                                                                                                                                                                                                                                               among 14 reddest BRC 27 sources in $JHK_s$ diagram, high \av\ likely (\S\ref{sec:nearirproperties})\\
 19&   070402.2-112542&     new YSO cand.& IRx& 0.13& flat&   bright in $r$ (\S\ref{sec:opticalproperties}); somewhat unusually shaped SED (see \S\ref{sec:sednotes}); 8 \mum\ bright from PAH emission? (\S\ref{sec:sednotes}); very bright, may be nebulous companion to bright 070402.7-112325=row 20 (\S\ref{sec:sednotes}); when SED slope calculated between 2 and 8 \mum, class changes from a Flat to Class I (\S\ref{sec:sedslopes})\\
 20&   070402.3-112539&                YSO&IRx&-0.07& flat&                                                                                                                                            bright in $r$ (\S\ref{sec:opticalproperties}); blue in $JHK_s$ (\S\ref{sec:nearirproperties}); `flat' class (\S\ref{sec:sedslopes} from fitting \ks\ to [8] but very bright star, small fractional IR excess, $K_s$ not at peak of SED\\
 21&   070402.7-112325&     new YSO cand.& IRx& 0.13& flat&                                                                                                                                                                                                                                                                                                                             reasonably deeply embedded SED (\S\ref{sec:sednotes})\\
 22&   070402.9-112337&     lit.~YSO~cand.&IRx&-0.07& flat&                                                                                                                                      two previously identified sources are unresolved (\S\ref{sec:litsrcs}); reasonably deeply embedded SED (\S\ref{sec:sednotes}); when SED slope calculated between 2 and 8 \mum, class changes from a Flat to Class II (\S\ref{sec:sedslopes})\\
 23&   070403.0-112350&                YSO&IRx&-0.18& flat&                                                                                                                                                                                                                        among 14 reddest BRC 27 sources in $JHK_s$ diagram, high \av\ likely (\S\ref{sec:nearirproperties}); reasonably deeply embedded SED (\S\ref{sec:sednotes})\\
 24&   070403.1-112327&     lit.~YSO~cand.&IRx&-0.62&   II&                                                                                                                                                            among 14 reddest BRC 27 sources in $JHK_s$ diagram, high \av\ likely (\S\ref{sec:nearirproperties}); literature optical points discontinuous with respect to rest of SED, not clear exactly why (\S\ref{sec:sednotes})\\
 25&   070403.9-112609&  lit.~YSO~cand.&no IRx&-2.77&  III&                                                                                                                                                                                  no detectable IRAC excess (\S\ref{sec:iracysos}) and undetected at [24]; no significant IR excess; bright in $r$ (\S\ref{sec:opticalproperties}); blue in $JHK_s$ (\S\ref{sec:nearirproperties})\\
 26&   070403.9-112326&     new YSO cand.& IRx& 0.37&    I&                                                                                                                                          among 4 reddest BRC 27 objects in $JHK_s$ diagram, high \av\ likely (\S\ref{sec:jhkysos}); SED suggests substantial disk (\S\ref{sec:sednotes}); SED slope within 0.1 of borderline between SED class I and flat (\S\ref{sec:sedslopes})\\
 27&   070404.2-112355&                YSO&IRx&-0.85&   II&                                                                                                                                                                                                                                                                                                                                                            {\em no special notes}\\
 28&   070404.5-112555&  lit.~YSO~cand.&no IRx&-1.52&   II&                                                                                   large errors on [5.8] and [8], which is the only indication of IR excess (\S\ref{sec:338ysos}); no significant IR excess; 5.8 and 8 \mum\ points discontinuous with rest of SED (\S\ref{sec:sednotes}); SED slope within 0.1 of borderline between SED class II and III (\S\ref{sec:sedslopes})\\
 29&   070404.7-112339&     lit.~YSO~cand.&IRx&-0.65&   II&                                                                                                                                                                                                                                                                                                                                                            {\em no special notes}\\
 30&   070405.1-112313&                YSO&IRx&-0.31&   II&                                                                                                                                                                                     among 14 reddest BRC 27 sources in $JHK_s$ diagram, high \av\ likely (\S\ref{sec:nearirproperties}); SED slope within 0.1 of borderline between SED class flat and II (\S\ref{sec:sedslopes})\\
 31&   070405.7-112123&     new YSO cand.& IRx&-0.67&   II&                                                                                                                                                                                                                                                                                             5.8 and 8 \mum\ points somewhat discontinuous with rest of SED (\S\ref{sec:sednotes})\\
 32&   070405.9-112358&     lit.~YSO~cand.&IRx&-0.54&   II&                                                                                                                                                                                                                                                                                                                                                            {\em no special notes}\\
 33&   070406.0-112128&    new YSO cand.& IRx:&-1.06&   II&   [3.6]$-$[4.5]=0, [5.8]$-$[8]=0.23 (\S\ref{sec:iracysos}); [3.6]$-$[24]$>$4 mags, though nebular contamination possible (\S\ref{sec:3324ysos}); uncertain IRx; bright in $r$ (\S\ref{sec:opticalproperties}); blue in $JHK_s$ (\S\ref{sec:nearirproperties}); when SED slope calculated between 2 and 8 \mum, class changes from a Class II to Class III (\S\ref{sec:sedslopes})\\
\tablebreak
 34&   070406.0-112315&     lit.~YSO~cand.&IRx&-0.18& flat&                                                                                                                                                                among 14 reddest BRC 27 sources in $JHK_s$ diagram, high \av\ likely (\S\ref{sec:nearirproperties}); when SED slope calculated between 2 and 8 \mum, class changes from a Flat to Class II (\S\ref{sec:sedslopes})\\
 35&   070406.4-112336&                YSO&IRx&-1.59&   II&                                                                                                                                                                                                                                                                                           SED slope within 0.1 of borderline between SED class II and III (\S\ref{sec:sedslopes})\\
 36&   070406.5-112227&    new YSO cand.& IRx:&-2.12&  III&                                                                                                                                                                                                                                                                      {\em possible} very small [5.8],[8] excess (\S\ref{sec:338ysos}), no [24]; very weak evidence for IR excess.\\
 37&   070406.5-112128&     new YSO cand.& IRx& 0.96&    I&                                                                                                                                                                                                                   $r,i$ optical detections particularly uncertain, and place below ZAMS (\S\ref{sec:opticalproperties}); somewhat unusually shaped SED (see \S\ref{sec:sednotes})\\
 38&   070406.5-112316&     lit.~YSO~cand.&IRx&-1.32&   II&                                                                                                                                                                                                                                                                                                                                                            {\em no special notes}\\
 39&   070407.9-112311&     lit.~YSO~cand.&IRx&-1.30&   II&                                                                                                                                                                                                           selected in \S\ref{sec:338ysos} for a $\sim$11$\sigma$ excess at [8] and no [24]; high \av\ likely; abrupt change in SED between $K_s$ and [3.6] (\S\ref{sec:sednotes})\\
 40&   070408.0-112354&                YSO&IRx&-0.60&   II&                                                                                                                                 bright in $r$ (\S\ref{sec:opticalproperties}); large excess at [24]; inner disk hole or contamination (\S\ref{sec:sednotes})?; when SED slope calculated between 2 and 8 \mum, class changes from a Class II to Class III (\S\ref{sec:sedslopes})\\
 41&   070408.1-112313&     new YSO cand.& IRx&-1.60&   II&                                                                                                                                                                                                                            multi-IRAC band weak excess and no [24] (\S\ref{sec:338ysos}); SED slope within 0.1 of borderline between SED class II and III (\S\ref{sec:sedslopes})\\
 42&   070408.1-112309&                YSO&IRx&-1.13&   II&                                                                                                                                                                                                                                                                                                               abrupt change in SED between $K_s$ and [3.6] (\S\ref{sec:sednotes})\\
   \cutinhead{BRC 34}
 43&   213314.5+580351&     new YSO cand.& IRx&-1.52&   II&                                                                                                identified from [8] excess (\S\ref{sec:338ysos}) with $>$10$\sigma$ significance; inner disk hole?; 5.8 and 8 \mum\ points somewhat discontinuous with rest of SED (\S\ref{sec:sednotes}); SED slope within 0.1 of borderline between SED class II and III (\S\ref{sec:sedslopes})\\
 44&   213315.6+580407&     new YSO cand.& IRx&-1.95&  III&                                                                                                                                                                                                                                                                                identified from [8] excess (\S\ref{sec:338ysos}) with $>$10$\sigma$ significance; inner disk hole?\\
 45&   213319.4+580406&     new YSO cand.& IRx&-1.56&   II&                                                                                                                                                                                       identified from [8] excess (\S\ref{sec:338ysos}) with $>$10$\sigma$ significance; inner disk hole?; SED slope within 0.1 of borderline between SED class II and III (\S\ref{sec:sedslopes})\\
 46&   213323.8+580632&     new YSO cand.& IRx&-1.06&   II&                                                                                                                               identified from [8] excess (\S\ref{sec:338ysos}) with $>$10$\sigma$ significance; inner disk hole?; located on ZAMS in $r,i$ (\S\ref{sec:opticalproperties}); 5.8 and 8 \mum\ points somewhat discontinuous with rest of SED (\S\ref{sec:sednotes})\\
 47&   213327.2+580413&     new YSO cand.& IRx&-1.71&  III&                                                                                                                                                                                                                                                                                                                                                            {\em no special notes}\\
 48&   213329.2+580250&                YSO&IRx&-0.68&   II&                                                                                                                                                                                                                                                                                                             this is the only literature YSO or literature YSO candidate in BRC 34\\
 49&   213332.2+580329&     new YSO cand.& IRx& 1.32&    I&                                                                                                                                                                               largest [3.6]$-$[4.5](=1.6) of both BRCs (\S\ref{sec:iracysos}); high \av\ likely (\S\ref{sec:iracysos}, \ref{sec:sednotes}); bright at [70]; reasonably deeply embedded SED (\S\ref{sec:sednotes})\\
 50&   213332.2+580558&     new YSO cand.& IRx&-1.89&  III&                                                                                                                                                                                                                   identified from [8] excess (\S\ref{sec:338ysos}) with $>$10$\sigma$ significance; in 4 reddest BRC34 sources in $JHK_s$, high \av\ likely (\S\ref{sec:jhkysos})\\
 51&   213334.8+580409&     new YSO cand.& IRx&-1.12&   II&                                                                                                                                                                                                               low [5.8]$-$[8] (=0.21) but high [3.6]$-$[4.5](=0.48) (\S\ref{sec:iracysos}); $H-K_s=1.08$, $J-H=2.3$, high \av\ likely (\S\ref{sec:jhkysos}, \S\ref{sec:sednotes})\\
 52&   213335.3+580647&     new YSO cand.& IRx&-1.68&  III&                                                                                                                                                                                                                                                                                           SED slope within 0.1 of borderline between SED class II and III (\S\ref{sec:sedslopes})\\
 53&   213336.2+580324&     new YSO cand.& IRx&-0.26& flat&                                                                                                                                                                           5.8 and 8 \mum\ points rather abruptly rise compared to rest of SED (\S\ref{sec:sednotes}); inner disk hole??; SED slope within 0.1 of borderline between SED class flat and II (\S\ref{sec:sedslopes})\\
 54&   213336.8+580329&     new YSO cand.& IRx&-1.15&   II&                                                                                                                                                                                                                                                                                   5.8 and 8 \mum\ points discontinuous with rest of SED (\S\ref{sec:sednotes}); inner disk hole??\\
 55&   213340.8+580626&     new YSO cand.& IRx&-0.75&   II&                                                                                                                                                                                                                                                                               in 4 reddest BRC34 sources in $JHK_s$, high \av\ likely (\S\ref{sec:jhkysos}, \S\ref{sec:sednotes})\\
 56&   213340.8+580631&     new YSO cand.& IRx&-1.42&   II&                                                                                                                                                                                                                                                                               in 4 reddest BRC34 sources in $JHK_s$, high \av\ likely (\S\ref{sec:jhkysos}, \S\ref{sec:sednotes})\\
\enddata       
\tablenotetext{a}{YSO status can be decoded as follows:
lit.~YSO~cand.= literature YSO candidate; YSO = literature likely YSO;
new YSO can.= new YSO candidate identified here.}
\tablenotetext{b}{IRx (IR excess) status can be decoded as follows:
IRx=IR excess detected here; IRx:=uncertain IR excess identified
here; no IRx=no IR excess detected here.}
\tablenotetext{c}{$\alpha$ is the slope of the SED between 2 and 24
\mum, obtained as described in the text.}
\tablenotetext{d}{SED class is obtained by binning up the SED slope
($\alpha$) values as described in the text into the classes defined in
Section~\ref{sec:evol}.}
\tablenotetext{e}{Notes on individual objects as described in the
text.}
\end{deluxetable}
\clearpage

\section{Properties of YSOs and YSO Candidates}
\label{sec:properties}

\subsection{Optical Properties}
\label{sec:opticalproperties}

\begin{figure*}[tbp]
\epsscale{1}
\plotone{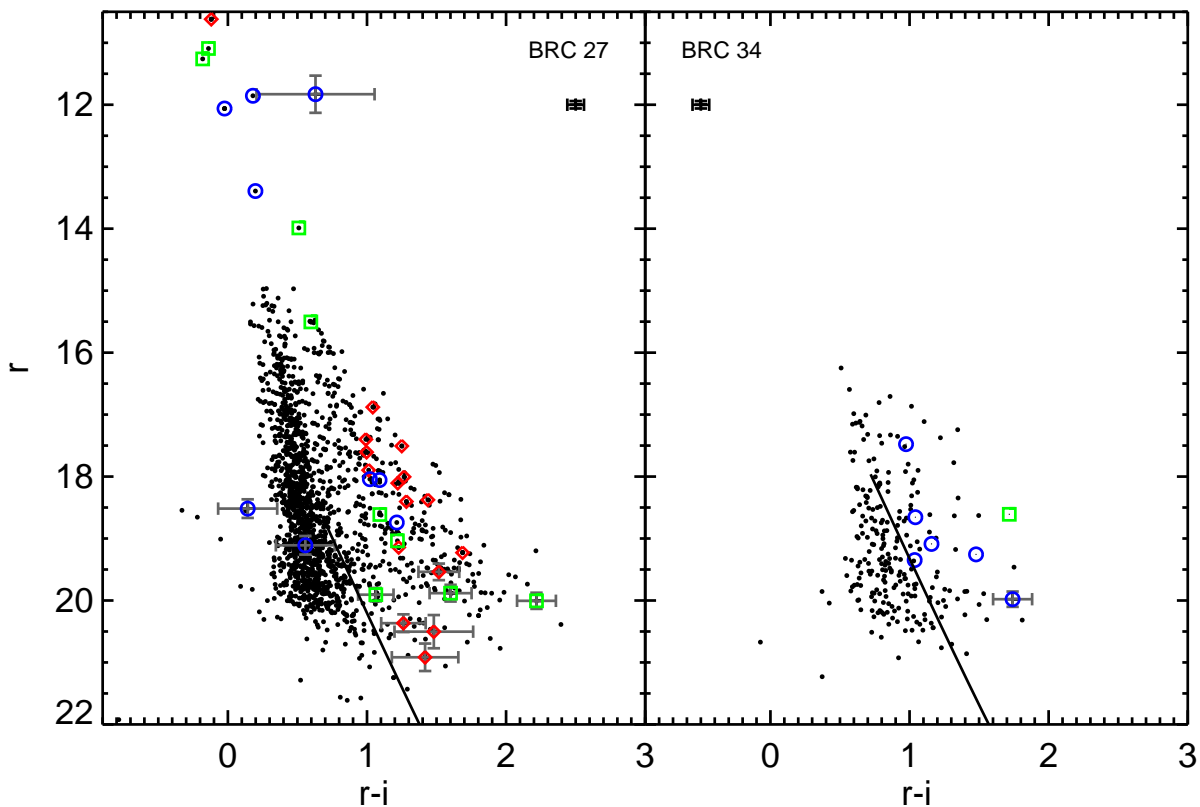}
\caption{$r$ vs.\ $(r-i)$ color-magnitude diagram for BRC 27 (left)
and BRC 34 (right). Note that these are AB magnitudes (not Vega
magnitudes). No comparison to CG4+Sa101 or SWIRE is provided
here, as optical color-magnitude diagrams are reasonably common.    
In each panel, small dots are objects in the catalog (i.e., objects
seen in the image); green squares are literature high-confidence YSOs,
red diamonds are literature candidate YSO, and blue circles are our
new candidate YSOs presented here.  
Typical error bars are indicated below the BRC names in the plots --
$\sim$0.06 mags in $r$ and $r-i$. Objects with exceptionally large
error bars have error bars overplotted in grey.  
The solid black line is the main sequence relation from Bochanski
\etal\ (2010), shifted to 1.2 kpc for BRC 27 and to 800 pc for BRC 34;
the relation is not provided for $r-i<0.62$.
Most of the YSOs and YSO candidates are in the location expected for
confirmed YSOs. Two new YSO candidates in BRC 27 seem anomalously
faint, as do three objects seen here only as limits; see text. }
\label{fig:optysos}
\end{figure*}

Optical photometric data can greatly aid in confirming or refuting YSO
candidacy because they provide constraints on the Wien side of the
SED. In Guieu \etal\ (2010), most of the IRAC-selected candidates in
IC~2118 proved to be too faint, most vividly in the optical, to be
likely cluster members. We obtained new optical data for our BRC 27
and 34 candidates as described in Section~\ref{sec:opticaldata}. An
optical color-magnitude diagram for our regions and candidates appears
in Figure~\ref{fig:optysos}; recall that these optical magnitudes are
in the AB system rather than the Vega system.  Few galaxies are
expected to be detected in these bands; most of the objects seen here
are stars, and most of the stars from the rest of the Milky Way Galaxy
in the background will fall below the zero-age main sequence (ZAMS)
placed at an appropriate distance for the BRCs we are considering. 
Legitimate young stars will be located above the ZAMS, but so will AGB
star contaminants and foreground stars. Legitimate young stars may
also suffer from reddening, which will slide the observed points back
along a line roughly parallel to the ZAMS. Some legitimate young stars
suffering very high extinction (notably edge-on disks where the
detected optical light is primarily scattered light) may appear below
the ZAMS (see, e.g., Rebull \etal\ 2010). 

Most of the literature YSOs, literature YSO candidates, and new YSO
candidates, at least those for which we have enough optical data to
plot them in the Figure, are in the expected location in the optical
CMD, meaning above the clump of objects in the rest of the Galaxy and
above the ZAMS. Many of the objects that have IRAC measurements but
were not selected as YSO candidates above (e.g., failing the
Gutermuth-style color cuts in Section~\ref{sec:coloroverview}) also
fall in the clump of objects from the rest of the Galaxy.   The YSOs
and candidate YSOs in BRC 27 between $r\sim16$ and $\sim$20 that are
above the ZAMS are even reasonably tightly correlated roughly along an
isochrone (which is roughly parallel to the ZAMS), consistent with
them all being members of a co-eval cluster (e.g., Orion in Rebull
\etal\ 2000). It is not as strongly correlated, as expected, at the
fainter end, because the photometry gets more uncertain.  There are
fewer objects in BRC 34, and they are less clearly clumped along an
isochrone; this may be an indication that there are more contaminants
in the BRC 34 sample, or there may be more of a range of apparent ages
in this region.

Nine of the BRC 27 objects, candidates and known YSOs, are very
bright, with $r\lesssim16$. A common contaminant in these kinds of
Spitzer-driven source selection is reddened background asymptotic
giant branch (AGB) stars. None of the SEDs for these objects (see
\S\ref{sec:seds}) resemble highly reddened objects, and they are not
bright enough to be nearby AGB stars. The bright known YSOs or
literature YSO candidates are of comparable brightness to the new
bright YSO candidates, so the new objects are not distinctly different
in optical properties.  None of the BRC 34 objects are as bright as
these brightest BRC 27 objects, despite the fact that BRC 34 is closer
at $\sim$800 pc (vs.\ 1.2 kpc for BRC 27). Follow-up spectra are
desirable, including line regions that can discriminate AGB stars
from YSOs.  These optically bright objects (also tagged in
Table~\ref{tab:ourysonotes}) are: 070353.5-112350 (=row 4),
070353.8-112341 (=row 6; this is the bright one with the large error in
Figure~\ref{fig:optysos}), 070358.4-112325 (=row 10), 070401.3-112334
(=row 15), 070402.2-112542 (=row 19), 070402.3-112539 (=row 20),
070403.9-112609 (=row 25), 070406.0-112128 (=row 33), and 
070408.0-112354 (=row 40).

Two of the new YSO candidate objects from BRC 27, 070401.6-112406 (=row
16) and 070406.5-112128 (=row 37), appear below the ZAMS in
Figure~\ref{fig:optysos}. The latter has large enough errors that it
could also be on (rather than below) the ZAMS, though even that would
still set it apart from most of the rest of the YSOs and candidates in
Figure~\ref{fig:optysos}. We examined the optical images for these
objects, and both objects can clearly be seen in the images. However,
070406.5-112128 (=row 37) is near a brighter star such that it falls
right on a diffraction spike. We have attempted to estimate the flux
density despite the spike, but there is most likely a larger
uncertainty on that measurement than we have estimated. Both SEDs are
somewhat unusual (see Section \ref{sec:sednotes} below, and
Table~\ref{tab:ourysonotes}), but they are located right on the bright
rim of the BRC (Section~\ref{sec:locationonthesky}). We have retained
these in the list of YSO candidates under consideration although there
is some uncertainty, in particular, for these objects.

One object from BRC 34 is on the ZAMS -- 213323.8+580632 (=row 46).
Notably, this object was one of the ones added based on the
[3.6]$-$[8] color in Section~\ref{sec:338ysos}, and was not
automatically selected by the Gutermuth-style color cuts
(Section~\ref{sec:coloroverview}). While this optical placement does
not rule out the selection of this object as a YSO candidate, there is
more uncertainty for this object than many of the rest of the YSO
candidates in BRC 34.

As mentioned in Section~\ref{sec:opticaldata} above, we sought out
optical detections or limits for those objects on our list of new YSO
candidates. Therefore, despite our optical completeness limits of 
$r\sim20$, $i\sim19$, some of the objects in Table~\ref{tab:ourysos}
have optical detections fainter than this; each of those have been
checked in the image. Many objects are too faint to be detected on the
image, which may be an indication of embeddedness, or it may be an
indication that the object is a contaminant.  Those objects with high
\av\ can be selected either from the shape of the SED (see
\S\ref{sec:seds}), or from the near-IR, which we now discuss.

\subsection{Near-IR Properties}
\label{sec:nearirproperties}
\label{sec:jhkysos}

\begin{figure*}[tbp]
\epsscale{1}
\plotone{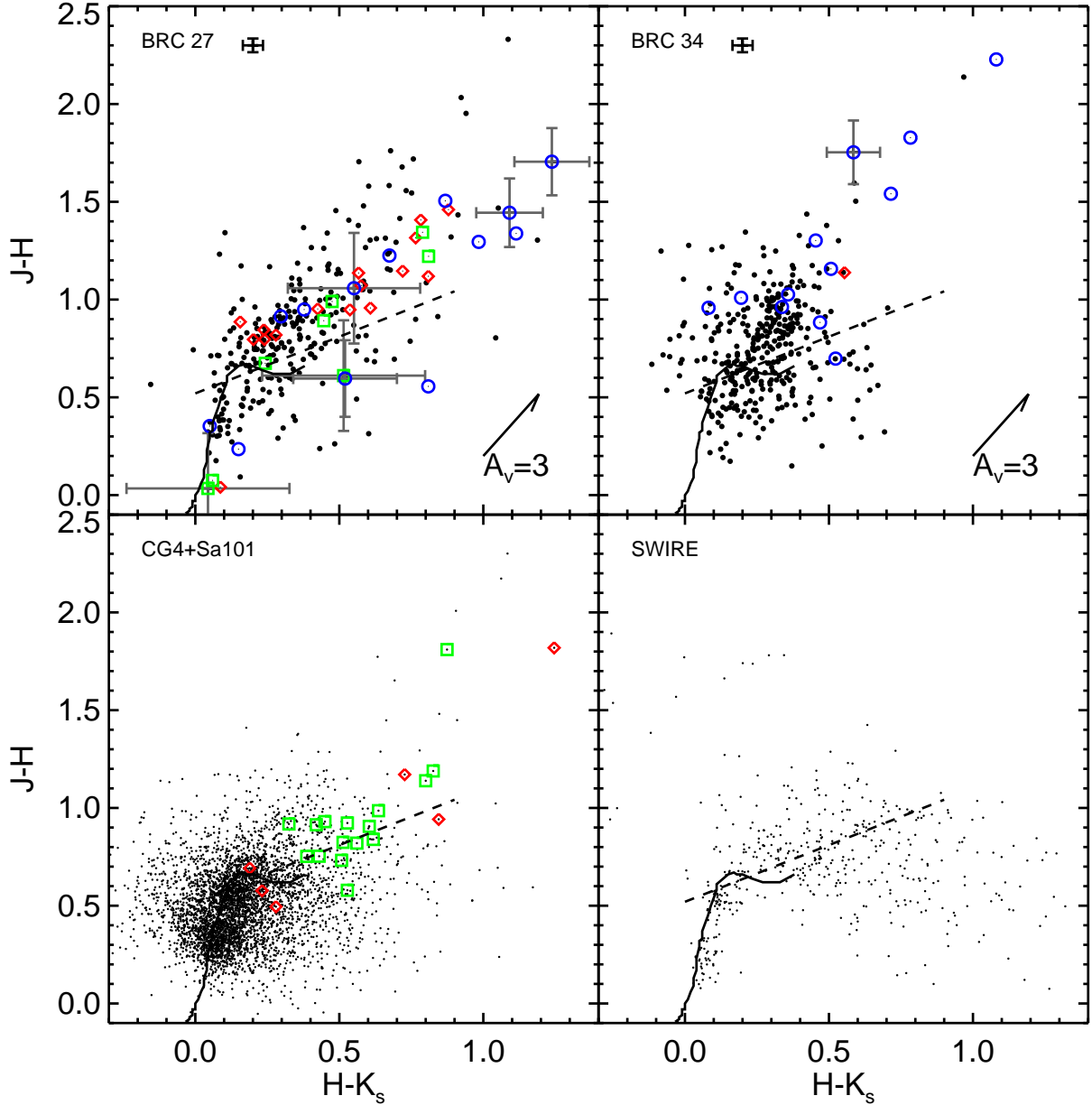}
\caption{$J-H$ vs.\ $H-K_s$ color-color diagram for BRC 27 (upper
left) and BRC 34 (upper right), with CG4+Sa101 (lower left) and SWIRE
(lower right) for comparison. In each panel, small dots are objects in
the catalog (i.e., objects seen in the image); green squares are
literature high-confidence YSOs, red diamonds  are literature
candidate YSO, and blue circles are our new candidate YSOs presented
here.  Sample error bars are indicated just to the right of the BRC
label in each of the BRC plots. Objects with exceptionally large error
bars have error bars overplotted in grey. A sample \av=3 (note not 30
as in the other plots) vector is included in the BRC 27 and 34 panels
for reference. The ZAMS is indicated by a solid line, and the
dashed line is the locus of dereddened young stars with a NIR excess
from Meyer \etal\ (1997).  Many of the known YSOs, literature YSO
candidates, and new YSO candidates have a NIR excess starting at
H-band with moderate reddening; see text for more discussion. }
\label{fig:jhkysos}
\end{figure*}

Near-IR photometric data can also aid in confirming or refuting YSO
candidacy, through adding points to the SED and through location of
the objects in the color-color diagram. Since we do not have spectral
types for most of our sources, it is difficult to estimate the degree
of reddening (\av) uniquely for each object, but $JHK_s$ data can help
us identify those objects with likely large \av.
Figure~\ref{fig:jhkysos} shows $J-H$ vs.\ $H-K_s$ for the sample, with
the data from CG4+Sa101 and SWIRE again included for comparison.
Dust-free and reddening-free photospheres will follow the main
sequence relation. SWIRE's few stars cluster around the ZAMS relation;
a large number of stars cluster around the ZAMS relation in CG4+Sa101.
Because 2MASS is relatively shallow, relatively few galaxies are
expected to be detected in the BRCs, though galaxies can appear in the
same portion of the diagram as YSOs (as seen in the SWIRE panel of
Fig.~\ref{fig:jhkysos}).  Stars (or, indeed, any objects) that are
simply reddened will be shifted along the reddening vector as
indicated to the upper right; if the star has no NIR excess due to a
circumstellar disk, an estimate of \av\ can be obtained by moving the
object back along the reverse direction of the \av\ vector until
intercepting the main sequence relation. However, there is a
degeneracy in this process in that one can move the star back until it
intercepts the low-mass portion of the ZAMS relation or the
higher-mass portion, so a spectral classification is needed to obtain
a good estimate of \av. Moreover, stars with large NIR excesses due to
circumstellar dust, when dereddened, cluster along a locus defined by
Meyer \etal\ (1997), so objects with high \av\ as well as a NIR excess
may not slide all the way back to the ZAMS relation.  

Figure~\ref{fig:jhkysos} suggests that many of our new YSO candidates
have an infrared excess with a moderate degree of reddening. In BRC
27, 30 of the 38 shown here (82\%) are above the locus from Meyer
\etal\ (1997); in BRC 34, 12 of the 13 shown here (92\%) are above the
locus from Meyer \etal\ (1997). Far fewer objects in CG4+Sa101 (and
essentially none in SWIRE) are subject to comparable levels of
reddening. 

The four most extreme red YSOs or candidate YSOs in BRC 34 are, in
order from reddest to bluest, 213334.8+580409 (=row 51 in the Tables),
213340.8+580631 (=row 56), 213332.2+580558 (=row 50), and
213340.8+580626 (=row 55). These objects have $H-K_s>0.56$ and
$J-H>1.2$. All four of these objects can be moved back to the main
sequence relation without invoking an excess at $K_s$ band, and their
SED shapes support this (Section~\ref{sec:seds}).  These are noted in
Table~\ref{tab:ourysonotes}.

In BRC 27, 14 YSOs or candidate YSOs have $H-K_s>0.6$ and $J-H>1.1$.
The four reddest are 070400.7-112323 (=row 11), 070401.2-112242 (=row
13), 070401.6-112406 (=row 16), 070403.9-112326 (=row 26), all of which
are new YSO candidates. The next ten are 070352.2-112100 (=row
1=Chauhan 109),  070353.2-112403 (=row 3=Ogura 3),  070401.2-112531
(=row 12), 070401.2-112233 (=row 14=Chauhan-anon), 070401.6-112132 (=row
17), 070402.1-112512 (=row 18), 070403.0-112350 (=row 23=Ogura
10=Chauhan 85), 070403.1-112327 (=row 24=Chauhan 107), 070405.1-112313
(=row 30=Ogura 15=Chauhan 88), and  070406.0-112315 (=row 34=Ogura
17=Chauhan 90).  These are noted in Table~\ref{tab:ourysonotes}.  
Most if not all of these can be moved back to the Meyer \etal\ (1997)
locus before reaching the ZAMS relation, suggesting that most of them
have an excess at $K_s$ band. Inspection of the SED shapes support
this in many cases (Section~\ref{sec:sednotes}).

At the blue end, none of the BRC 34 YSOs or candidate YSOs but up to
five of the YSOs or candidate YSOs in BRC 27 are likely high-mass
objects (B or A stars) with little or no \av.  These objects may have
formed in or near BRC 27, or they may be foreground objects.  The only
two previously identified objects in BRC 27 that have spectral types
are B and A stars, and are among these bluest objects. These bluest
YSOs and YSO candidates are 070353.5-112350 (=row 4=Shevchenko 90, type
A0),  070358.4-112325 (=row 10), 070402.3-112539 (=row 20=Shevchenko 99,
Gregorio 75, type B3-5), 070403.9-112609 (=row 25=Shevchenko 102), and
070406.0-112128 (=row 33). These are also noted in
Table~\ref{tab:ourysonotes}.

\subsection{Spectral Energy Distributions}
\label{sec:seds}

\begin{figure*}[tbp]
\epsscale{1}
\plotone{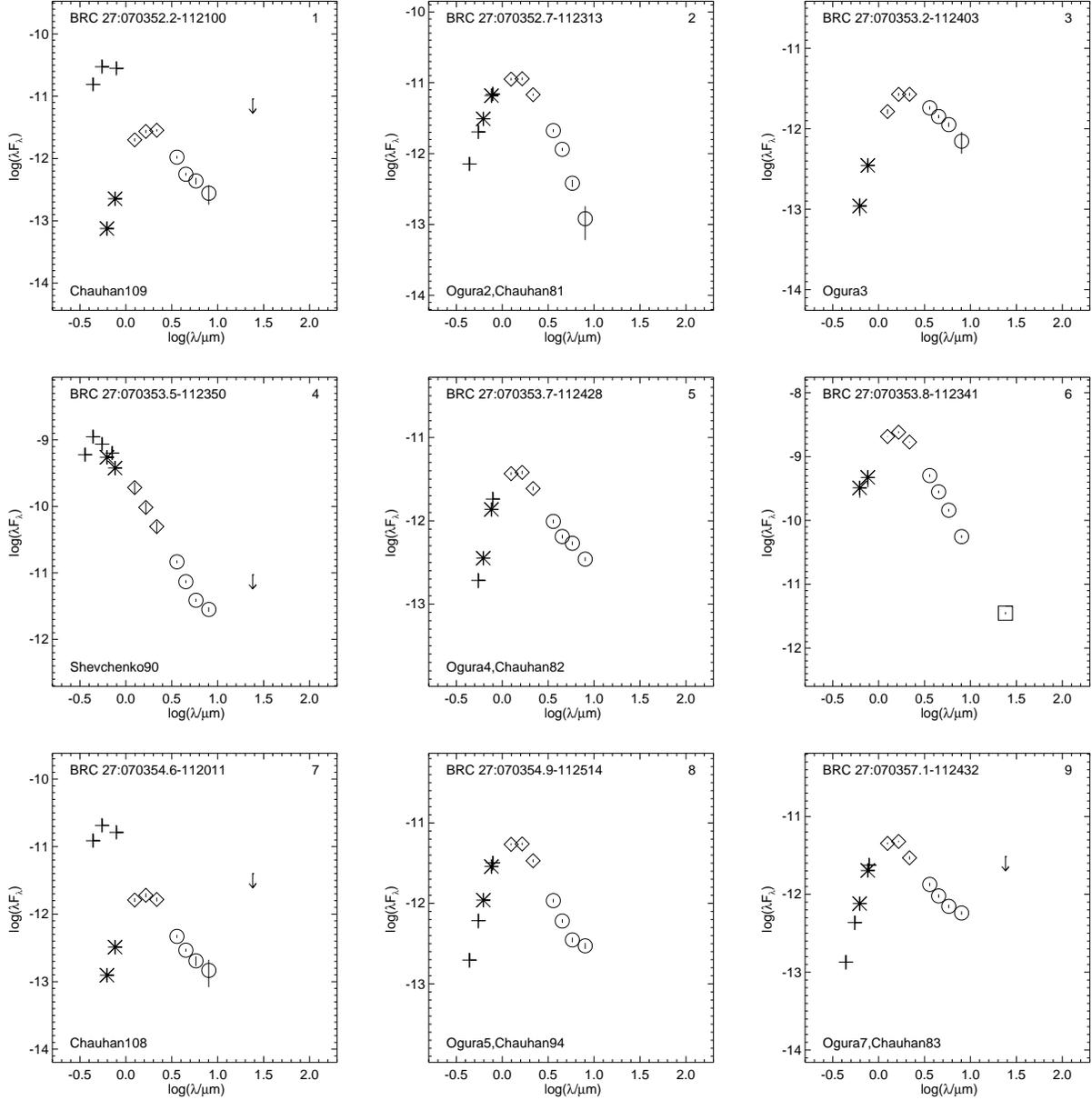}
\caption{Spectral Energy Distributions (SEDs) for the YSOs and YSO
candidates discussed here.   Units of $\lambda F_{\lambda}$  as
presented are erg s$^{-1}$ cm$^{-2}$, and $\lambda$ is in microns. $+$
symbols are optical data from the literature, asterisks are our new
optical data, diamonds are 2MASS (NIR) data, circles are IRAC data,
and squares are MIPS data.  Arrows are upper limits.   The error bars (most
frequently far smaller than the size of the symbol) are indicated at
the center of the symbol. Catalog numbers appear in the upper left,
row numbers (from the Tables) appear in the upper right, and a prior
identification, if it exists, is in the lower left.}
\label{fig:seds1}
\end{figure*}

\begin{figure*}[tbp]
\epsscale{1}
\plotone{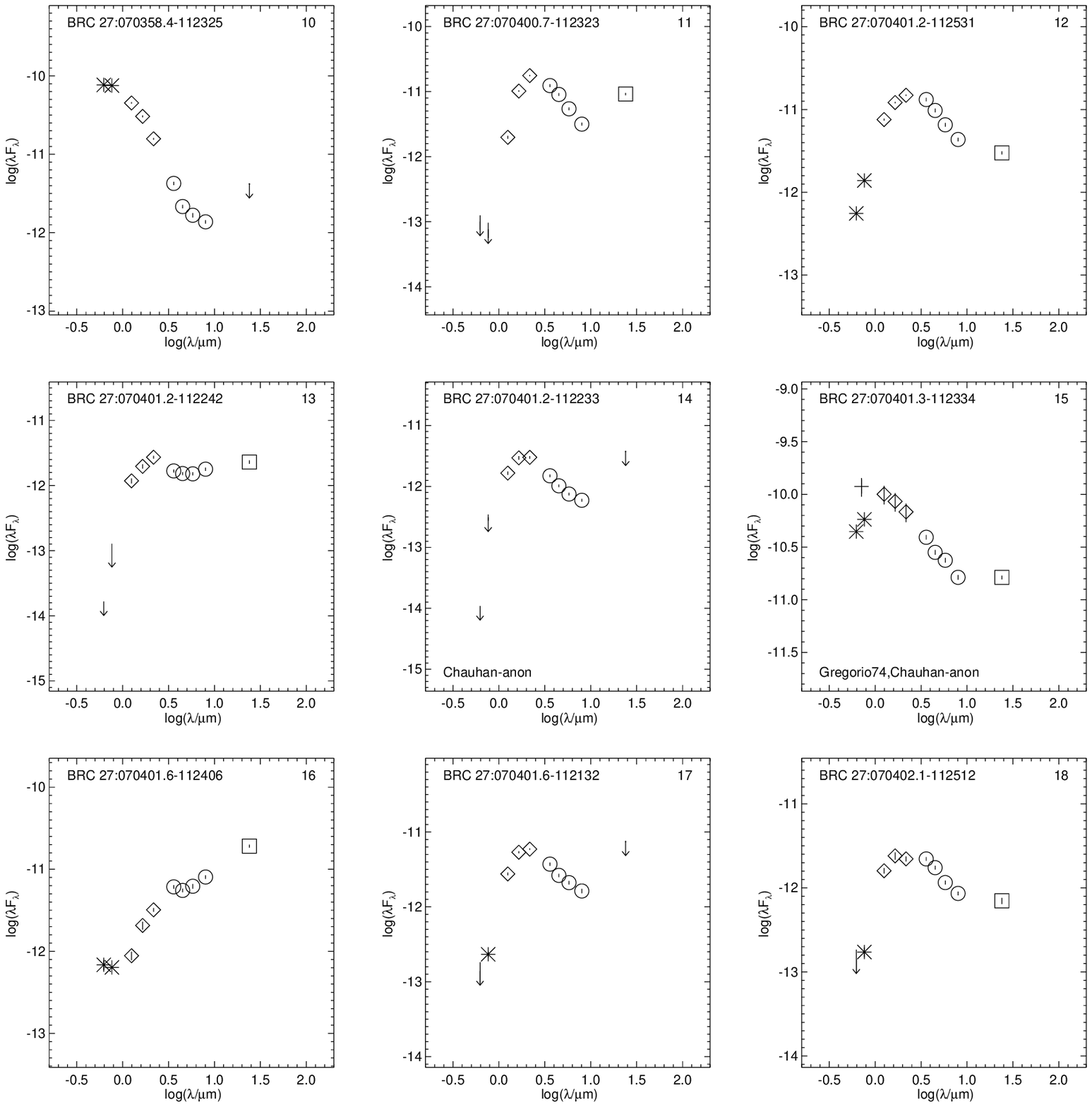}
\caption{SEDs, continued. Notation as in previous figure.}
\label{fig:seds2}
\end{figure*}

\begin{figure*}[tbp]
\epsscale{1}
\plotone{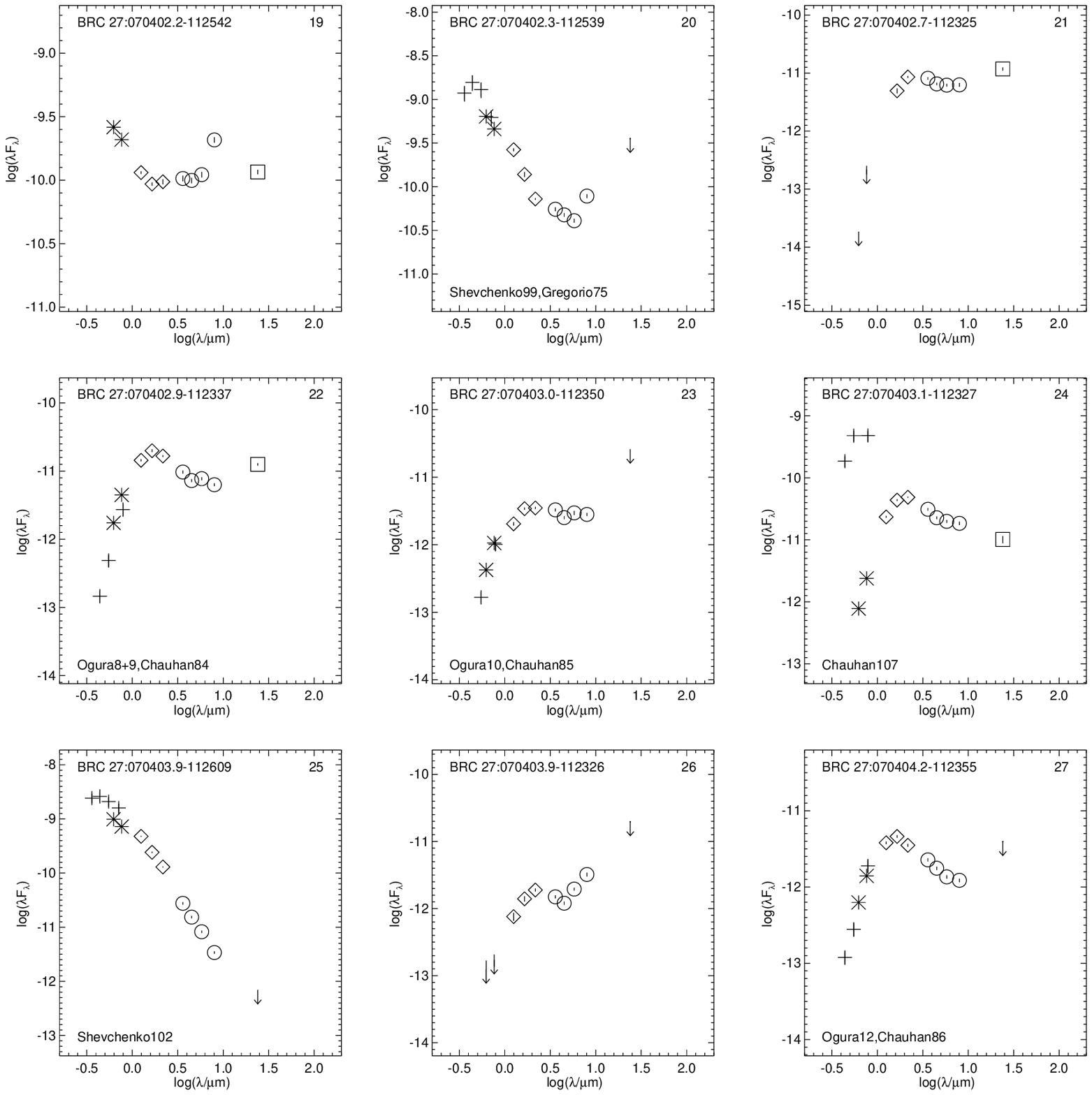}
\caption{SEDs, continued. Notation as in previous figure.}
\label{fig:seds3}
\end{figure*}

\begin{figure*}[tbp]
\epsscale{1}
\plotone{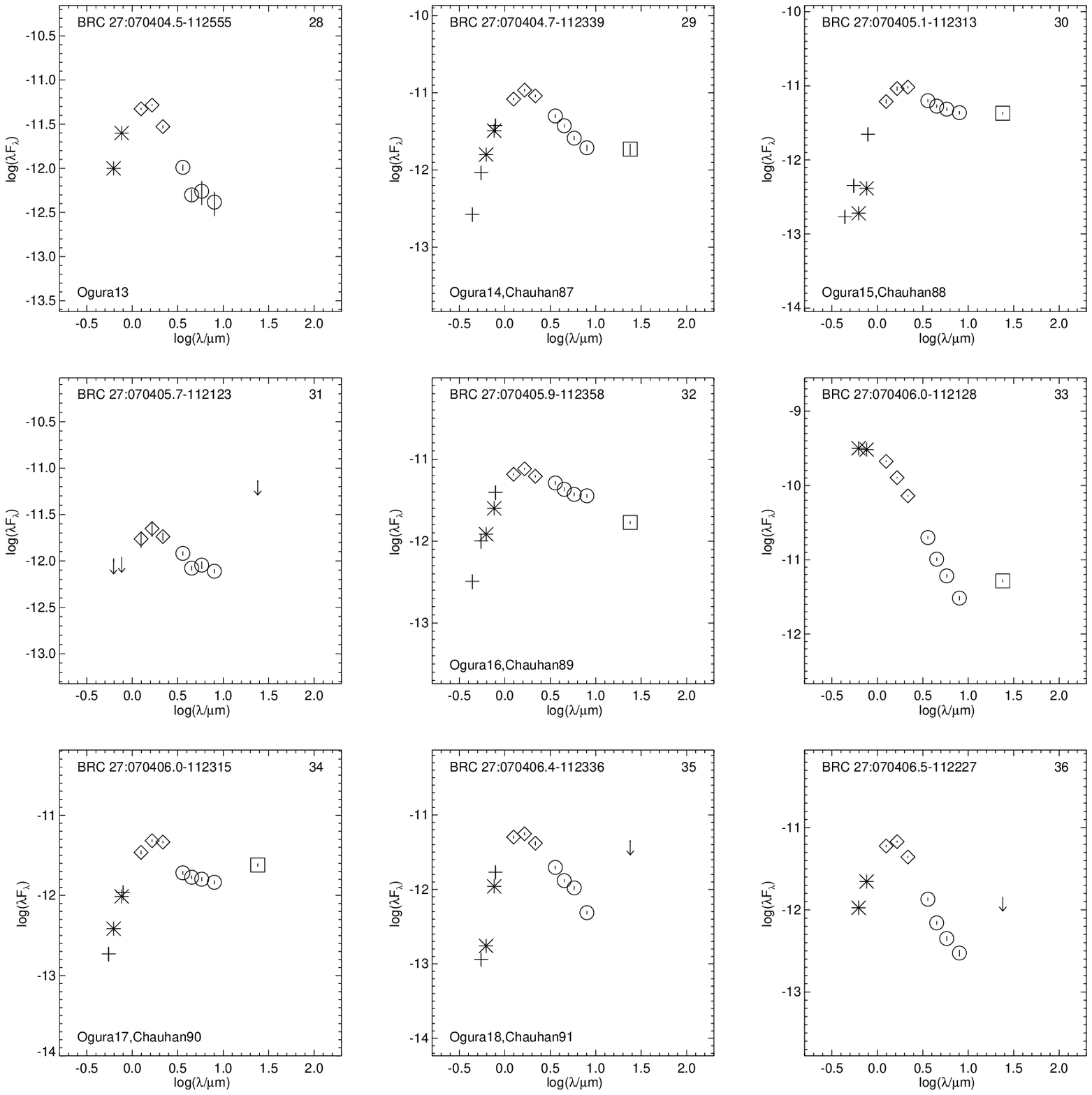}
\caption{SEDs, continued. Notation as in previous figure.}
\label{fig:seds4}
\end{figure*}

\begin{figure*}[tbp]
\epsscale{1}
\plotone{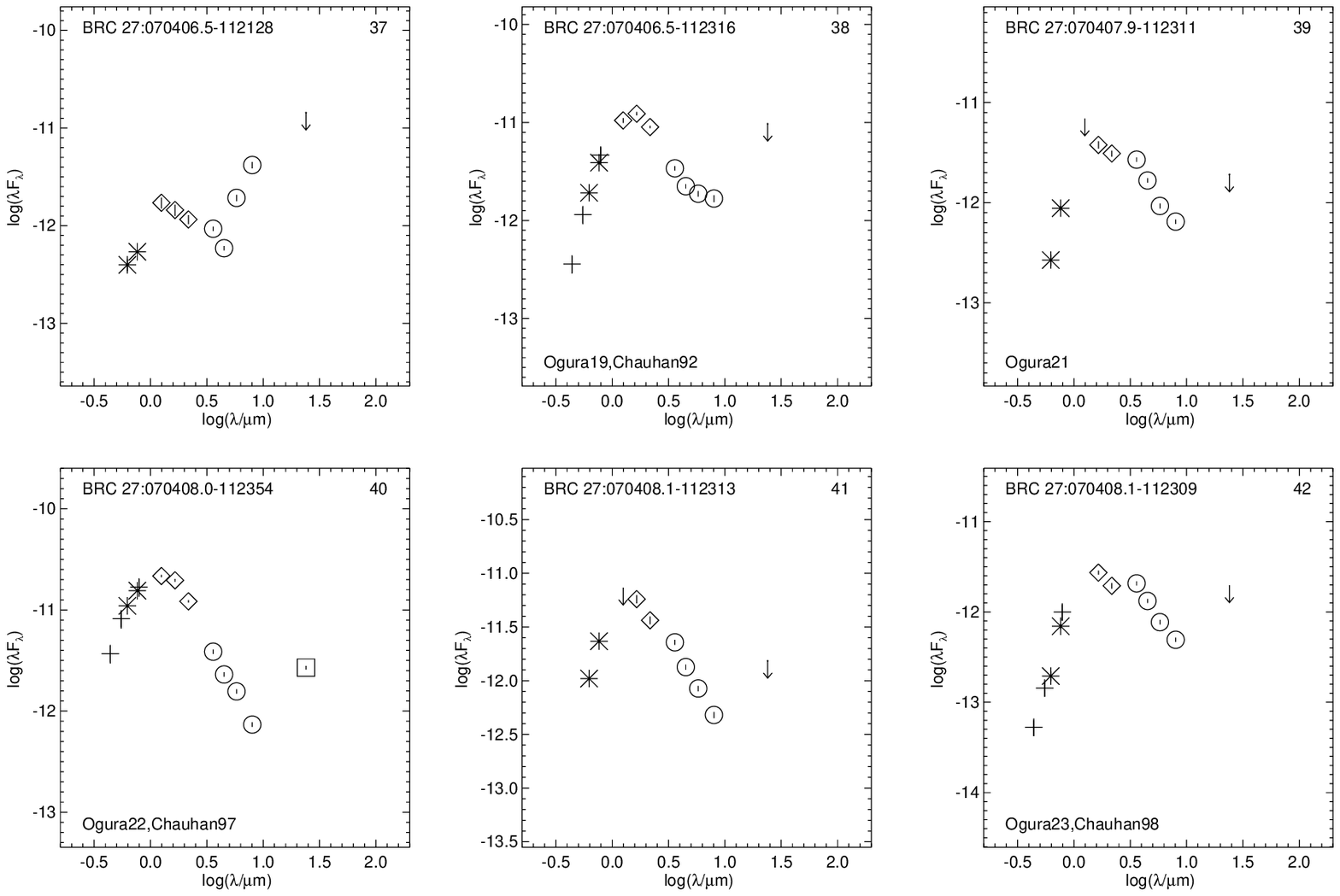}
\caption{SEDs, continued. Notation as in previous figure.}
\label{fig:seds5}
\end{figure*}

\begin{figure*}[tbp]
\epsscale{1}
\plotone{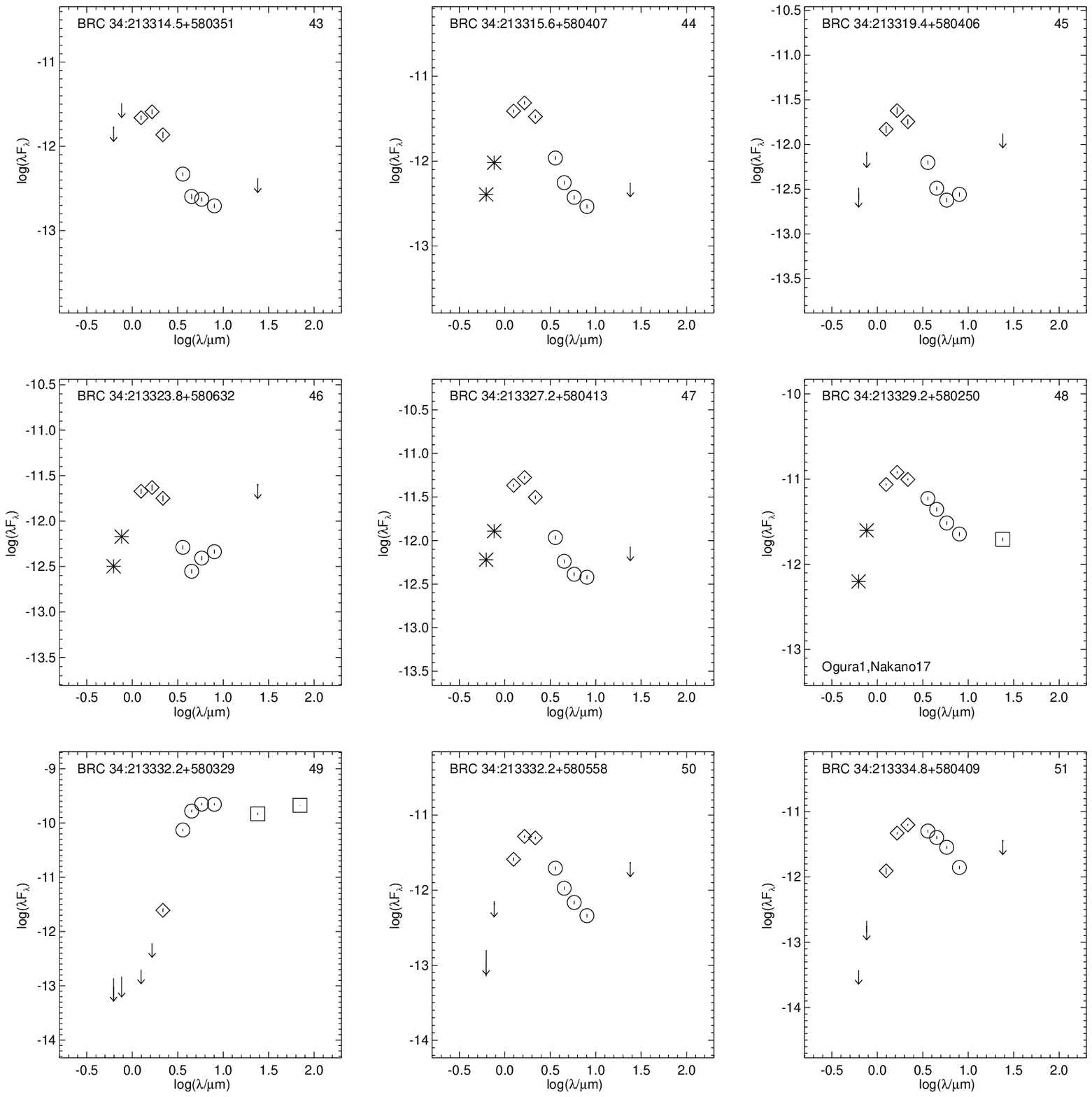}
\caption{SEDs, continued. Notation as in previous figure.}
\label{fig:seds6}
\end{figure*}

\begin{figure*}[tbp]
\epsscale{1}
\plotone{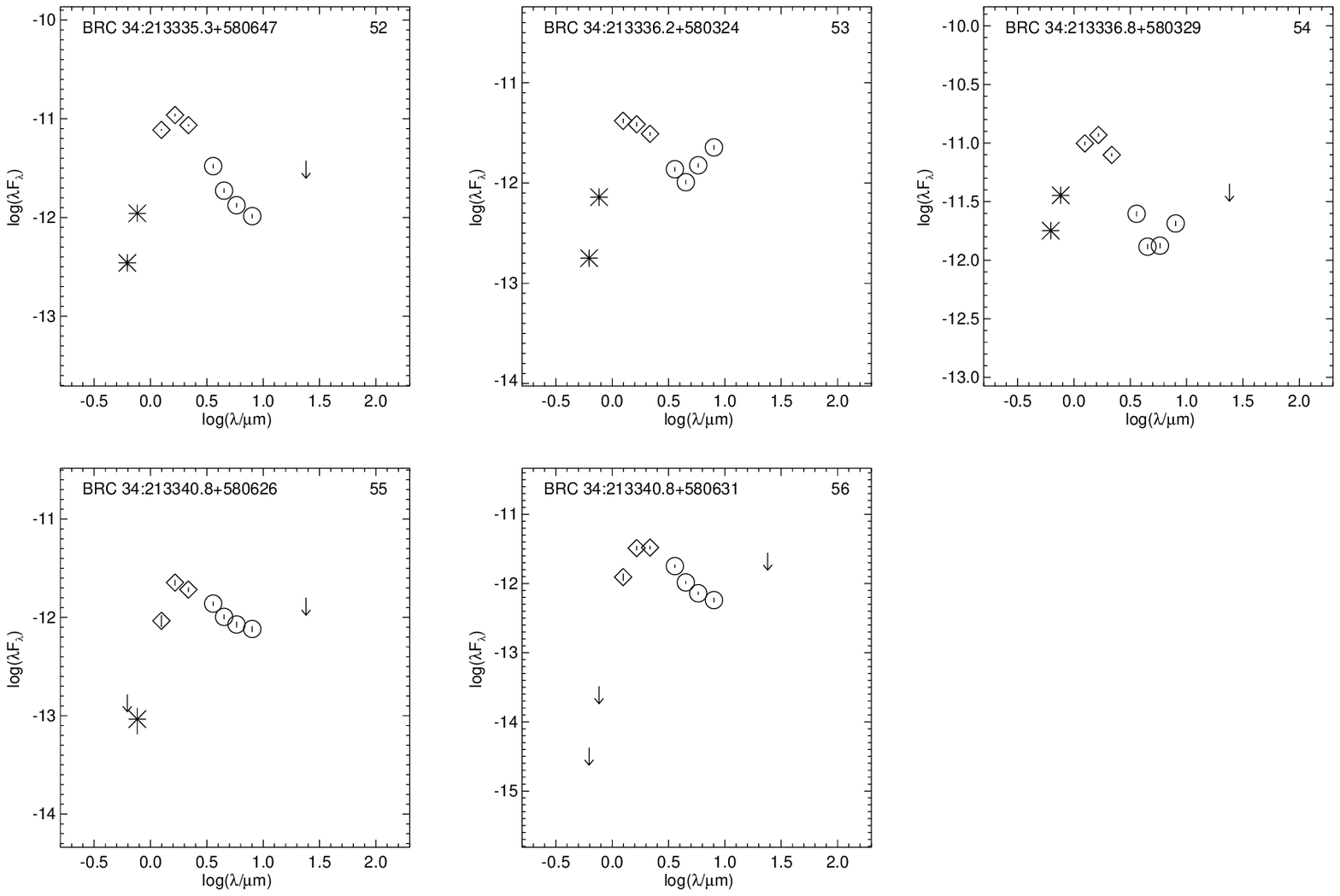}
\caption{SEDs, continued. Notation as in previous figure.}
\label{fig:seds7}
\end{figure*}

\begin{deluxetable}{lll}
\tablecaption{Numbers and Fractions of YSO Classes in BRC 27 and BRC
34\label{tab:ysostats}}
\tablewidth{0pt}
\tablehead{
\colhead{type}  & \colhead{BRC 27} & \colhead{BRC 34} }
\startdata
Class I   &  3 (7\%$^{+6}_{-3}$)  &  1 (7\%$^{+13}_{-2}$) \\
Flat      &  7 (17\%$^{+7}_{-4}$) &  1 (7\%$^{+13}_{-2}$) \\
Class II  & 23 (55\%$^{+7}_{-8}$) &  8 (57\%$^{+11}_{-13}$) \\
Class III &  9 (21\%$^{+8}_{-5}$) &  4 (29\%$^{+14}_{-9}$) \\
\enddata       
\end{deluxetable}

\subsubsection{Comments on Individual SEDs}
\label{sec:sednotes}

Figures~\ref{fig:seds1}--\ref{fig:seds7} are the SEDs for the YSOs and
YSO candidates discussed here.  All of the YSOs and YSO candidates
have SEDs that resemble SEDs from other young stars (see, e.g., Rebull
\etal\ 2010). We now highlight a few SEDs for discussion here; these
notes are summarized in Table~\ref{tab:ourysonotes}.

Three objects (070352.2-112100=Chauhan 109=row 1 in the Tables, 
070354.6-112011=Chauhan 108=row 7, and 070403.1-112327=Chauhan 107=row
24) have literature optical points that are not consistent with the
rest of the SED, including the optical data that we report here. While
young stars are expected to vary in the optical, even if these are all
legitimate young stars, the discrepancy is considerably larger than
would be expected ($\sim$2 orders of magnitude in the SED). These
optical data all come from Chauhan \etal\ (2009). Not all of the
objects with data from Chauhan \etal\ (2009) are inconsistent with the
rest of the respective SED, but all three of these are from Chauhan
\etal\ (2009). As we note above in Section~\ref{sec:litbrc27},
extended emission near these objects can be seen in the 2MASS image,
suggesting that perhaps the optical data from Chauhan \etal\ (2009)
could have included some component due to extended emission. 

One more source, 070401.3-112334 (=Gregorio 74, Chauhan-anon=row 15),
has one optical data point (not from Chauhan \etal\ 2009 but from
Gregorio-Hetem \etal\ 2009) that is not quite aligned with the rest of
the SED. However, it is in the upper range of expected variability for
young stars in the optical, and, since this $R$ photometry reported in
Gregorio-Hetem \etal\ (2009) traces back to the USNO catalog from the
digitization of the POSS (R) plates, uncertainties comparable to this
are possible if not likely.

Four sources from the literature are identified in
Section~\ref{sec:findthem} above as not having significant excesses,
and their SED shapes bear that out. Object 070352.7-112313 (=Ogura 2,
Chauhan 81=row 2) and 070403.9-112609 (=Shevchenko 102=row 25) do not
obviously have an IR excess. Object 070353.5-112350 (=Shevchenko
90=row 4) can be seen in the SED to be very bright; there might be a
weak excess, though as the calculations above bear out, it is not very
significant. Finally, object 070404.5-112555 (=Ogura 13=row 28) can be
seen to have 5.8 and 8 \mum\ points with considerably larger errors
than are in the other SEDs, and a line through those points does not
particularly smoothly join with the rest of the SED, consistent with
the fact that these points are indeed quite uncertain. It could be
that the 4.5 \mum\ point is too low; however, the large errors on the
longer wavelength points lead us to suspect that they are more likely
to be in error. 

One new YSO candidate in BRC 27 and four new YSO candidates in BRC 34
have SEDs very similar to 070404.5-112555 (=row 28) in that the 5.8
and 8 \mum\ points abruptly rise in comparison to the 3.6 and 4.5
\mum\ points; in each of these cases, it could be (also or instead)
that the 4.5 \mum\ point is too low.  For object 070405.7-112123 (=row
31),  the formal errors on the 5.8 and 8 \mum\ are larger than usual,
but not as large as 070404.5-112555 (=Ogura 13=row 28). The four
similar objects from BRC 34 are 213314.5+580351 (=row 43),
213323.8+580632 (=row 46), 213336.2+580324 (=row 53), 213336.8+580329
(=row 54). All of these are not as discontinuous as 070404.5-112555
(=Ogura 13=row 28), and they do not have as large errors on 5.8 and 8
\mum, but the SED shapes are similar. 

Three more objects from BRC 27 have an abrupt rise in the longest
wavelength point available, which could be indicative of a large inner
disk hole. However, the spatial resolution at the longest bands is
worse than at the shorter bands, and this longest wavelength point
could be subject to contamination from the nebula or a nearby (in
projection) background object.  Object 070406.0-112128 (=row 33) and
070408.0-112354 (=Ogura 22, Chauhan 97=row 40) may have their 24 \mum\
measurements contaminated;  070402.3-112539 (=Shevchenko 99, Gregorio
75=row 20) has no 24 \mum\ data, but its 8 \mum\ data may similarly be
subject to contamination.

There is one last set of objects with abrupt changes between two
points adjacent in wavelength. Objects 070407.9-112311 (=Ogura 21=row
39) and  070408.1-112309 (=Ogura 23, Chauhan 98=row 42) both have a
discontinuity between $K_s$ and 3.6 \mum. It is unclear what the
physical origin of such a discontinuity might be, except for intrinsic
stellar variations, and/or errors in the photometry.

Object 070353.8-112341 (=row 6) was mentioned above because it has no
IRAC excess (\S\ref{sec:iracysos}), a marginal 24 \mum\ excess
(\S\ref{sec:3324ysos}), and was bright in $r$ though with large errors
(\S\ref{sec:opticalproperties}). The SED is consistent with all of
these observations; the optical portion of the SED suggests that
either there is a small $r$ excess, or that the measurement is in
error by being too bright. It is unlikely that a YSO with such a small
IR excess would be accreting at a high enough rate to affect $r$ via
veiling. Multiband optical photometry (and spectroscopy, of course)
will clarify what is going on with this object.

Two objects from BRC 27 were called out in Section~\ref{sec:338ysos}
as having small excesses; their SEDs are consistent with that. They
are 070354.9-112514 (=Ogura 5, Chauhan 94=row 8) and 070406.5-112227
(=row 36).

Two objects from BRC 27 and three from BRC 34 particularly seem, from
their SEDs, to be subject to high \av. All of them were also
identified in Section~\ref{sec:jhkysos} as having large \av. They are
070400.7-112323 (=row 11), 070401.2-112233 (=Chauhan-anon=row 14),
213334.8+580409 (=row 51), 213340.8+580626 (=row 55), and
213340.8+580631 (=row 56).

Four YSOs and YSO candidates in BRC 27 and one in BRC 34 have SEDs
characteristic of reasonably deeply embedded YSOs. They are object
070401.2-112242 (=row 13), 070402.7-112325 (=row 21), 070402.9-112337
(=Ogura 8+9, Chauhan 84=row 22), 070403.0-112350 (=Ogura 10, Chauhan
85=row 23), and (in BRC 34) 213332.2+580329 (=row 49). This last one
in BRC 34 is the only object in either BRC detected at 70 \mum, and it
is very bright, with [70]$\sim-2$. We suspect that this will turn out
to be a legitimate YSO and likely the youngest object of the ensemble
we discuss here. It also seems to be centered within the globule of
dust being illuminated to form the bright rimmed cloud (see
Figure~\ref{fig:brc34mips12} or \ref{fig:brc34-3color}). It is also
subject to so much reddening that there is no $J$ or $H$ detected for
this object. Follow-up of this object will be difficult given the high
\av. The objects in BRC 27 have roughly similarly shaped SEDs. Object
070401.2-112242 (=row 13), a new YSO candidate, is not detected at $r$
or $i$ but does have $JHK_s$ measurements. Object 070402.7-112325
(=row 21), another new YSO candidate, is not detected at $r$, $i$, or
$J$. Object 070402.9-112337 (=Ogura 8+9, Chauhan 84=row 22) is a
literature YSO candidate consisting of two sources unresolved in the
2MASS and IRAC data. This one is detected at the optical and NIR
bands, but the overall SED shape is roughly similar to the others.
Object 070403.0-112350 (=Ogura 10, Chauhan 85=row 23), similarly, is
detected at the optical and NIR bands.

Somewhat similarly to the objects in the prior paragraph,
070403.9-112326 (=row 26) seems to have a substantial IR excess, and
is not detected in the optical. It is possible in this case that are
are detecting photosphere at $JHK_s[3.6]$, but if that were the case,
substantial extinction would be needed. It could be a nearly edge-on
disk seen mostly in scattered light at the shorter bands. However,
this SED is also consistent with an background galaxy subject to
extinction.
Follow-up spectra would be very helpful in determining the nature of
this object.

Three SEDs are somewhat unusually shaped. Objects  070401.6-112406
(=row 16), 070402.2-112542 (=row 19), and 070406.5-112128 (=row 37)
are all new YSO candidates from BRC 37. Two of them, 070401.6-112406
(=row 16) and 070406.5-112128 (=row 37), have optical detections that
place them below the ZAMS (Section~\ref{sec:opticalproperties}), but
they are located right on the bright rim itself
(Section~\ref{sec:locationonthesky}), a suggestive location for young
stars. Object 070401.6-112406 (=row 16) is a more or less steadily
rising SED, which could be a background galaxy subject to extinction,
or a YSO. This object also, in the $J$ through [4.5] images, has
nearby emission that could be from the nebula or could be from nearby
point sources. The morphology of the emission around this source is
complex, particularly in the [3.6] and [4.5] bands, where there seems
to be a mixture of point and extended emission within 2-10$\arcsec$;
higher spatial resolution observations would be very useful for this
object.  In contrast to 070401.6-112406 (=row 16), 070402.2-112542
(=row 19) is bright in $r$ (Section \ref{sec:opticalproperties}). It
has a point at 8 \mum\ that is higher than the 5.8 and 24 \mum\ points
in the SED, very suggestive of PAH emission that could be a background
galaxy (see ``8 micron pop-ups'' in Rebull \etal\ 2010), but it could
also be a young star. It is very bright in the image, though fainter
than another nearby bright star (070402.3-112539=Shevchenko 99,
Gregorio 75=row 20), so photometry is difficult. This object may be a
legitimate infrared-bright  companion to this bright object (like in
WL 20, Ressler \& Barsony 2001). Follow-up observations are warranted.

All the rest of the YSOs and YSO candidates have SEDs that are
completely consistent with young stars having IR excesses, but not
deserving of special comment here.

\subsubsection{SED Classifications}
\label{sec:sedslopes}

Following the discussion in Section~\ref{sec:evol}, in the spirit of
Wilking \etal\ (2001; see also Lada \& Wilking 1984,  Lada 1987, 
Greene \etal\ 1994, and Bachiller 1996), we define the near- to mid-IR
(2 to 24 \mum) slope of the SED, $\alpha = d \log \lambda
F_{\lambda}/d \log  \lambda$,  where  $\alpha > 0.3$ for a Class I,
0.3 to $-$0.3 for a flat-spectrum  source, $-$0.3 to $-$1.6 for a
Class II, and $<-$1.6 for a Class III.  For each of the YSOs and
candidate YSOs in our sample, we performed a simple ordinary least
squares linear fit to all available photometry (just detections, not
including upper or lower limits) as observed between 2 and 24 $\mu$m,
inclusive.  Note that: (a) the formal errors on the infrared points
are so small as to not affect the fitted SED slope; (b) the fit is
performed on the observed SED, meaning that no reddening corrections
are applied to the observed photometry before fitting; (c) the fit is
performed on the observed SED, meaning that if there is no 24 \mum\
data point, then that point is necessarily not included in the fit. 
In the literature, the precise definition of $\alpha$ can vary, which
may result in different classifications for certain objects.
Classification via this method is provided specifically to enable
comparison within this paper (and to CG4+Sa101) via internally
consistent means. Note that the formal classification puts no lower
limit on the colors of Class III objects (thereby including those with
SEDs resembling bare stellar photospheres, and allowing for other
criteria to define youth).  By searching for IR excesses, we are
incomplete in our sample of Class III objects.   The SED slopes and
classes for the YSO and YSO candidates discussed here appear in
Table~\ref{tab:ourysonotes}. 

Of all the evolutionary stages of YSOs among the YSOs or YSO
candidates discussed here, Class I is the shortest lived, and
therefore the rarest of all the SED classes.  There are four objects,
all new YSO candidates, that have an SED class of I.  Three of these
objects are in BRC 27, and one is in BRC 34. Of the three objects in
BRC 27, two are possible (if not likely) background objects (see
Section~\ref{sec:sednotes}); these are 070401.6-112406 (=row 16) and
070406.5-112128 (=row 37), and they are the ones below the main
sequence in Figure~\ref{fig:optysos} (and
Section~\ref{sec:opticalproperties}.  The other two new Class I
candidates are 070403.9-112326 (=row 26) and 213332.2+580329 (=row
49), the latter of which is the bright source seen at [70] in BRC 34. 

Several objects are on the borderline between SED classes, meaning
that their fitted SED slope is within 0.1 of the dividing line between
the classes as defined just above.  These objects may be members of
adjacent classes at a different inclination (see discussion above in 
Section~\ref{sec:evol}); the addition of a new point at 24 \mum\ (or a
new detection where any source confusion is resolved) may also change
the classification of some of these objects.  There are six of these
borderline objects in BRC 27:
070403.9-112326 (=row 26, on the borderline between Class I and Flat),
070401.7-112323 (=row 11, on the borderline between Flat and Class II; in
this case, a different [24] could make a big difference),
070405.1-112313 (=row 30=Ogura 15, Chauhan 88, on the borderline between Flat and Class II),
070404.5-112555 (=row 28=Ogura 13, on the borderline between Class II and III; we have
tagged this one as not having an IR excess due to the uncertainty of the 5.8 and 8 \mum\ points,
so this ends up as a Class II [rather than a III] with no excess),
070406.4-112336 (=row 35=Ogura 18, Chauhan 91, on the borderline between Class II and III),
and
070408.1-112313 (=row 41, on the borderline between Class II and III).
In BRC 34, there are four borderline objects:
213336.2+580324 (=row 53, on the borderline between Flat and Class II),
and the remaining three are all on the borderline between Class II and III:
213314.5+580351 (=row 43),
213319.4+580406 (=row 45, and in this case a different measurement at [8] might
directly impact the slope), and
213335.3+580647 (=row 52).  
When the SED slope fit is formally made to all available measurements
between 2 and 8 \mum, rather than all available points between 2 and
24 \mum, the slopes change enough to change the classifications for
just 6 of the 56 total objects (11\%) considered here (out of the 19
with 24 \mum detections, 32\%), all of which are in BRC 27, and none
of which are listed as borderline cases above. These are: 
070401.2-112242 (=row 13, which changes from a Flat to Class II),
070402.2-112542 (=row 19, which changes from a Flat to Class I)
070402.9-112337 (=row 22=Ogura 8+9=Chauhan 84, which changes from a Flat to Class II)
070406.0-112128 (=row 33, which changes from a Class II to Class III)
070406.0-112315 (=row 34=Ogura 17=Chauhan 90, which changes from a Flat to Class II)
070408.0-112354 (=row 40=Ogura 22=Chauhan 97, which changes from a
Class II to Class III).

Taking all of the slopes and classifications at face value,
Table~\ref{tab:ysostats} summarizes the fraction of Class I, Flat,
Class IIs and Class IIIs (such as we know them) for each of the two
BRCs. The errors on these disk fractions as tabulated in
Table~\ref{tab:ysostats} were calculated using the binomial
distribution, as per Burgasser \etal\ (2003).  The largest calculated
fraction of objects in BRC 27 is Class II, at $\sim$55\%. The largest
fraction of objects in BRC 34 is also Class II objects, at a quite
comparable $\sim$57\%, but with so many fewer YSOs in BRC 34, the
uncertainty on this fraction is larger than for BRC 27.  For both
BRCs, the next largest fraction of objects is Class III, at 21\% and
29\% for BRC 27 and 34, respectively. These are consistent with each
other within the errors, and also certainly our sample of Class III
objects is incomplete in both BRCs. The fraction of Class I+Flat
spectrum objects in BRC 27 (10/42, or 24\%; incorporating errors, this
value could be between 19 and 31\%) is consistent within errors with
that for BRC 34 (2/14, or 14\%, incorporating errors, this value could
be between 9 and 28\%). It is very difficult to draw any conclusions
about relative ages of the two BRCs. They are identical within
small-number statistics; moreover our samples are incomplete, and
include unconfirmed YSOs. Further spectroscopy is desirable.

In CG4+Sa101, there are only 7 YSOs or YSO candidates near the BRC
itself (CG4); there are 15 more nominally associated with Sa101.
Considering the ensemble of objects in CG4+Sa101, 73\% (incorporating
errors, this value could be between  62\%-80\%) are Class II objects;
for just CG4, 71\% (incorporating errors, this value could be between
51\%-82\%) are Class II objects.  For either sample, this is a
slightly larger fraction than found in either BRC 27 or BRC 34,
suggesting that CG4+Sa101 could be slightly older.

\subsection{Location on the Sky}
\label{sec:locationonthesky}

\begin{figure*}[tbp]
\epsscale{1}
\plotone{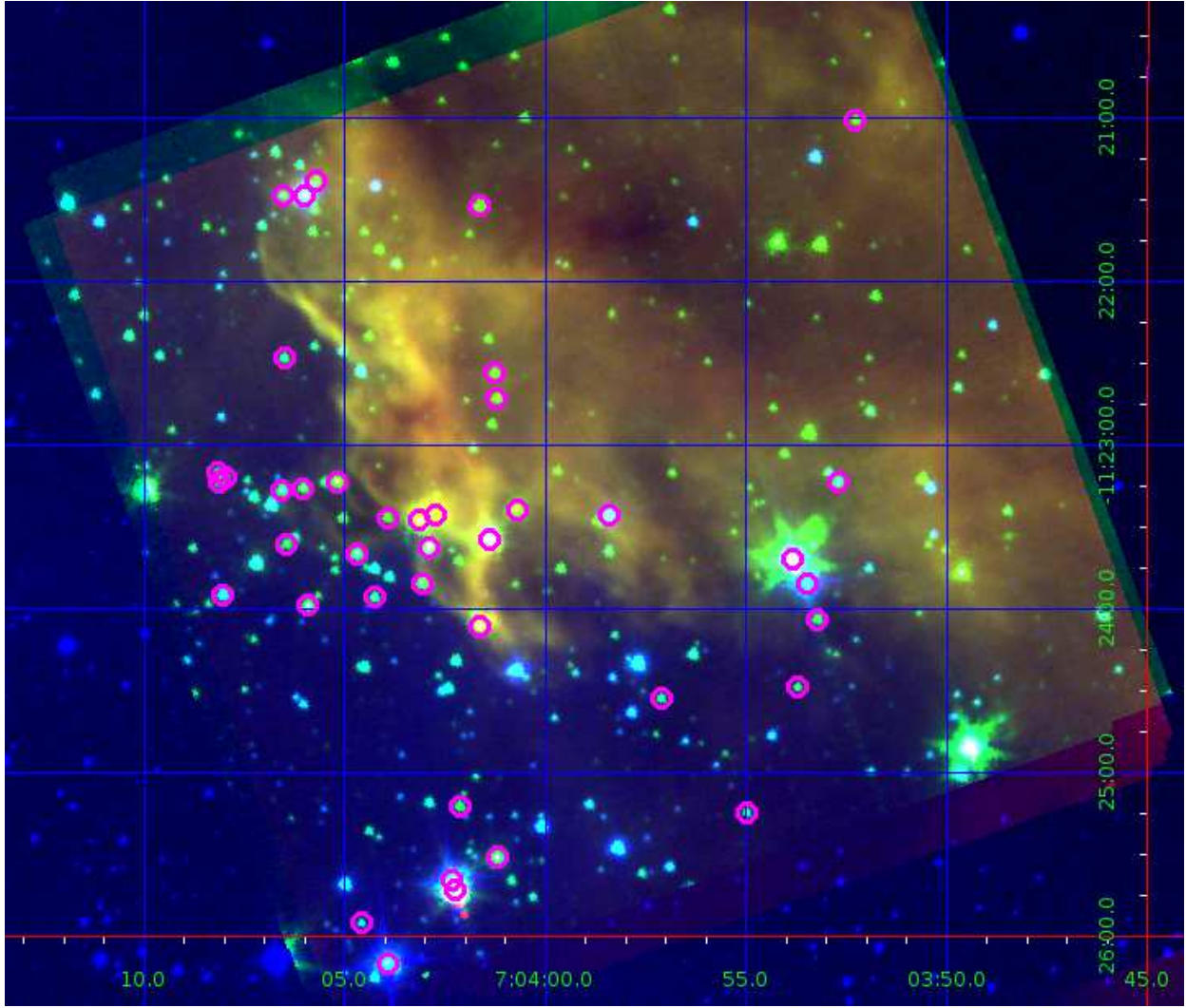}
\caption{Three-color image of BRC 27 (blue=$i$, green=3.6 \mum, red=8
\mum).  The YSOs (known and candidate) are indicated by
additional magenta circles. North is up, and coordinates are
indicated. Young stars are statistically more likely to be clustered,
and associated with the nebulosity, than background or foreground
contaminants. }
\label{fig:brc27-3color}
\end{figure*}

\begin{figure*}[tbp]
\epsscale{1}
\plotone{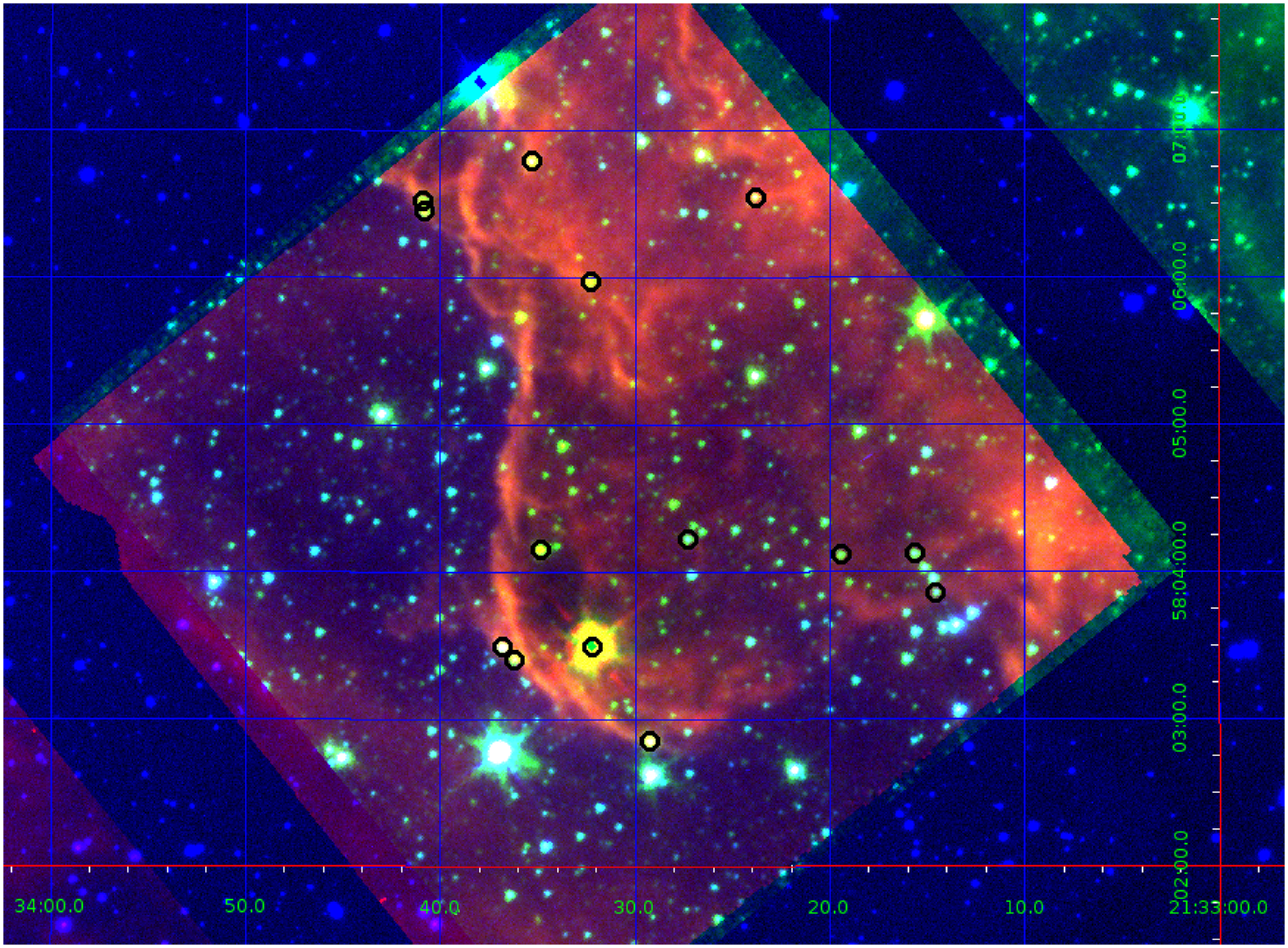}
\caption{Three-color image of BRC 34 (blue=$i$, green=3.6 \mum, red=8
\mum).  The YSOs (known and candidate) are indicated by
additional black circles. North is up, and coordinates are
indicated. Young stars are statistically more likely to be clustered,
and associated with the nebulosity, than background or foreground
contaminants.}
\label{fig:brc34-3color}
\end{figure*}

\begin{figure*}[tbp]
\epsscale{1}
\plotone{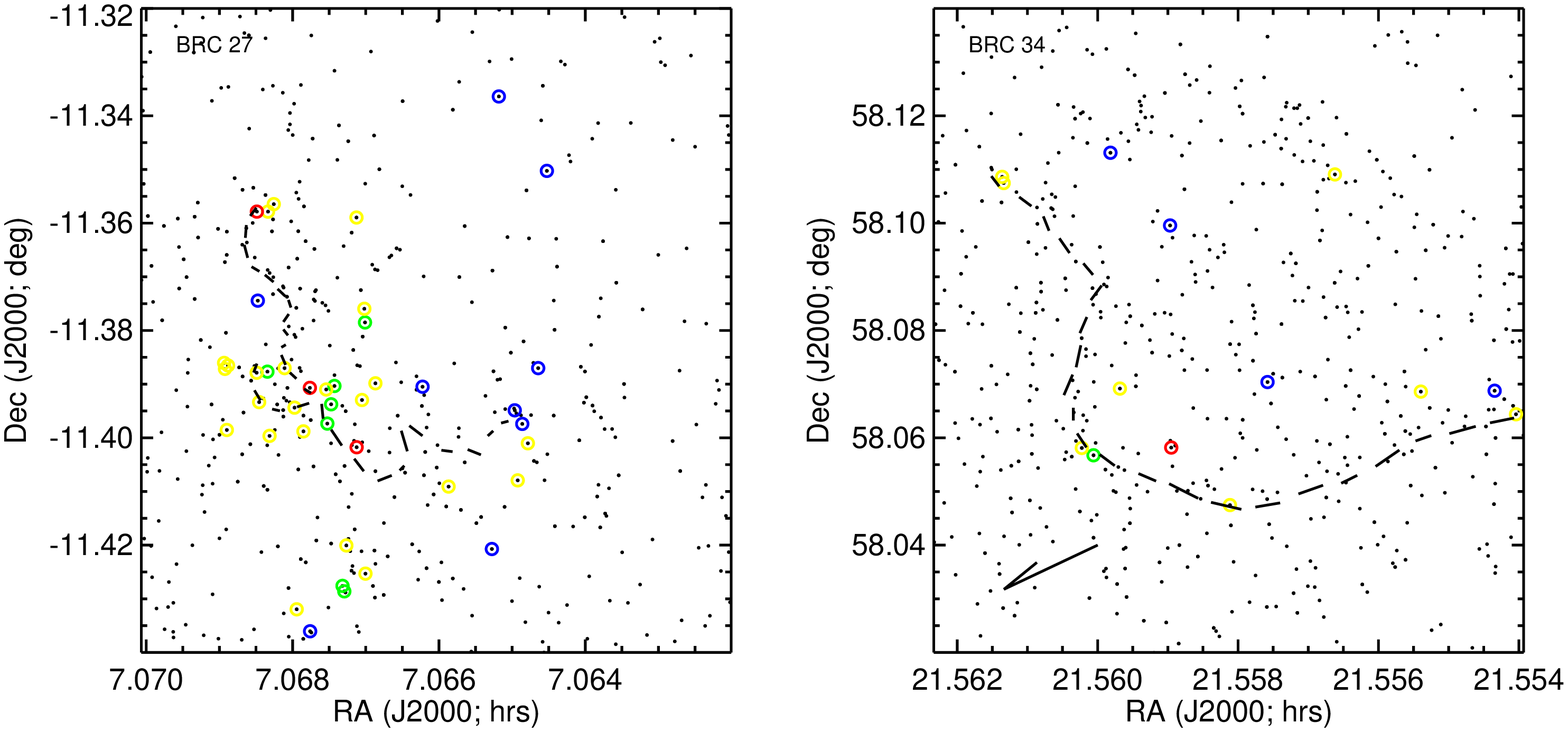}
\caption{Positions of the objects in both BRCs on the sky, color-coded
by shape of the SED -- blue is Class III, green is Class II, yellow is
Flat, and red is Class I. Small black dots are objects in the 2MASS
point source catalog. the black dashed lines are the approximate edges
of the BRC, as seen in Figure~\ref{fig:brc27-3color} and
\ref{fig:brc34-3color}. While the source of the shock front is
uncertain for BRC 27, it is probably to the Southeast.  The source is
known for BRC 34; the source (HD 206267) is about a degree away in the
direction of the arrow as shown. There is some evidence in the
literature that more embedded objects (earlier classes) should be
further from the exciting source. There are too few sources here to
assess this. }
\label{fig:positions}
\end{figure*}

Figure~\ref{fig:brc27-3color} and \ref{fig:brc34-3color} both show
3-color images with the selected set of YSOs indicated. Young stars
are statistically more likely to be clustered, and associated with the
nebulosity, than background or foreground contaminants. Certainly,
this is not a strict rule; young stars can appear off the cloud and
contaminants can be seen through the cloud, superimposed on the cloud,
and/or be clustered.

Figure~\ref{fig:positions} shows the position of the objects with
color coding corresponding to the shape of the SED.  
The source of the shock front is not known in BRC 27, but it probably
to the Southeast, given the shape of the nebula.  The source of the
shock is known in BRC 34 to be HD 206267, which is about a degree
away, also to the Southeast. The eponymous ``bright rim'' is more
obvious in the MIR in BRC 34 than BRC 27.  The rim of BRC 27 is less
well-defined and has more texture in the images, perhaps indicating
the effects of more than one exciting OB star, or a distribution of
densities in the molecular matter composing the BRC. 

In BRC 27, there are 3 times more objects than BRC 34, with a YSO
surface density of $\sim$1.7 objects arcmin$^{-1}$, as compared with
$\sim$0.6 objects arcmin$^{-1}$ in BRC 34.   In BRC 27, there seems to
be a cluster of objects just off the dark cloud, near $(\alpha,
\delta)$=(106.02$\arcdeg$, $-$11.40$\arcdeg$)=(07$^h$05$^m$,
$-$11$\arcdeg$23$\arcmin$05$\arcsec$). They may have been recently
revealed by the action of the shock front; they are largely Class II
objects, consistent with this idea, but not conclusive proof of it.
All of the Class I candidate objects are on or behind the bright rim,
again consistent with (but not proof) that younger objects are found
within the dark cloud.  If the shock front is triggering star
formation in the clump, one might expect that all of the Class I
objects would be deep inside the cloud, with the least embedded
objects having been revealed by the ionization front in front of the
cloud. Both clouds have Class III objects projected onto the cloud,
though the Class III objects are the least complete sample and most
uncertain membership.   In BRC 34, many of the YSOs are along the
bright rim itself, whereas in BRC 27, they are more dispersed with
respect to the distribution of bright ISM, with more YSOs or
candidates off the dark cloud, on the excitation side of the bright
rim. If this is, in both cases, a wave of star formation that moves
through the BRC, this would be consistent with BRC 27 being slightly
older than BRC 34. However, with so few objects, and with the various
uncertainties that have gone into selecting our sample, we cannot make
any clear statements about whether there is, in fact, an age gradient
through the BRCs, or evidence for small-scale sequential star
formation.

In CG4+Sa101, the YSOs or candidate YSOs formally associated with
Sa101 are relatively tightly clustered, with a median nearest neighbor
distance of 62$\arcsec$; in the CG4 subregion, the median nearest
neighbor distance is five times larger. A similar kind of calculation
in the BRCs is more difficult, because the maps are so much smaller,
and because there is not as obvious a clumping in the image. For both
BRCs, the median nearest neighbor distance is $\sim$10$\arcsec$.
Limiting this calculation to just the clump of objects in BRC 27
between 7.0665 and 7.069 hrs of RA and $-$11.38 and $-$11.40 deg of
Dec obtains approximately the same median separation, presumably
because this clump dominates the statistics. Surveys of a larger
region around these BRCs will help reveal any clustering.

\section{Conclusions}
\label{sec:concl}

We used Spitzer Space Telescope data from the Spitzer Heritage Archive
to search for new candidate young stars in two BRC regions, BRC 27
(part of CMa R1) and BRC 34 (part of the IC 1396 complex). These regions both
appear to be actively forming young stars, perhaps triggered by the
proximate OB stars.

We have presented Spitzer and optical data for 42 YSOs, literature YSO
candidates, and new YSO candidates in BRC 27. Out of those, we
identify 22 of the 26 literature YSOs or literature YSO candidates as
having an IR excess, though one of them has an uncertain IR excess.
There are 16 new YSO
candidates that we have identified from their Spitzer colors, although
3 of them have somewhat uncertain IR excesses.

Similarly, we have presented Spitzer and optical data for 14 YSO
candidates (including one known YSO) in BRC 34. The one
known YSO in BRC 34 has a clear IR excess, and there are
13 additional new objects we have identified as candidate YSOs from
their Spitzer colors. Of those 13, 3 have somewhat uncertain IR
excesses; one is a likely Class I and is the only 70 \mum\ point
source detection in either BRC. 

As far as we can determine, these objects have properties in the
optical, NIR, and MIR that suggest that they are YSOs. However,
follow-up spectroscopy is needed to affirm or refute their YSO status.

Assuming that these YSO candidates are all legitimate YSOs, in BRC 27,
there are 3 times more objects than BRC 34, with a YSO surface density
of $\sim$1.7 objects arcmin$^{-1}$, as compared with $\sim$0.6 objects
arcmin$^{-1}$ in BRC 34.   Considering the entire ensemble, both BRCs
are likely of comparable ages, based primarily on SED class ratios.
Within small-number statistics, and the fact that our samples are
probably incomplete and include unconfirmed YSOs, no definitive
statement can be made about the relative ages of the ensemble of YSOs
in these BRCs. However, they both seem to have more objects at an
earlier evolutionary stage than another BRC, CG4+Sa101 (BRC 48).
Similarly, no clear conclusions can be drawn about any possible age
gradients that may be present. 

We plan to continue this project using Wide-field Infrared Survey
Explorer (WISE; Wright \etal\ 2010) to investigate the YSO surface
density in a wider area around these BRCs, as well as pursue follow-up
spectroscopy.

\acknowledgements 

This work was performed as part of the NASA/IPAC Teacher Archive
Research Program (NITARP; http://nitarp.ipac.caltech.edu), class of
2011 (Round 5).  We acknowledge here all of the students, scientists,
and staff who contribute their time and energy to NITARP. 

This research has made use of NASA's Astrophysics Data System (ADS)
Abstract Service, and of the SIMBAD database, operated at CDS,
Strasbourg, France.  This research has made use of data products from
the Two Micron All-Sky Survey (2MASS), which is a joint project of the
University of Massachusetts and the Infrared Processing and Analysis
Center, funded by the National Aeronautics and Space Administration
and the National Science Foundation.  These data were served by the
NASA/IPAC Infrared Science Archive, which is operated by the Jet
Propulsion Laboratory, California Institute of Technology, under
contract with the National Aeronautics and Space Administration.  This
research has made use of the Digitized Sky Surveys, which were
produced at the Space Telescope Science Institute under U.S.
Government grant NAG W-2166. The images of these surveys are based on
photographic data obtained using the Oschin Schmidt Telescope on
Palomar Mountain and the UK Schmidt Telescope. The plates were
processed into the present compressed digital form with the permission
of these institutions. This research has made use of the NASA/IPAC
Extragalactic  Database (NED) which is operated by the Jet Propulsion 
Laboratory, California Institute of Technology, under  contract with
the National Aeronautics and Space Administration. 

The research described in this paper was partially carried out at the
Jet Propulsion Laboratory, California Institute of Technology, under
contract with the National Aeronautics and Space Administration.

\end{document}